\newcommand{\prc}{{Phys.~Rev.~C}}

\documentclass[epj,plus]{svjour}
\usepackage{graphics}
\usepackage{graphicx}
\usepackage{upgreek}
\usepackage{color}
\begin{document}
\title{Neutron-induced cross sections}
\subtitle{From raw data to astrophysical rates}
\author{
Ren\'e Reifarth \inst{1} \fnmsep \thanks{\email{reifarth@physik.uni-frankfurt.de}} \and
Philipp~Erbacher \inst{1} \and
Stefan~Fiebiger \inst{1} \and
Kathrin~G\"obel \inst{1} \and
Tanja~Heftrich \inst{1} \and
Michael~Heil \inst{2} \and
Franz~K\"appeler \inst{3} \and
Nadine~Klapper \inst{1} \and
Deniz~Kurtulgil \inst{1} \and
Christoph~Langer \inst{1} \and
Claudia~Lederer-Woods \inst{4}  \and
Alberto Mengoni \inst{5,6}  \and
Benedikt~Thomas \inst{1} \and
Stefan~Schmidt \inst{7} \and
Mario~Weigand \inst{1}  \and
Michael~Wiescher \inst{8} 
}                     % Do not remove
\institute{Goethe University Frankfurt am Main, Frankfurt, Germany \and
GSI Helmholtzzentrum f\"ur Schwerionenforschung, Darmstadt, Germany \and
Karlsruhe Institute of Technology, Karlsruhe, Germany \and
School of Physics and Astronomy, University of Edinburgh, Edinburgh, UK \and
ENEA, Bologna, Italy \and
INFN, Sezione di Bologna, Italy \and
Frankfurt Institute for Advanced Studies, Frankfurt, Germany \and
University of Notre Dame, Notre Dame, IN, USA
}
\date{Received: date / Revised version: date}
\abstract{
Neutron capture cross sections are one of the most important nuclear inputs to models of stellar nucleosynthesis of the elements heavier than iron. The activation technique and the time-of-flight method are mostly used to determine the required data experimentally. Recent developments of experimental techniques allow for new experiments on radioactive isotopes. Monte-Carlo based analysis methods give new insights into the systematic uncertainties of previous measurements. We present an overview over the state-of-the-art experimental techniques, a detailed new evaluation of the $^{197}$Au(n,$\gamma$) cross section in the keV-regime and the corresponding re-evaluation of 63 more isotopes, which have been measured in the past relative to the gold cross section.
\PACS{
	{29.87.+g}{Nuclear data compilation} \and
    {28.20.Fc}{Neutron absorption}   \and
    {29.25.Dz}{Neutron sources} \and      
    {29.30.Kv}{X- and g-ray spectroscopy} \and
    {28.20.Np}{Neutron capture g-rays}      
    } % end of PACS codes
} %end of abstract
\maketitle

\tableofcontents

\section{Introduction}
\label{intro}

In astrophysics the neutron energy range between 1~keV and 1~MeV is most important, 
because 
it corresponds to the temperature regimes of the relevant sites for synthesizing 
all nuclei between iron and the actinides \cite{RLK14}. In this context (n,$\gamma$) cross sections 
for unstable isotopes are requested for the s process related to stellar helium and carbon
burning as well as for the r and p processes related to explosive nucleosynthesis. In the s process, these data are required for analyzing 
branchings in the reaction path, which can be interpreted as diagnostic tools for 
the physical state of the stellar plasma \cite{KTG16}. Most of the nucleosynthesis reactions during 
the r and p processes occur outside the stability valley, involving rather 
short-lived nuclei. Here, the challenge for (n,$\gamma$) data is linked to the freeze-out 
of the final abundance pattern, when the remaining free neutrons are captured as 
the temperature drops below the binding energy \cite{HAC18,SuE01,GGK15}. Since 
many of these nuclei are too short-lived to be accessed by direct measurements, it 
is essential to obtain as much experimental information as possible 
off the stability line in order to assist theoretical extrapolations of nuclear 
properties towards the drip lines. 

Apart from the astrophysical motivation there is continuing interest on neutron 
cross sections for technological applications, i.e. with respect to the neutron 
balance in advanced reactors, which are aiming at high burn-up rates, as well as 
for concepts dealing with transmutation of radioactive wastes.

The astrophysical reaction rate $r$ as a function of the number density $N_x, N_y$ of (non-identical) particles is \cite{RaT00} 

\begin{equation}
           r = N_x N_y \langle\sigma v\rangle
\end{equation}
with the reaction rate per particle pair $\langle\sigma v\rangle$ depending on the reaction cross section $\sigma(v)$ and the velocity distribution $\phi(v)$

\begin{equation}
    \langle\sigma v\rangle = \int_0^\infty{\sigma(v) v \phi(v) \mbox{d}v}~.
\end{equation}
The velocity $v$ of neutrons and nuclei in thermalized environments can be described with the 
Maxwell-Boltzmann distribution

\begin{equation}\label{maxwell-boltzmann}
\phi_\mathrm{MB}(v)\mbox{d}v = \left(\frac{m}{2\pi kT}\right)^{3/2}~ \exp{\left(-\frac{mv^2}{2kT}\right)} 4\pi v^2 \mbox{d}v~.
\end{equation}
It follows for the Maxwellian Averaged Cross Section ($MACS$) 
\begin{equation}
   MACS = \frac{\langle\sigma v\rangle_\mathrm{MB}}{v_T}=\frac{\int_0^\infty{\sigma(v) v \phi_\mathrm{MB}(v) \mbox{d}v}}{v_T}
\end{equation}  
with 
\begin{equation}
   \frac{3}{2}kT = \bar{E}_\mathrm{kin}=\frac{m}{2}v_T^2
\end{equation}  
or as a function of energy \cite{RLK14}:

\begin{equation}\label{MACS}
   MACS = \frac{2}{\sqrt{\pi}} \frac{1}{(kT)^{2}}\cdot \int_0^\infty E \sigma(E) \cdot \exp{\left(-\frac{E}{kT}\right)} \mbox{d}E
\end{equation}  
The $MACS$ is needed for temperatures $kT$ between about 5~keV and 100~keV. Therefore the energy-dependent cross sections $\sigma(E)$ are required between about 1~keV and 1~MeV. 

In this article, the two main techniques to determine neutron-induced cross sections are described. Integral methods are usually based on the activation technique (section \ref{sec:integral}) while the determination of energy-differential cross section is in most cases based on the time-of-flight method (section~\ref{sec:differential}). 
Current challenges and developments are discussed. In section~\ref{sec:au} we discuss the possible solution to a long-standing puzzle, the disagreement between two high-precision measurements of the important neutron capture cross section of $^{197}$Au. One of the measurements was based on the activation of gold in a standardized neutron field while the others are time-of-flight measurements. We recommend a new Maxwellian-averaged cross section for $^{197}$Au(n,$\gamma$). Since many of the cross section measurements in the past used gold as a reference, we present a re-evaluation of 63 neutron capture measurements based on the time-of-flight method performed at Forschungszentrum Karlsruhe between 1990 and 2000, (section~\ref{sec:renormalization}). Section~\ref{sec:indirect} covers some experimental methods, which are helping to bridge the gap between isotopes where the direct methods can be applied and the astrophysically driven needs for data on (short-lived) radioactive isotopes. 

\section{Integral measurements} \label{sec:integral}

\subsection{General idea}

The neutron capture on a nucleus can be expressed as

\begin{equation}
^{A}X + \mathrm{n} \rightarrow ^{A+1}X^{*} 
\end{equation}

where $^{A}X$ stands for the isotope with mass $A$ of the element $X$. The star in the reaction product symbolizes the fact that the nucleus will be in an excited state after the fusion with the neutron. If it de-excites via $\gamma$-emission,

\begin{equation}
^{A+1}X^{*}  \rightarrow ^{A+1}X + \gamma,
\end{equation}

the neutron is captured. The detection of those promptly emitted $\gamma$-rays is the main idea of the time-of-flight method (TOF) described in section \ref{sec:differential}. If the freshly produced nucleus $^{A+1}X$ is radioactive, it will decay following the exponential decay law. The particles emitted during the delayed decay, e.g.

\begin{equation}
^{A+1}X \rightarrow ^{A+1}Y^{*} + \beta^{-} \rightarrow ^{A+1}Y + \gamma,
\end{equation}

can be detected after the neutron irradiation. This approach is called the activation technique - it always consists of two distinctly different phases: irradiation of the sample and detection of the freshly produced nuclei.

\begin{figure}
\begin{center}
\renewcommand{\baselinestretch}{1}
  \includegraphics[width=0.49\textwidth]{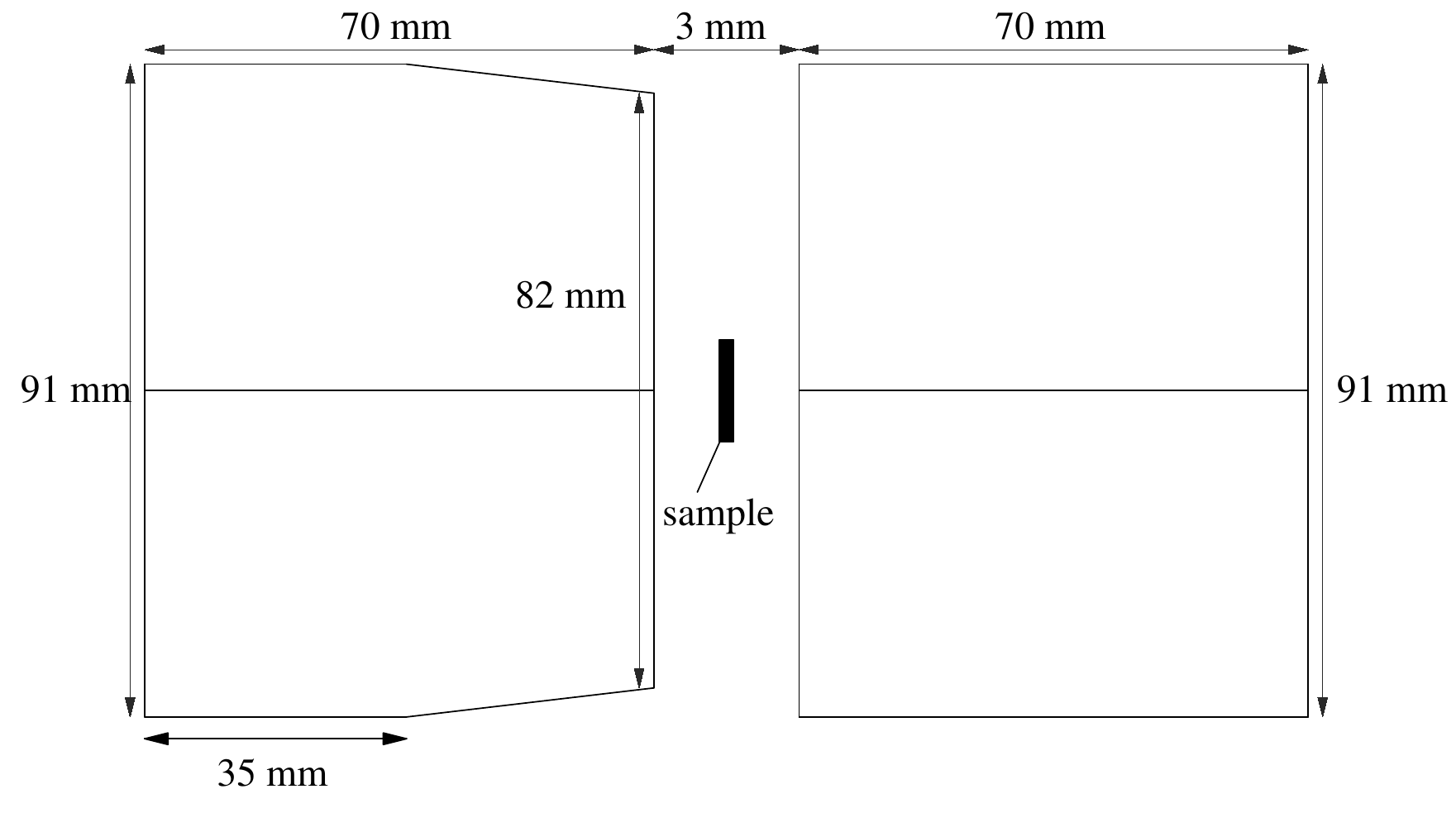}
  \caption{\label{fig:clover_setup} Schematic view of a possible $\gamma$-detection system consisting of two
Clover-type detectors in close geometry \cite{RDH12}. }
\end{center}
\end{figure}

There are several huge advantages of the activation technique over the TOF method. First, the neutron flux at the sample is typically about 5 orders of magnitude higher, because the sample can be very close to the neutron source and the neutron production does not need to be pulsed. Second, the detection setup can be in a low-background environment with very sensitive equipment. 
An example is the use of a 4$\pi$-setup of high purity germanium 4-fold clover detectors for $\gamma$-detection, see Fig.~\ref{fig:clover_setup} \cite{RDH12}. An additional advantage are the low demands on the sample purity. Usually sample material with natural composition can be used. Very often, more than one isotope can be investigated simultaneously with one sample as in the example of the activation of natural Zn \cite{RDH12}. The last years witnessed enormous progress in the field of data acquisition. The combination of traditional detectors with state-of-the-art electronics allows the processing of much higher count rates \cite{SHG14}. Samples with higher intrinsic decay rate can therefore be used.

The disadvantage is that the neutron energies are not known anymore at the time of the activity measurement. Only spectrum-averaged cross sections ($SACS$) can be determined, therefore it is called integral measurement:

\begin{equation}
SACS = \frac{\int{\sigma(E)\Phi(E) \mathrm{d}E}}{\int{\Phi(E) \mathrm{d}E}}
\end{equation}

The number of atoms produced during the activation ($N_{\mathrm{activation}}$) can be written as:

\begin{equation}
N_{\mathrm{activation}} = N_{\mathrm{sample}} \Phi_{\mathrm{n}} \sigma t_{\mathrm{a}}~,
\end{equation}
where $\Phi_n$ is the energy-integrated neutron flux (cm$^{-2}$s$^{-1}$).
If the activation time ($t_{\mathrm{a}}$) is short compared to the half-life time ($t_{1/2}$) of the radioactive neutron-capture product, the freshly produced activity ($A_{\mathrm{activation}}$) increases linearly with the activation time:

\begin{equation}
A_{\mathrm{activation}} \approx \lambda N_{\mathrm{activation}} = \frac{\ln{2}}{t_{1/2}} N_{\mathrm{sample}} \Phi_{\mathrm{n}} \sigma t_{\mathrm{a}}
\end{equation}

Small cross sections $\sigma$ or small samples $N_{\mathrm{sample}}$ can therefore partly be compensated with longer activation time or increased neutron flux. The amount of nuclei, which decays before the activity measurement can be accounted for, see section \ref{sec:time_corrections}. 
Samples smaller than 1~$\upmu$g could be investigated with neutrons in the keV-regime with this very sensitive method. In some cases, the sample itself was already radioactive. This limited the amount of sample material to 28~ng of $^{147}$Pm \cite{RAH03} and about 1~$\upmu$g of $^{60}$Fe \cite{URS09}. 

Even if no $\gamma$-rays are emitted, the method can be applied. The detection setup and sample preparation however are more demanding because very thin samples are required in order to detect the emitted particles. Examples are the successful detection of delayed emitted electrons after activation of $^{34}$S using a silicon $\beta$-spectrometer \cite{RSK00} or of $\alpha$-particles following the $^{7}$Li(n,$\gamma$)$^{8}$Li reaction using an ionization chamber \cite{HKW98}.

If the half-life of the product is very long, it might not be feasible to determine the activity of the capture product. In some of those cases, e.g. $^{62}$Ni(n,$\gamma$) \cite{NPA05}, it is possible to count the number of produced atoms with accelerator mass spectroscopy. If, however, the half-life is very short, it might be necessary to repeatedly irradiate and count the decays \cite{BRW94}. This can be carried out as an automated cyclic activation as in the case of $^{14}$C(n,$\gamma$) \cite{RHF08}.

\subsection{Correction for nuclei decayed during the activation}\label{sec:time_corrections}

If the length of the activation can not be neglected compared to the half-life of the activation product, some of the freshly produced nuclei decay already during the irradiation phase.  
With

\begin{itemize}
\item{$t=0$ .. the beginning of activity counting}
\item{$N_0=N(t=0)$ .. number of produced nuclei at the beginning of activity counting}
\item{$N(t) = N_0 \exp{(-\lambda t)}$ .. no feeding, just radioactive decays}
\item{$t_{\mathrm{m}}$ .. real time of activity counting}
\item{$t_{\mathrm{w}}$ .. time between end of activation and beginning of activity counting}
\item{$C_{\gamma}$ .. the number of counts in the line corresponding to the observed $\gamma$}
\item{$\epsilon_{\gamma}$ .. the detection efficiency of the observed $\gamma$}
\item{$I_{\gamma}$ .. the line intensity - number of emitted $\gamma$-rays per decay}
\end{itemize}

the number of decays during activity measurement can be expressed as

\begin{eqnarray}
	\Delta N &=& N_{0} - N(t_{\mathrm{m}}) \\
	         &=& N_{0} \left(1-\mbox{e}^{-\lambda t_{\mathrm{m}}}\right) \\
	         &=& \frac{C_{\gamma}} {\epsilon_{\gamma} I_{\gamma} }
\end{eqnarray}

hence:

\begin{equation}
	N_{0} = \frac{C_{\gamma}}{\epsilon_{\gamma} I_{\gamma}} \frac{1}{ 1-\mbox{e}^{-\lambda t_{\mathrm{m}}} }
\end{equation}
Therefore the number of nuclei at the end of activation is:

\begin{eqnarray}
	N(-t_{\mathrm{w}}) &=& N_{0} \mbox{e}^{\lambda t_{\mathrm{w}}} \\
	          &=& \frac{C_{\gamma}}{\epsilon_{\gamma} I_{\gamma}} \frac{1}{ 1-\mbox{e}^{-\lambda t_{\mathrm{m}}}}
	              \frac{1}{ \mbox{e}^{-\lambda t_{\mathrm{w}}}}
\end{eqnarray}
$N(-t_{w})$ is the number of produced nuclei reduced by the decays that occurred already during the activation. The number of atoms present during the activation follows as:

\begin{equation}
	\dot{N}(t) = P(t) - \lambda N(t)
\end{equation}
Assuming a constant production rate $P(t)=P$, and $N(t_{\mathrm{start}})=0$, the solution is:

\begin{equation}
	N(t) =  \frac{P}{\lambda} \left( 1 - \mbox{e}^{-\lambda (t-t_{\mathrm{ start}})} \right)
\end{equation}
Hence the number of nuclei at the end of the activation ($t_{\mathrm{end}}=-t_{\mathrm{w}}; t_{\mathrm{a}}=t_{\mathrm{end}}-t_{\mathrm{start}}$):

\begin{eqnarray}\label{number-at-end}
	N(t_{\mathrm{end}}) &=&  \frac{P}{\lambda} \left( 1 - \mbox{e}^{-\lambda (t_{\mathrm{end}}-t_{\mathrm{start}})} \right) \\
			   &=&  \frac{P}{\lambda} \left( 1 - \mbox{e}^{-\lambda t_{\mathrm{a}}} \right)
\end{eqnarray}
However, the number of produced nuclei is:
\begin{equation}\label{eq:activated}
	N_{\mathrm{activation}} = \int_{t_{\mathrm{start}}}^{t_{\mathrm{end}}} P \mbox{d}t = P t_{\mathrm{a}}
\end{equation}
It therefore follows for the $f_{\mathrm{b}}$-factor:

\begin{equation}\label{fb}
	f_{\mathrm{b}} := \frac{N(t_{\mathrm{end}})}{N_{\mathrm{activation}}} = \frac{1 - \mbox{e}^{-\lambda t_{\mathrm{a}}} }{\lambda t_{\mathrm{a}}} 
\end{equation}
With the equations above, one finds for the number of produced nuclei:

\begin{eqnarray}
	N_{\mathrm{activation}} &=& N(-t_{\mathrm{w}}) / f_{\mathrm{b}} \\
	               &=& \frac{C_{\gamma}}{\epsilon_{\gamma} I_{\gamma}} \frac{1}{ 1-\mbox{e}^{-\lambda t_{\mathrm{m}}}}
	                   \frac{1}{ \mbox{e}^{-\lambda t_{\mathrm{w}}}} \frac{1}{f_{\mathrm{b}} } \\
	               &=& \frac{C_{\gamma}}{\epsilon_{\gamma} I_{\gamma}} \frac{1}{ 1-\mbox{e}^{-\lambda t_{\mathrm{m}}}}
	                   \frac{1}{ \mbox{e}^{-\lambda t_{\mathrm{w}}}} \frac{\lambda t_{\mathrm{a}}}{1 - \mbox{e}^{-\lambda t_{\mathrm{a}}} } 
\end{eqnarray}

If the production rate during the activation is not constant, because the irradiation is changing or the activation is interrupted, eq. (\ref{fb}) can easily be generalized. Under real experimental conditions, it is very often appropriate to assume a production rate, which is constant over some periods of time. This occurs either, because several activations are performed or the production rate (proportional to neutron flux or beam current) is constant over time intervals, which are small compared to the half life of the produced isotope. Under those assumptions, the activation can be treated as a sequence of several smaller activations with constant production rates $P_{i}$, activation times $t_{\mathrm{a},i}$, starting times $t_{\mathrm{start},i}$ and ending times $t_{\mathrm{end},i}$. The time between the end of each activation until the end of the last activation will be called $t_{\mathrm{w},i}$, while $t_{\mathrm{w}}$ is the time between the end of the last activation and the beginning of the activity counting. 
Then eq.~(\ref{eq:activated}) becomes:
\begin{eqnarray}
	N_{\mathrm{activation}} &=& \sum_{i}\int_{t_{\mathrm{start},i}}^{t_{\mathrm{end},i}} P_{i} \mbox{d}t \\
	               &=& \sum_{i} P_{i} t_{\mathrm{a},i}
\end{eqnarray}
and eq.~(\ref{number-at-end}) becomes:

\begin{eqnarray}
	N_{i}(t_{\mathrm{end},i}) &=&  \frac{P_{i}}{\lambda} \left( 1 - \mbox{e}^{-\lambda t_{\mathrm{a},i}} \right)
\end{eqnarray}
Therefore the nuclei after the last activation are the sum of all left-overs of all activations:
\begin{eqnarray}
	N(t_{\mathrm{end}}) &=& \sum_{i} N_{i}(t_{\mathrm{end},i}) \mbox{e}^{-\lambda t_{\mathrm{w},i}} \\
	      &=& \sum_{i} \frac{P_{i}}{\lambda} \left( 1 - \mbox{e}^{-\lambda t_{\mathrm{a},i}}\right) \mbox{e}^{-\lambda t_{\mathrm{w},i}}
\end{eqnarray}
and the $f_{\mathrm{b}}$-factor can be written as:
\begin{equation}\label{eq:fb-general}
	f_{\mathrm{b}} = \frac{N(t_{\mathrm{end}})}{N_{\mathrm{activation}}} = \frac{\sum_{i} P_{i} \left( 1 - \mbox{e}^{-\lambda t_{\mathrm{a},i}}\right) \mbox{e}^{-\lambda t_{\mathrm{w},i}}}{\lambda \sum_{i} P_{i} t_{\mathrm{a},i}} 
\end{equation}

It is interesting to look into some special cases of this equation. First, if the individual activation times are short compared to the half-life of the activation product, for instance if the neutron flux or beam current is recorded over short time intervals, eq.~(\ref{eq:fb-general}) becomes:

\begin{equation}\label{fb-short-intervalls}
	f_{\mathrm{b}} = \frac{\sum_{i} P_{i} t_{\mathrm{a},i} \mbox{e}^{-\lambda t_{\mathrm{w},i}}}{\sum_{i} P_{i} t_{\mathrm{a},i}} 
\end{equation}
Further, if the activation times are equal for all activations ($t_{\mathrm{a},i} == t_{\mathrm{a}}$), one finds:

\begin{equation}\label{fb-short-constant-intervals}
	f_{\mathrm{b}} = \frac{\sum_{i} P_{i} \mbox{e}^{-\lambda t_{\mathrm{w},i}}}{\sum_{i} P_{i}} 
\end{equation}
Let the production rate be:

\begin{equation}
	P(t) = \sigma N_{\mathrm{sample}} \Phi(t)
\end{equation}
then eq.~(\ref{fb-short-constant-intervals}) becomes:

\begin{equation}
	f_{\mathrm{b}} = \frac{\sum_{i} \Phi_{i} \mbox{e}^{-\lambda t_{\mathrm{w},i}}}{\sum_{i} \Phi_{i}} 
\end{equation}

\subsection{Partial cross sections to isomers and ground state}
One additional advantage of activation measurements is the possibility to determine the partial cross sections populating isomeric states or the ground state of the reaction product. 

If only one isomeric state is of importance and if it decays partly to the ground state, the neutron capture cross section feeding the isomeric state can be determined via the same
$\gamma$-lines as for the captures to the ground state. The advantage of this 
method is that all uncertainties caused by detection efficiency, 
$\gamma$-ray and neutron self-absorption in the sample, cascade corrections, and
line intensities are canceling out in a relative measurement of the 
cross sections. The only systematic uncertainties left are due to time correlated 
measurements. The time dependence of the isomeric state is a simple exponential behavior:

\begin{equation} 
 N^{\mathrm{m}}(t) = N^{\mathrm{m}}(0) =e^{-\lambda^{\mathrm{m}} t} 
\end{equation}
while the one of the ground state abundance is given by the differential equation:      
\begin{equation} 
 \frac{dN^{\mathrm{g}}}{dt}(t) = -\lambda^{\mathrm{g}} \cdot N^{\mathrm{g}}(t)+ f_i \cdot \lambda^{\mathrm{m}} \cdot N^{\mathrm{m}}(t), 
\end{equation}
where $\lambda$ is the decay constant and $f_i$ the part of internal decays 
feeding the ground state.
The solution is:
\begin{equation} 
 N^{\mathrm{g}}(t) = N^{\mathrm{g}}(0) \cdot e^{-\lambda^{\mathrm{g}} t} + \frac{\lambda^{\mathrm{m}}}{\lambda^{\mathrm{g}}-\lambda^{\mathrm{m}}} f_i \cdot N^{\mathrm{m}}(0)
                \left (e^{-\lambda^{\mathrm{m}} t}-e^{-\lambda^{\mathrm{g}} t} \right ) 
\end{equation}
The time dependence of the number of ground state nuclei and via
$A(t)=\lambda N(t)$ the activity of the ground state decay is a sum of two 
exponential decays. If the half-life of the ground state is much 
shorter than the one of the isomeric state 
($\lambda^{\mathrm{g}} \gg \lambda^{\mathrm{m}}$), the activity immediately after the activation is
determined by the decay activity of the ground state 

\begin{equation}
A^{\mathrm{g}}(t) = \lambda^{\mathrm{g}} \cdot N^{\mathrm{g}}(0) \cdot e^{-\lambda^{\mathrm{g}} t}
\end{equation}
and the isomer decay determines the time dependence at much later times

\begin{equation}
 A^{\mathrm{g}}(t) =  \frac{\lambda^{\mathrm{g}} \lambda^{\mathrm{m}}}{\lambda^{\mathrm{g}}-\lambda^{\mathrm{m}}} f_i \cdot N^{\mathrm{m}}(0) 
           \cdot e^{-\lambda^{\mathrm{m}} t} ~.
\end{equation}

An example of such an analysis is the activation of natural Te. Four isotopes with a total of 4 isomers and 3 ground state decays could be analyzed after the activation with keV-neutrons \cite{ReK02}.

\subsection{Neutron spectra}

The $^7$Li(p,n) reaction as a neutron source in combination with a Van de Graaff accelerator was used for almost thirty 
years at Forschungszentrum Karlsruhe. 
The development of new accelerator technologies \cite{CGK18}, in particular the development of radiofrequency quadrupoles (RFQ) has 
provided much higher proton currents than previously achievable. The additional 
development of liquid-lithium target 
technology to handle the target cooling opens a new era of activation experiments thanks to the enormously 
increased neutron flux. 
Projects like SARAF \cite{TPA15} and FRANZ \cite{WCD10,RCH09},
underline this statement. 

\begin{figure}
\begin{center}
 \renewcommand{\baselinestretch}{1}
 \includegraphics[width=0.49\textwidth]{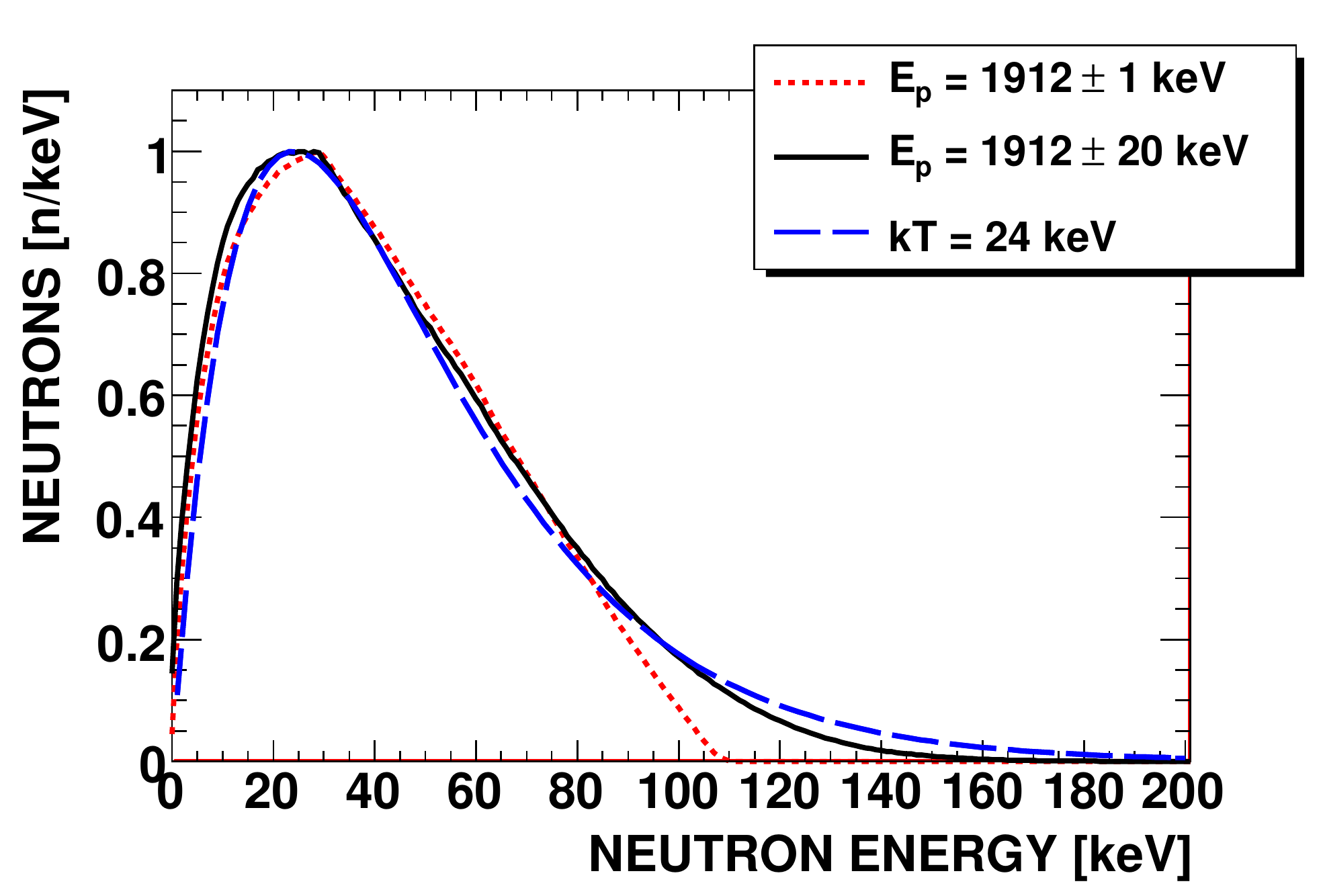}
 \caption{\label{fig:25kev_spectra} Comparison of the number of angle-integrated neutrons per linear energy
bin for simulations using PINO that contain weighting and include a Gaussian proton-energy
profile \cite{RHK09}. A sample of 10 mm radius and a Li-spot of 3 mm radius was assumed. All
simulated spectra are normalized to a common maximum of 1.}
\end{center}
\end{figure}

While other neutron-energy distributions were used on occasion \cite{RSK00,RHF08}, 
the quasi-stellar neutron spectrum, which can be obtained by bombarding a 
thick metallic Li target with protons of 1912~keV slightly above 
the reaction threshold at 1881~keV, was the working horse at Karlsruhe 
\cite{BBK00}. Under such conditions, the $^7$Li(p,n)$^7$Be reaction 
yields a continuous energy distribution with a high-energy 
cutoff at $E_{\mathrm{n}}$~=~106~keV. The produced neutrons are emitted in a 
forward cone of 120$^\circ$ opening angle. The angle-integrated 
spectrum closely resembles a spectrum necessary to measure the Maxwellian averaged cross
section at $kT$~=~25~keV \cite{RaK88,BeK80} (Fig.~\ref{fig:25kev_spectra}) i.e.,

\begin{equation}
  \frac{dN}{dE} \propto E \cdot e^{-\frac{E}{kT}} \propto \sqrt{E}\cdot \phi_{{\mathrm{MB}}},
\end{equation}

where $\phi_{{\mathrm{MB}}}$ is the Maxwellian distribution for a thermal energy of 
$kT$~=~25~keV, see eq. (\ref{maxwell-boltzmann}).

\begin{figure}
\begin{center}
 \renewcommand{\baselinestretch}{1}
 \includegraphics[width=0.49\textwidth]{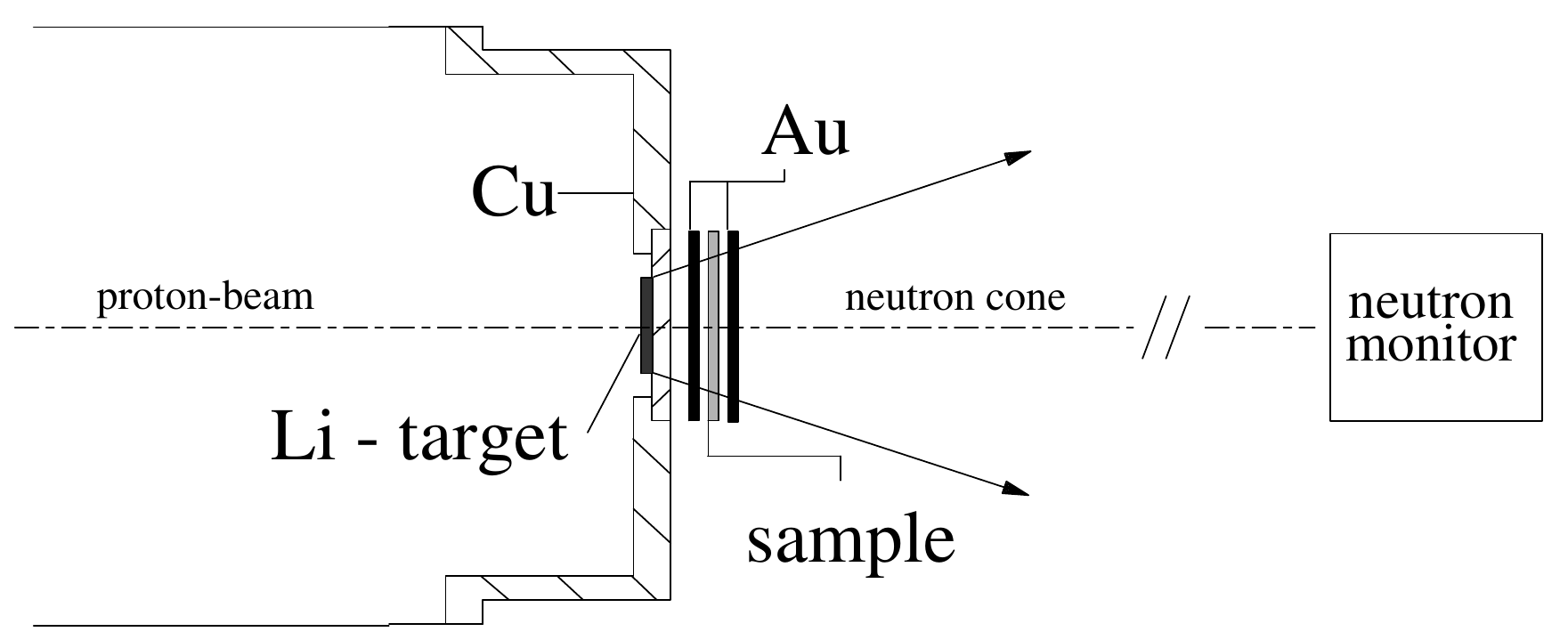}
 \caption{\label{fig:activation_setup} Typical activation setup. Neutrons are
produced via the $^7$Li(p,n) reaction just above the production threshold. The
emitted neutrons are then kinematically focussed into a cone with an opening
angle of 120$^\circ$. The sample is usually sandwiched by two gold foils in order to
determine the neutron flux just before and behind the sample.}
\end{center}
\end{figure}

The samples are typically sandwiched between gold foils and placed 
directly on the backing of the lithium target. A typical setup is 
sketched in Fig.~\ref{fig:activation_setup}. 
The simultaneous activation of 
the gold foils provides a convenient tool for measuring the 
neutron flux, since both the stellar neutron capture cross section
of $^{197}$Au \cite{RaK88} and the parameters of the 
$^{198}$Au decay \cite{Aub83} are accurately known, see also section~\ref{sec:au}.

While the neutron spectrum for the standard case is very well understood, a tool 
for extrapolation to different experimental conditions is desired. Such changes of the
standard setup typically include differences in the angle coverage of the sample, a different
thickness of the lithium layer, or different proton energies. The extrapolation is, while 
conceptually obvious, not straight forward. After impinging onto the lithium layer,
the protons are slowed down until they either leave the lithium layer (in case of a very thin layer)
or fall below the (p,n) reaction threshold and do not contribute to the neutron production anymore.
The double-differential (p,n) cross section changes significantly during this process, 
especially in the energy regime close to the production threshold. Additionally the kinematics of
the reaction is important during the process. Since the Q-value of the reaction is positive, the 
reaction products, and the neutrons in particular, are emitted into a cone in the direction 
of the protons (Fig.~\ref{fig:activation_setup}). This effect becomes less and less pronounced as the 
proton energy increases. If the proton energy in the center-of-mass system is above 2.37~MeV, a
second reaction channel $^7$Li(p,n)$^7$Be$^\star$ opens, which leads to a second neutron group 
at lower energies. To model these processes quantitatively, a tool to simulate the neutron spectrum resulting 
from the $^7$Li(p,n) reaction with a Monte-Carlo approach is indispensable. Therefore the  
highly specialized program PINO - Protons In Neutrons Out - was developed \cite{RHK09}. The power of this approach
was demonstrated during an activation of $^{14}$C with different settings and correspondingly different average neutron 
energies between 25~keV and almost 1~MeV \cite{RHF08}.    
  
\begin{figure}
\begin{center}
 \renewcommand{\baselinestretch}{1}
 \includegraphics[width=0.49\textwidth]{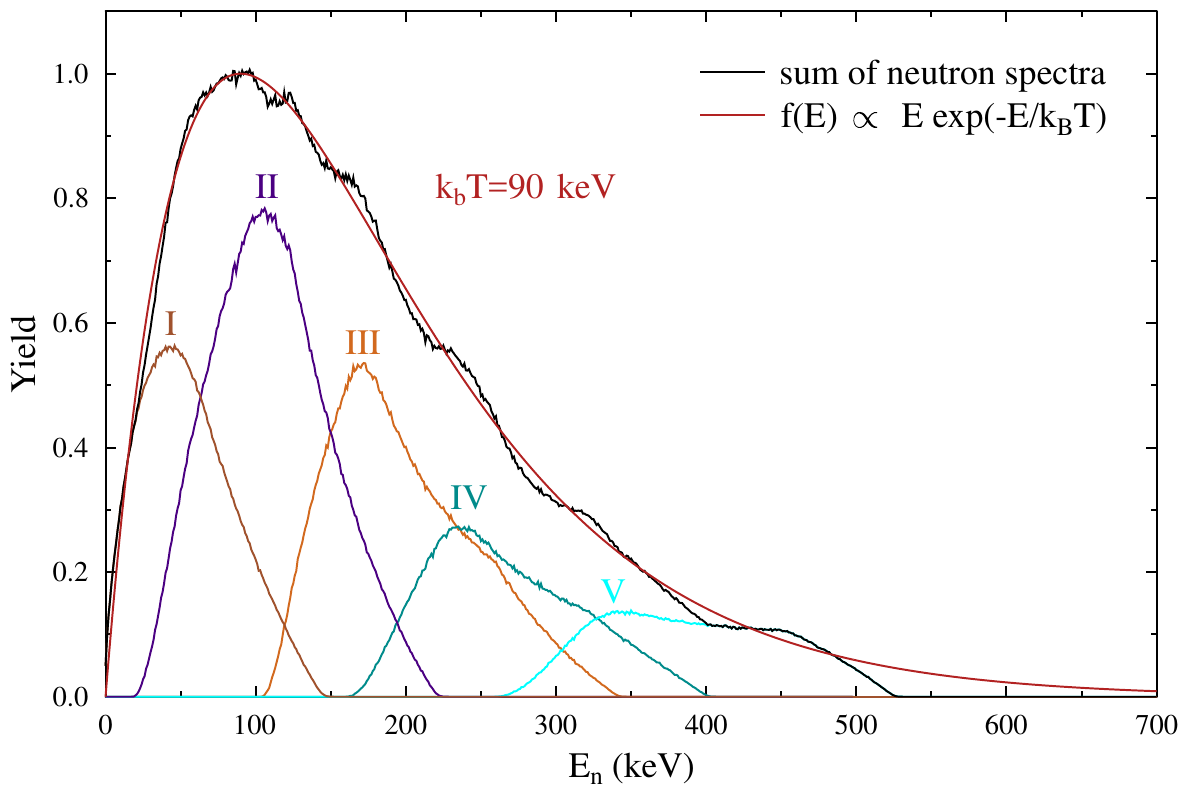}
 \caption{\label{fig:90keV_MACS} Superposition of different neutron spectra to approximate a spectrum corresponding to $kT=90~$keV. The neutron spectra were simulated using the online tool PINO \cite{RHK09}. The corresponding beam parameters are summarized in Table~\ref{Tab:macs_90}.}
\end{center}
\end{figure}

While a temperature of $kT=25$~keV is very well suited for many nucleosynthesis simulations, other temperatures are of interest too. Since extrapolations from activations at a given energy become increasingly uncertain if the temperature is very different, activations with spectra corresponding to other energies are desirable. One approach is the superposition of different spectra. Fig.~\ref{fig:90keV_MACS} and Table~\ref{Tab:macs_90} show the results of a PINO-based study to emulate a spectrum corresponding to $kT=90$~keV. The final spectrum is a linear combination of five components.

\begin{table}
 \caption{Beam parameter used for the spectra in Fig.~\ref{fig:90keV_MACS}. 
          The thickness of the lithium target was always 8~$\upmu$m, which corresponds to a proton-energy loss of 2~keV. The beam spot had a radius of 3~mm. 
          The sample was a disc with a radius of 10~mm and negligible thickness.}
   \label{Tab:macs_90}
\renewcommand{\arraystretch}{1.5} % enlarge line spacing
\begin{tabular}{cccc}
\hline
spectrum& $E_{\textrm{p}}$ 	& distance 	& scaling factor\\
ID		& [keV]				& [mm]		&\\
\hline
\hline
I 	& 1936 & 6 & 0.563 \\
II 	& 1993 & 5 & 0.783 \\
III & 2093 & 3 & 0.535 \\
IV 	& 2145 & 4 & 0.273 \\
V 	& 2257 & 4 & 0.138 \\
\hline
\end{tabular}
\end{table}

A very important ingredient is the double-differential $^{7}$Li(p,n) cross section in particular close to the reaction threshold. The current version of PINO contains data from Liskien and Paulson \cite{LiP75} above a proton energy of $E_p=1950$~keV and below an energy dependence as described by Lee and Zhou \cite{LeZ99}:

\begin{equation}
\sigma(E_{\mathrm{p}}) = \frac{A}{E_{\mathrm{p}}}\frac{x}{(1+x)^{2}}
\end{equation}
with
\begin{equation}
x = c_0 \sqrt{1-\frac{E_\mathrm{threshold}}{E_{\mathrm{p}}}}
\end{equation}
and  
\begin{eqnarray*}
c_0 & = & 6	\\
A   & = & 164.913 \\
E_\mathrm{threshold} & = & 1.8804~\mathrm{MeV}.
\end{eqnarray*}
%We basically used Eq. (15)
%  s(E) = A * x(E) / (E * (1 + x(E))^2)
%  with x(E) = c * sqrt(1 - Eth/E)
%from
%  C. L. Lee, X.-L. Zhou
%  Nucl. Instr. and Meth. in Phys. Res. B 152 (1999) 1--11
%  doi:10.1016/S0168-583X(99)00026-9
%with c = 6, A = 164.913
%Eth = 1.8804289890248638 MeV
Measurements of the $^{7}$Li(p,n) cross section very close to the threshold are difficult. An alternative method is the use of the reverse reaction, $^{7}$Be(n,p) \cite{BAM18}.

\subsection{Impact of backing material}

The neutrons produced in the $^{7}$Li(p,n) reactions have to pass a backing supporting the thin layer of lithium before they reach the sample. Depending on the backing material and thickness, this can not only reduce the number of neutrons but also alter the spectrum. A plain, neutron-energy-independent reduction would not alter the results of an activation measurement as described so far, since it would affect the sample in the same way as the neutron monitors. However, absolute measurements can be affected, see discussion in section~\ref{sec:au}.

\begin{figure}
\begin{center}
 \renewcommand{\baselinestretch}{1}
 \includegraphics[width=0.49\textwidth]{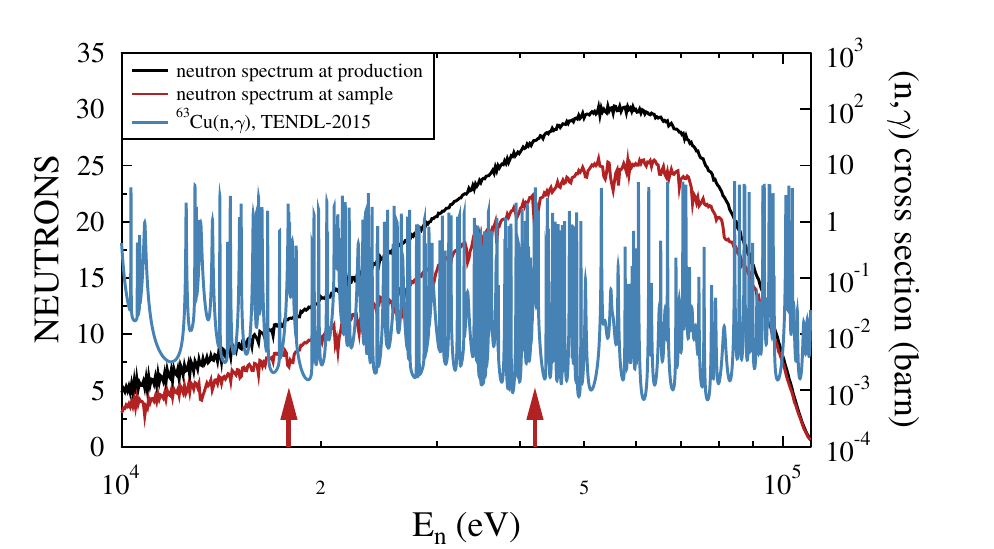}
 \caption{\label{fig:cu_backing_effect_focus_point_63Cu} A 1~mm thick copper backing as commonly used during activation measurements has little effects on the neutron spectrum seen by the sample. However, in the case of a $^{63}$Cu sample,  flux reductions are at the same energy as the resonances in the sample (see arrows for clear examples). In total $10^{11}$ neutrons were simulated.}
\end{center}
\end{figure}

\begin{figure}
\begin{center}
 \renewcommand{\baselinestretch}{1}
 \includegraphics[width=0.49\textwidth]{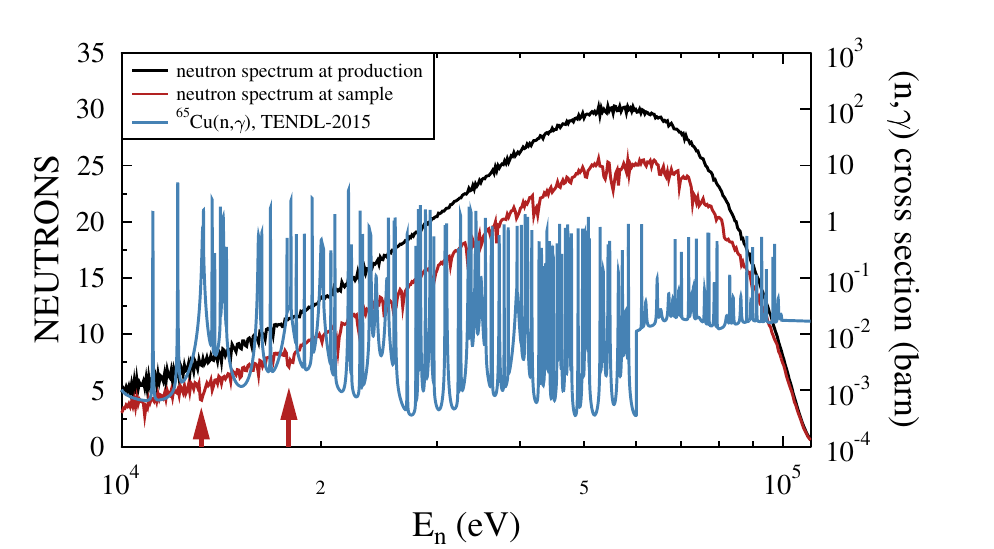}
 \caption{\label{fig:cu_backing_effect_focus_point_65Cu} A 1~mm thick copper backing as commonly used during activation measurements has little effects on the neutron spectrum seen by the sample. However, in the case of a $^{65}$Cu sample,  flux reductions are at the same energy as the resonances in the sample(see arrows for clear examples). In total $10^{11}$ neutrons were simulated.}
\end{center}
\end{figure}

\begin{figure}
\begin{center}
 \renewcommand{\baselinestretch}{1}
 \includegraphics[width=0.49\textwidth]{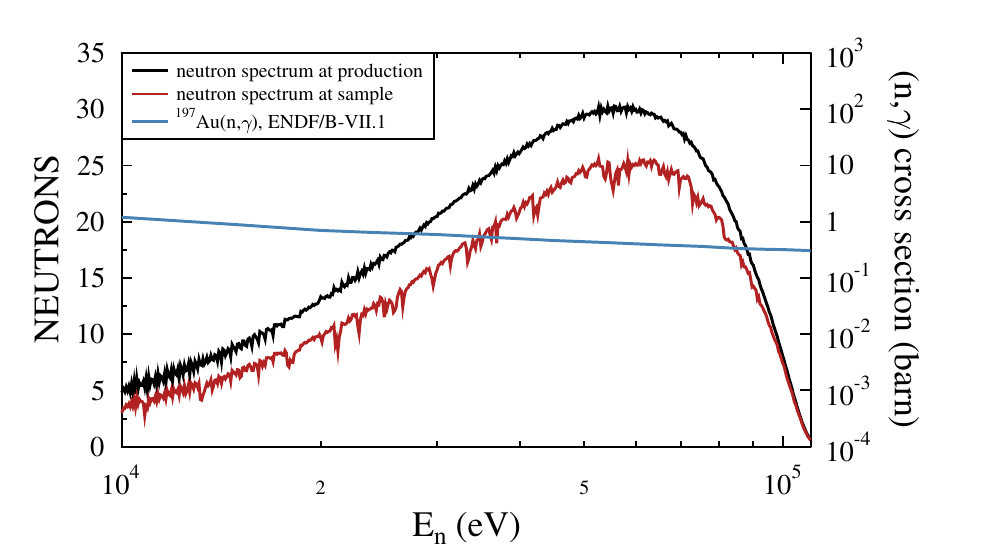}
 \caption{\label{fig:cu_backing_effect_focus_point_197Au} A 1~mm thick copper backing as commonly used during activation measurements has little effects on the neutron spectrum seen by the sample. Almost no effect is expected for a $^{197}$Au sample. In total $10^{11}$ neutrons were simulated.}
\end{center}
\end{figure}

The shape of the neutron spectrum is usually only slightly affected by the backing. Only under the very rare circumstances that the resonances in the sample are at the same energy as the flux reductions caused by the backing, the effect on the sample is different than the effect on the neutron monitor. This was the unfortunate case of an activation of a Cu sample behind a Cu backing (Figs.~\ref{fig:cu_backing_effect_focus_point_63Cu}-\ref{fig:cu_backing_effect_focus_point_197Au}) \cite{HKU08,WBC17}.

New and ongoing investigations based on Monte-Carlo simulations suggest that this effect can be neglected for almost all activations carried out by the Van de Graaff group at Karlsruhe. The only exceptions found so far are $^{63,65}$Cu~\cite{HKU08} and $^{62}$Ni~\cite{NPA05}.

\section{Differential cross section measurements} \label{sec:differential}

\subsection{General idea}

Neutron-induced cross sections usually show a strong resonant structure, caused by the existence of excited levels in the compound nucleus. The excitation function for a reaction can accordingly be divided into three regions, the resonance region, where the experimental setup allows to identify individual resonances, the unresolved resonance region, where the average level spacing is still larger than the natural resonance widths, and the continuum region, where resonances start to overlap. The border between the first two regions is determined
by the average level spacing and by the neutron-energy resolution of the experiment.

\begin{figure}
\begin{center}
\renewcommand{\baselinestretch}{1}
  \includegraphics[width=0.49\textwidth]{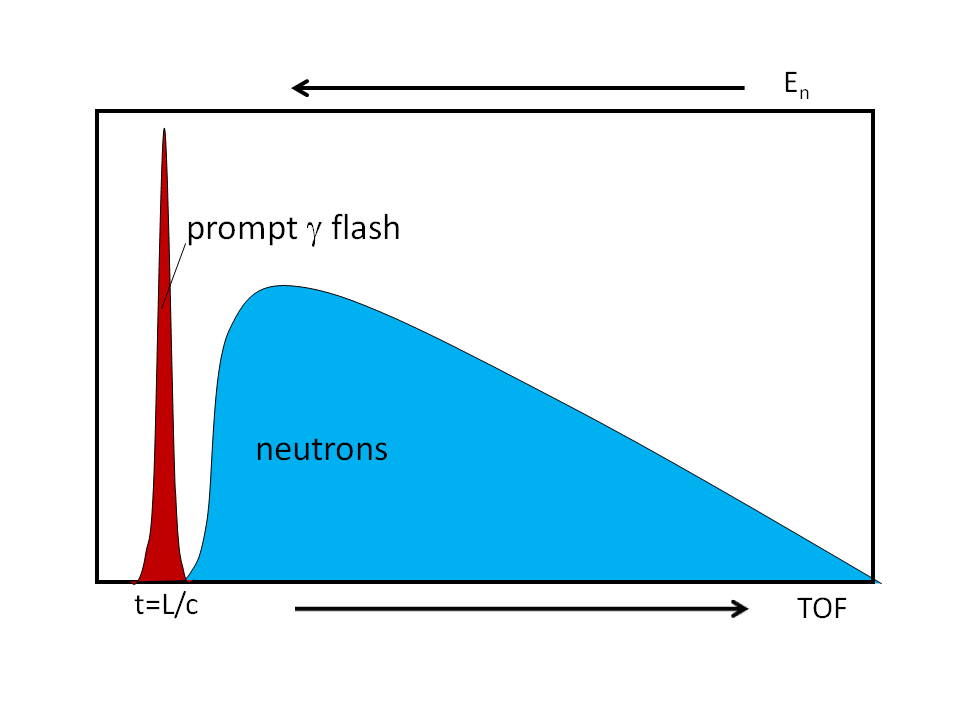}
  \caption{\label{fig:tof_spectrum_schematic} Schematic time-of-flight spectrum. The sharp peak at $t=L/c$ is caused 
by prompt photons produced by the impact of a particle pulse on the target. Neutrons reach the 
measurement station at later times and give rise to a broad distribution depending on their initial 
energies.}
\end{center}
\end{figure}

The time-of-flight (TOF) method enables cross section measurements as a function of neutron energy. 
Neutrons are produced quasi-simultaneously by a pulsed particle beam, thus allowing
one to determine the neutron flight time $t$ from the production target to the sample 
where the reaction takes place. For a flight path $L$, the neutron energy is 
\begin{equation}
 E_{{\mathrm{n}}}=m_{\mathrm{n}}c^{2}(\gamma-1)
\label{tofformularel}
\end{equation}
where $m_{\mathrm{n}}$ is the neutron mass and $c$ the speed of light. The relativistic correction 
$\gamma=\left(\sqrt{1-(L/t)^2/c^2}\right)^{-1}$ can usually be neglected in the neutron 
energy range of interest in nucleosynthesis studies and Eq. \ref{tofformularel}
reduces to 
\begin{equation}
 E_{{\mathrm{n}}}=\frac{1}{2}m_{\mathrm{n}}\left(\frac{L}{t}\right)^2.
\label{tofformula}
\end{equation}
The TOF method requires that the neutrons are produced at well defined times. This is achieved by
irradiation of an appropriate neutron production target with a fast-pulsed beam from particle 
accelerators. The TOF spectrum measured at a certain distance from the target is sketched
in Fig.~\ref{fig:tof_spectrum_schematic}. The essential features are a sharp peak at $t=L/c$, the so-called 
$\gamma$-flash due to prompt photons produced by the impact of a particle pulse on the 
target, followed by a broad distribution of events when the neutrons arrive at the sample position,
corresponding to the initial neutron energy spectrum.

Neutron TOF facilities are mainly characterized by two features, the energy resolution $\Delta E_{\mathrm{n}}$ 
and the flux $\phi$. The neutron energy resolution is determined by the uncertainties of the flight path 
$L$ and of the neutron flight time $t$:

\begin{equation}
\frac{\Delta E_{\mathrm{n}}}{E_{\mathrm{n}}}=2\sqrt{\frac{\Delta t^{2}}{t^{2}}+\frac{\Delta L^{2}}{L^{2}}} 
\end{equation}

The neutron energy resolution can be improved by increasing the flight path, but only at the expense of
the neutron flux, which scales with $1/L^2$. The ideal combination of energy resolution and neutron 
flux is, therefore, always an appropriate compromise. The energy resolution is affected by the Doppler 
broadening due to the thermal motion of the nuclei in the sample, by the pulse width of the particle 
beam used for neutron production, by the uncertainty of the flight path including the size of 
the production target, and by the energy resolution of the detector system. 

\subsection{Detection of the neutron-induced reaction}

In TOF measurements, capture cross sections are determined via the prompt $\gamma$-ray cascade emitted 
in the decay of the compound nucleus. Total absorption calorimeters or total energy detection systems are the most common detection principles for measuring neutron capture
cross sections.

The  total energy technique is based on a device with a $\gamma$-ray detection efficiency,
($\varepsilon_\gamma$) proportional to $\gamma$ energy ($E_\gamma$): 

\begin{equation}
\varepsilon_\gamma = k E_\gamma
\end{equation}

Provided that the overall 
efficiency is low and that no more than one $\gamma$ is detected per event, the efficiency for detecting 
a capture event becomes independent of cascade multiplicity and de-excitation pattern, 
but depends only on the excitation energy of the compound nucleus, which is equal to the sum of neutron separation energy and kinetic energy in the center of mass before the formation of the compound nucleus. It can be shown 
that with the given assumptions the probability $\epsilon_{\mathrm{casc}}$ 
to detect any $\gamma$-ray out of a cascade of $n$ $\gamma$-rays
can be written as:

\begin{equation}
\varepsilon_{\mathrm{casc}} = \sum_{i=1}^{n} \varepsilon_\gamma^{i} = \sum_{i=1}^{n} k E_\gamma^{i} =  k \sum_{i=1}^{n} E_\gamma^{i} = k (Q+E_{CM})
\end{equation}

A detector with an intrinsic proportionality of $E_\gamma$ and $\varepsilon_\gamma$ was first 
developed by Moxon and Rae \cite{MoR63} by combining a $\gamma$-to-electron converter
with a thin plastic scintillator. Because of this conversion, Moxon-Rae detectors are essentially insensitive 
to low-energy $\gamma$ rays and were, therefore, used in TOF measurements on radioactive samples 
\cite{WiK78,WiK79a}. The efficiency of Moxon-Rae detectors for capture events is typically less than a few 
percent.

Higher efficiencies of about 20\% can be obtained by an extension of the Moxon-Rae principle originally 
proposed by Maier-Leibnitz \cite{Rau63,MaG67b}. In these total energy detection systems the proportionality 
$E_\gamma$ - $\varepsilon_\gamma$ is extrinsically realized by an $a~posteriori$ treatment of the recorded 
pulse-height. This Pulse Height Weighting technique is commonly used with liquid scintillation detectors 
about one liter in volume, small enough for the condition to detect only one $\gamma$ per cascade. Present
applications at neutron facilities n\_TOF (CERN, Switzerland) and at GELINA (IRMM, Belgium) are using deuterated 
benzene (C$_6$D$_6$) as scintillator
because of its low sensitivity to scattered neutrons. Initially, the background due to scattered neutrons had 
been underestimated, resulting in overestimated cross sections, particularly in cases with large 
scattering-to-capture ratios as pointed out by Koehler {\it et al.} \cite{KWG00} and Guber {\it et al.} 
\cite{GLS05}. With an optimized design, an extremely low neutron sensitivity of 
$\varepsilon_\mathrm{n}/\varepsilon_\gamma\approx3\times10^{-5}$  has been obtained at n\_TOF \cite{PHK03},
which is especially important for light and intermediate-mass nuclei, where elastic scattering usually 
dominates the capture channel.

A total absorption calorimeter consists of a set of detectors arranged in 4$\pi$ geometry, thus covering the 
maximum solid angle. Because the efficiency for a single $\gamma$-ray of the capture cascade is usually 
close to 100\% in such arrays, capture events are characterized by signals corresponding essentially to the 
Q-value of the reaction. Provided good resolution in $\gamma$ energy, gating on the Q-value represents,
therefore, a possibility of significant background suppression. 

Total absorption calorimeters exist at several TOF facilities. Most are using BaF$_2$ as scintillator material, 
which combines excellent timing properties, fairly good energy resolution, and low sensitivity to neutrons 
scattered in the sample. In fact, neutron scattering dominates the background in calorimeter-type detectors, 
because the keV-cross sections for scattering are typically 10 to 100 times larger than for neutron capture.
In measurements at moderated neutron sources this background is usually reduced by an absorber surrounding 
the sample. Such a detector has been realized first at the Karlsruhe Van de Graaff accelerator \cite{WGK90a}.
This design, which consists of 42 crystals, is also used at the n\_TOF facility at CERN \cite{GAA09}, while a
geometry with 160 crystals has been adopted for the DANCE detector at Los Alamos \cite{HRF01,RBA04}. 
There are also $4\pi$ arrays made of NaI crystals \cite{MAA85,BDS94}, but in the astrophysically important 
keV region these detectors are suffering from the background induced by scattered neutrons, which are easily 
captured in the iodine of the scintillator.

\subsection{White spectra}

\begin{figure}
\begin{center}
\renewcommand{\baselinestretch}{1}
  \includegraphics[width=0.49\textwidth]{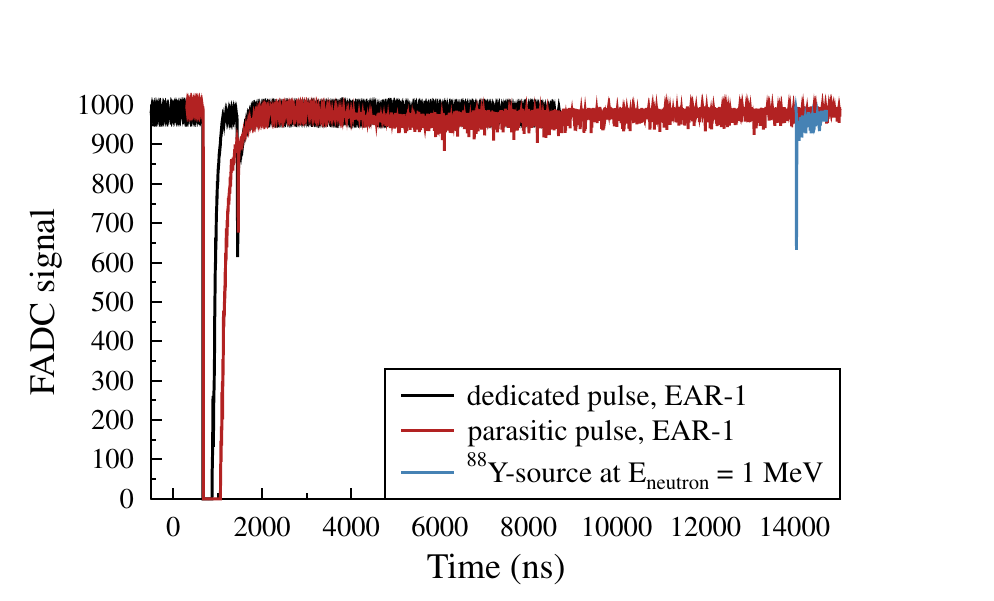}
  \caption{\label{fig:ntof_waveform} Output signal of a photomultiplier coupled to a BaF$_2$ crystal. The unit was positioned close to the beam line of experimental area 1 (EAR-1) at n\_TOF about 200~m away from the spallation source \cite{GTB13}.The $\gamma$-flash occurs about 600~ns after the proton pulse hits the lead target producing the neutrons via spallation reactions. The flash-ADC-based data acquisition is saturated during the flash. However, neutrons with an energy of 1~MeV would arrive about 14~$\upmu$s after that. The detector would be able to accept events at that time, as shown by an exemplary waveform taken with a $^{88}$Y calibration source. Dedicated n\-TOF pulsed are typically larger and cleaner than parasitic pulses, which are provided much more irregularly.} 
\end{center}
\end{figure}

At white neutron sources, the highest neutron energy for which the neutron capture cross section can be 
determined is limited by the recovery time of the detectors after the $\gamma$~flash. While the accessible 
neutron energy range is practically not restricted for C$_6$D$_6$ detectors, BaF$_2$ arrays are more 
sensitive, depending on the respective neutron source. At n\_TOF, for example, the BaF$_2$ calorimeter 
has been used only up to few keV so far \cite{GCM12}, whereas there are no strong limitations at DANCE at Los Alamos \cite{ERB08}. Recent tests suggest however that the NAUTILUS detector, which is strongly optimized for the handling of the $\gamma$-flash \cite{WGR18,RDF18}, can be used at n\_TOF up to about 1~MeV, see Fig. \ref{fig:ntof_waveform}. 

Independent 
of the detection system, measurements at higher neutron energies are increasingly difficult because 
the (n,$\gamma$) cross section decreases with neutron energy, while at the same time competing reaction 
channels, e.g. inelastic neutron scattering, are becoming stronger. Nevertheless, present techniques are 
covering the entire energy range of astrophysical relevance up to about 500~keV  with sufficient accuracy.

\subsection{Tailored spectra}

In specific 
neutron spectra, e.g. in measurements with the Karlsruhe array, where the maximum neutron energy was 
about 200~keV, scattered neutrons can be partially discriminated via TOF between sample and scintillators
because the capture $\gamma$ rays reach the detector before the scattered neutrons \cite{WGK90a}. The idea is that neutrons scattered on the sample reach the detector later than the prompt $\gamma$-rays following the neutron capture events. This idea is particularly powerful if the flight path is short. While most of the experiments (see chapter \ref{sec:renormalization}) with the Karlsruhe array were performed with a flight path of 80~cm, even shorter flight paths are necessary to measure (n,$\gamma$) reactions on radioactive samples.

\begin{figure}
\begin{center}
  \includegraphics[width=.49\textwidth]{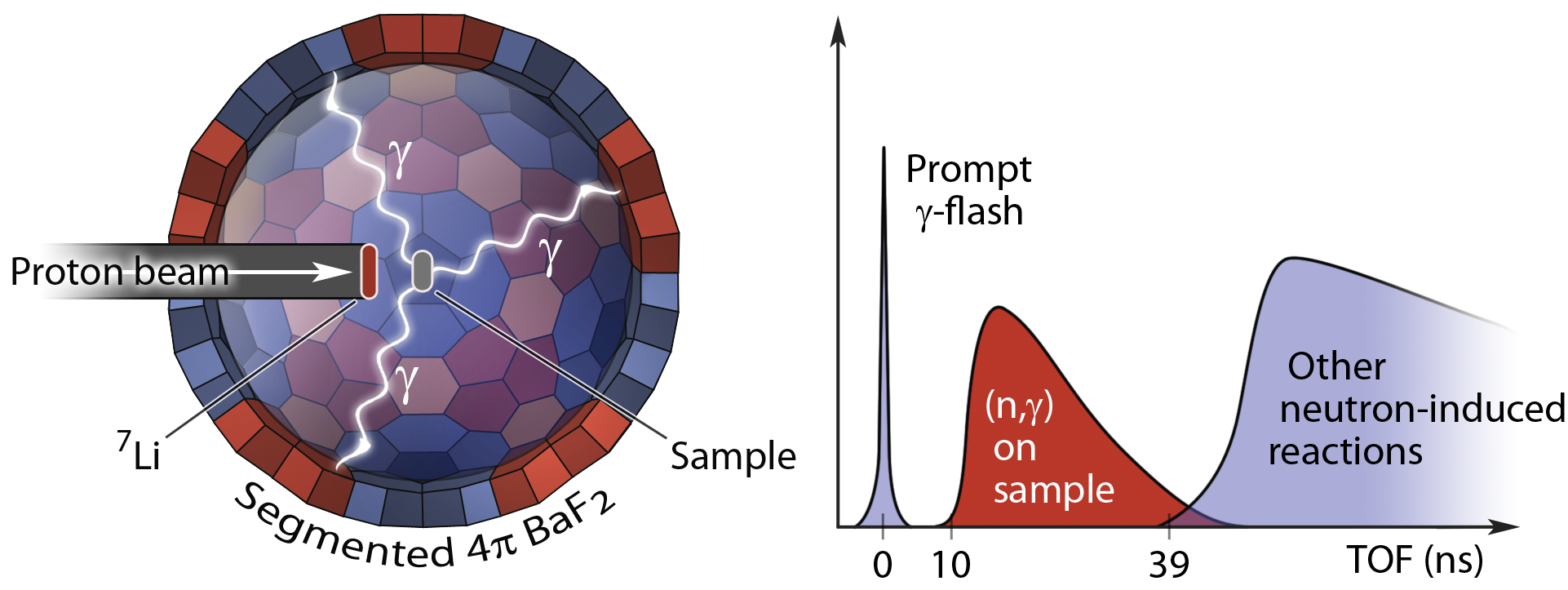}
\end{center}
  \caption{Left: Schematic setup of the planned neutron capture experiment with an ultra-short flight path of only 4~cm. Right: Anticipated time-of-flight spectrum \cite{RHH04,RDF18}.
  \label{fig:nautilus}}
\end{figure}

The present layout of the FRANZ facility at the Goethe University Frankfurt barely provides the high neutron fluxes needed to perform measurements on radioactive isotopes with comparably hard $\gamma$-ray emission like $^{85}$Kr \cite{RCH09,CoR07}. Since the neutron production is already at the limits of the current technology, one option is to get closer to the neutron production target to increase the solid angle covered by the sample material. Such a TOF measurement can be performed with sufficient accuracy even with a flight path as short as a few centimeters (Fig.~\ref{fig:nautilus}, left) \cite{RDF18}.

In this case, the only feasible solution is to produce the neutrons inside a spherical $\gamma$-detector and distinguish between background from interactions with the detector material and the signal from neutron captures on the sample based on the time after neutron production as illustrated in Fig.~\ref{fig:nautilus}, right. 
First, an initial $\gamma$-flash, occurring when the protons hit the neutron production target, is detected. Then the prompt $\gamma$-rays produced in the (n,$\gamma$) reaction at the sample induce a signal in the detector. Later, the neutrons from other reactions, such as scattering in the detector material, arrive in the detector with sufficient delay as they are traveling at much slower speed than the $\gamma$-rays, and produce background.
 
A detailed investigation of the geometry of the setup at an ultra-short flight path has been performed \cite{WGR18} within the framework of the NAUTILUS project. In contrast to the calorimeters used in such TOF experiments so far \cite{RLK14,RBA04,WGK90a}, the calorimeter shell has to be much thinner in order to allow the neutrons to escape quickly enough. The geometry is based on the DANCE array \cite{RBA04,HRF01}, which was designed as a high efficiency, highly segmented 4$\pi$ BaF$_2$ array. The NAUTILUS array consists of up to 162 crystals of 4 different shapes, each covering the same solid angle. The high segmentation distributes the envisaged high count rate over many channels, leading to a significant increase of the maximal tolerable total count rate that can be processed by the DAQ. The NAUTILUS array has an inner radius of 20~cm and a thickness of 12~cm. 

The advantage of this setup is the greatly enhanced neutron flux. Because of the reduction of the flight path from 1~m to 4~cm, the neutron flux will be increased by almost 3 orders of magnitude. The reduced time-of-flight resolution resulting in a reduced neutron-energy resolution is still sufficient for astrophysical and applied purposes. A comparison to the DANCE setup shows that, despite the much shorter flight path (4~cm vs. 20~m), a much better time resolution (1~ns vs. 125~ns) will be achieved at the proposed setup. Because of the different time  structure of the proton beam at the FRANZ facility the energy resolution will almost be the same for both setups.

\section{$^{197}$Au(n,$\gamma$) - a cross section standard \label{sec:au}}

Standard cross sections are important quantities in neutron experiments, because 
they allow to circumvent difficult absolute flux determinations by measuring 
simply cross section ratios. Therefore, a set of standard cross sections has been 
established and is periodically updated with improved data characterized by higher 
accuracies and/or wider energy ranges. A review of the most recent activity on 
neutron cross section standards can be found in \cite{CPH17}. 

Considered as an official standard only at thermal neutron energy (25.3 meV) 
and between 200~keV and 2.5~MeV \cite{BZC07,CPS09}, the (n,$\gamma$) cross section 
of gold is commonly also used in the keV region as a reference for neutron capture 
measurements related to nuclear astrophysics as well as for neutron flux 
determinations in reactor and dosimetry applications. 

From the experimental point of view $^{197}$Au exhibits most favorable 
features. Mechanically it is a monoisotopic metal available in very high purity 
that can easily be shaped to any desired sample geometry. Its nuclear properties 
are equally interesting: A strong resonance at 4.9 eV allows for the 
determination of the neutron flux or for the normalization of capture yields 
in TOF measurements by means of the "saturated resonance technique" 
\cite{SBD12}. The (n,$\gamma$) cross section in the keV region is rather 
large, thus facilitating the measurement of cross section ratios. And it is also 
easily applicable for neutron activation studies due to the decay of $^{198}$Au 
with the emission of an intense 412 keV $\gamma$-ray line. Both the decay 
rate ($\lambda=0.25718(7)$~d$^{-1}$) and the intensity per decay 
($I_{\gamma}=95.58(12)$\%) \cite{Chu02} are accurately known and 
perfectly suited for practical applications.

\subsection{Measurements}

In 1988, a direct measurement of the $^{197}$Au(n,$\gamma$) cross 
section based on the $^{7}$Li(p,n)$^{7}$Be reaction, which used the 
$^{7}$Be activity for an absolute determination of the neutron exposure, 
i.e. without reference to another standard, claimed a very low systematic 
uncertainty of 1.5\% \cite{RaK88}. This result, which  referred to the average 
cross section over a quasi-Maxwellian neutron spectrum for a thermal energy 
of $kT=25$ keV, was found in very good agreement with the value calculated 
on the basis of the energy-dependent cross section \cite{Mac81}, but was 
systematically lower by 5 to 7\% than the evaluated $^{197}$Au(n,$\gamma$) 
cross section based on a variety of data sets, including other reaction channels. 

To clarify this issue, the energy-dependent cross section has been measured  
at n\_TOF and GELINA in an effort to provide accurate new data in the resolved 
resonance region \cite{MDV10}, and at keV neutron energies \cite{LCD11,MBD14}. 
Both measurements were performed with C$_{6}$D$_{6}$ detectors, but used 
different neutron flux standards, a combination of the $^{6}$Li(n,t) 
and $^{235}$U(n,f) reactions at n\_TOF and the $^{10}$B(n,$\alpha$) 
cross section at GELINA. In each case, the capture yield was self-normalized to the 
saturated gold resonance at 4.9 eV. The combination of improved detection 
systems with detailed simulations and analysis techniques has yielded data sets 
with $MACS$ uncertainties slightly above 3\% at n\_TOF and between 1 and 2\% 
at GELINA. The new results agree with each other within systematic 
uncertainties and confirm the difference of about 5\% relative to the activation result 
of Ref. \cite{RaK88}.

In parallel to the TOF results from n\_TOF and GELINA, also the quasi-stellar 
neutron spectrum at a thermal energy $kT=25$ keV were remeasured at the 7-MV 
Van de Graaff laboratory at JRC Geel \cite{FFK12} and using the PIAF facility at 
the 3.75 MV Van de Graaff accelerator of PTB Braunschweig  \cite{LKM12}. Both 
measurements confirmed the neutron field reported previously \cite{RaK88} and 
showed that substantial effects related to slight shifts in the proton energy or to the
spectral broadening of the proton beam could be excluded as the cause of the 
difference to the new TOF data. Instead, an activation performed in addition to 
the spectrum measurements at JRC-Geel \cite{FFK12} found a 5\% higher cross section than 
Ref. \cite{RaK88}.

\begin{figure}
\begin{center}
\renewcommand{\baselinestretch}{1}
  \includegraphics{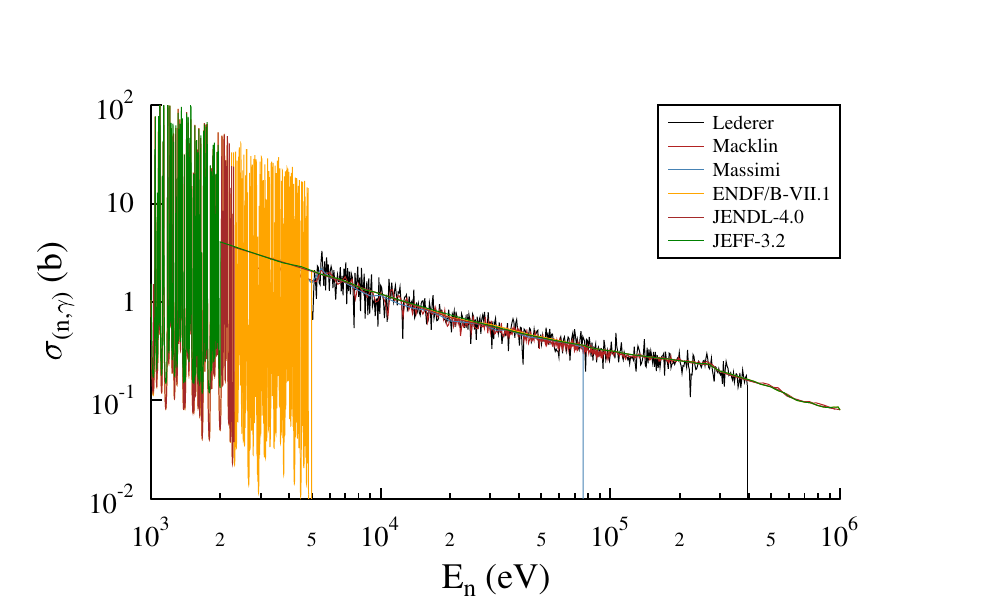}
  \caption{\label{fig:197au_ng_overview} The main TOF data in the 
keV-neutron-energy regime compared with the evaluated data sets 
in the ENDF/B-VII.1 \cite{CHO11}, JEFF-3.2 \cite{JEFF30,JEFF31}, 
and JENDL-4.0 \cite{KSO11} libraries. }
\end{center}
\end{figure}

\subsection{Monte Carlo simulations}

The analyses of the new measurements all benefit from detailed Monte Carlo 
simulations of the involved corrections, whereas the earlier activation had tried to 
find an experimental access to these corrections. Therefore, Monte Carlo (MC) 
simulations of the experimental situation in Ref. \cite{RaK88} were performed in 
an attempt to localize the reason for the above discrepancy. For easier comparison 
of TOF and activation results all data sets were averaged over the quasi-Maxwellian 
spectrum centered at 25 keV used in \cite{RaK88}. In the following these 
spectrum-averaged cross sections are denoted as $SACS$.

\begin{figure}
\begin{center}
\renewcommand{\baselinestretch}{1}
  \includegraphics[width=0.49\textwidth]{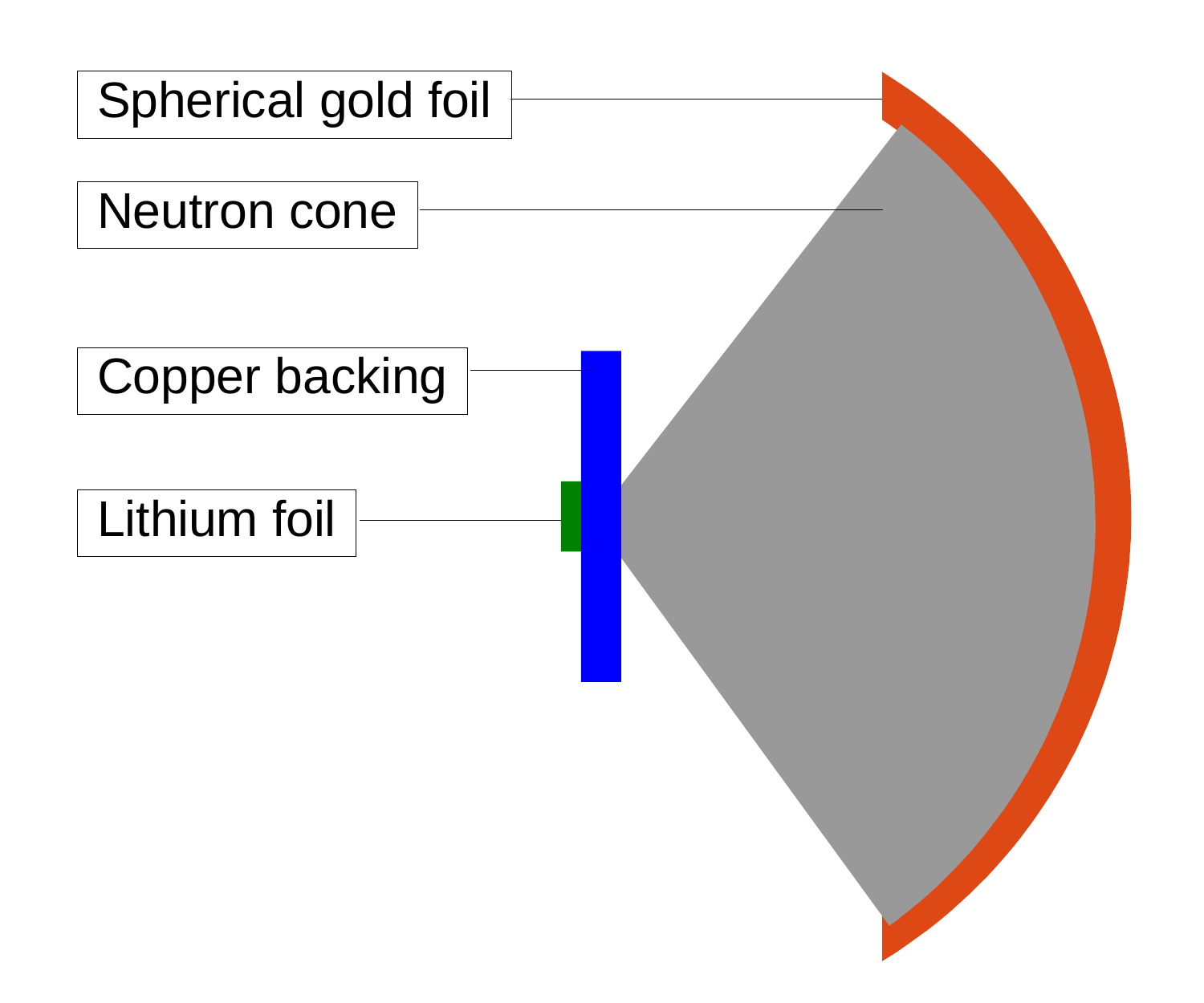}
  \caption{\label{fig:ratynski_setup} A spherical gold foil covering the entire neutron cone was activated during the $^{197}$Au(n,$\gamma$) measurement by Ratynski \& K\"appeler \cite{RaK88}. Different backing materials and thicknesses were used. Only the backing and the gold foil were included in the simulations. The real setup was similar to the one in Fig.~\ref{fig:activation_setup} except for the spherical sample.}
\end{center}
\end{figure}

The simulations were carried out by considering two volumes - the thin gold foil 
shaped as a section of a sphere and the backing of the $^7$Li target, which the 
neutrons have to pass in order to reach the gold foil, see Fig.~\ref{fig:ratynski_setup}.   
The neutrons were tracked 
according to the elastic scattering and capture cross sections adopted from the 
data libraries ENDF/B-VII.1, JEFF-3.2, and JENDL-4.0. In each case the Cu and Au 
cross sections were consistently taken from the same library. The neutrons 
were scattered  isotropically, but the energy loss during elastic scattering was taken 
into account. Different backing thicknesses were simulated for direct comparison 
with the original activation data as given in Table~III of Ref. \cite{RaK88}.

Some components of the experimental uncertainties for different backing thickness 
provided in Ref. \cite{RaK88} are correlated. The same detector and line intensities were used for all activations. In order to better observe the trend when varying the backing thickness, it is better to compare the ratios of the different setups to a given setup, e.g. to the lithium target with the 1-mm Cu backing. Then all correlated variables cancel out and their uncertainty does not contribute.

Table~\ref{Tab:sacs_backing_ratios} shows the results of this endeavor. The 
experimental data and the MC-based predictions with the evaluated cross sections 
of the data libraries are now in agreement within the experimental uncertainties - 
at least for the Cu-backing. It is not clear, whether the results for the Ag-backing 
disagree at this magnitude due to a problem with the Ag cross sections, because
there is even a large scatter quoted in Ref. \cite{RaK88}. The simulations are offering a plausible explanation for the difference between the activation of \cite{RaK88} and the newest TOF measurements \cite{LCD11,MBD14}, namely that the effect of the backing was not properly taken into account during the activations. However, the simulation are not sufficiently consistent with the data for a reliable posterior correction. Nevertheless, they demonstrate that the new TOF measurements provide a reliable basis for establishing the (n,$\gamma$) cross section of $^{197}$Au  with an uncertainty of 1\%, sufficient for re-considering gold as a neutron capture standard in the keV region.   

\begin{table}[htb]
 \caption{Ratio of Monte Carlo simulations of $SACS$ (mb) for $E_{\mathrm{proton}}=1912$~keV, each relative to the case of a 1-mm Cu-backing. 
These are compared to the experimental results obtained with different backing materials and thicknesses by Ratynski and K{\"a}ppeler \cite{RHK09,RaK88}. Only uncorrelated 
uncertainties were considered for the experimental data. See Fig. \ref{fig:197au_ng_overview} for the differential cross sections.}
   \label{Tab:sacs_backing_ratios}
   \renewcommand{\arraystretch}{1.5} % enlarge line spacing
   \begin{tabular}{rcccc}
    Backing    				& \multicolumn{4}{c}{Ratio to 1 mm Cu backing}\\
             				& ENDF  				& JEFF				& JENDL 		& Ratynski \\
    \hline
    1.0 mm Cu					&   1					&	1				&	1			& 1               \\
    0.7 mm Cu				&   1.016				&	1.016			&	1.025		& 1.028 $\pm$ 0.012\\
    0.5 mm Cu				&   1.032				&	1.029			&	1.031		& 1.042 $\pm$ 0.012\\
    0.2 mm Ag				&   1.057				&	1.051			&	1.062		& 1.003-1.037\\
    no backing				&   1.067				&	1.064			&	1.071		& \\
   \end{tabular}
\end{table}
 
\subsection{Evaluated cross section in data libraries}

The evaluated cross sections in the data libraries ENDF/B-VII.1 \cite{CHO11},
JEFF-3.2 \cite{JEFF30,JEFF31}, and JENDL-4.0 \cite{KSO11} (which are considered 
as representative of similar compilations) are essentially based on the main
TOF-measurements as illustrated in Fig.~\ref{fig:197au_ng_overview} and are, 
therefore, not affected by the discrepancy with the activation result of Ref. \cite{RaK88}.
The experimental data are best represented by the ENDF/B-VII.1 evaluation, while 
the cross sections given in JEFF-3.2 and JENDL-4.0 are exceeding the ENDF/B-VII.1 
data on average by about 3.5 and 2.4\% between 10 and 100 keV, respectively.

\subsection{Maxwellian average cross sections at stellar temperatures}
\label{sec_MACS}

The (n,$\gamma$) cross section of gold has been extensively used as a reference 
for numerous measurements devoted to studies of stellar nucleosynthesis in 
the slow neutron capture process (s process), which is associated with the He and C burning episodes of late evolutionary phases. The neutron spectrum typical of the various s-process sites discussed in nuclear
astrophysics (see e.g. \cite{KGB11}) is 
described by a Maxwell-Boltzmann distribution, because neutrons are 
quickly thermalized in the dense stellar plasma, and the effective 
stellar reaction cross sections are obtained by 
averaging the experimental data over that spectrum. The resulting
Maxwellian averaged cross sections ($MACS$)
are commonly compared for a thermal energy of $kT=30$ keV, but
for realistic s-process scenarios a range of thermal energies 
has to be considered, from about 8 keV in the so-called $^{13}$C 
pocket, a thin layer in the He shell of thermally pulsing low mass AGB 
stars, to about 90 keV during carbon shell burning in massive stars. 

To cover this full range, (n,$\gamma$) cross sections 
$\sigma(E)$ are needed at least in the energy window 1~keV and 1~MeV. Whenever experimental data are available only for part of this 
range, cross section calculations are required for filling the gaps. 
In this context, theoretical cross sections obtained via the Hauser-Feshbach 
approach are indispensable \cite{Rau14}.    

With equation \ref{MACS} the $MACS$ of $^{197}$Au was calculated by separating the required neutron energy range into three regions:

\begin{eqnarray}\label{eq:MACS_exp_lib}
   MACS       & = &  \frac{2}{\sqrt{\pi}} \frac{1}{(kT)^{2}}\cdot \large(I_{\mathrm{low}} + I_{\mathrm{exp}} + I_{\mathrm{high}} \large) \\
   I_{\mathrm{low}}    & = & \int_0^{E_{\mathrm{low}}} \eta\cdot \sigma_{\mathrm{lib}}(E) \cdot E \cdot \exp{\left(-\frac{E}{kT}\right)} \mbox{d}E \\
   I_{\mathrm{exp}}    & = & \int_{E_{\mathrm{low}}}^{E_{\mathrm{high}}} \sigma_{\mathrm{exp}}(E) \cdot E \cdot \exp{\left(-\frac{E}{kT}\right)} \mbox{d}E\\
   I_{\mathrm{high}}   & = & \int_{E_{\mathrm{high}}}^\infty \eta\cdot \sigma_{\mathrm{lib}}(E) \cdot E \cdot \exp{\left(-\frac{E}{kT}\right)} \mbox{d}E
\end{eqnarray}  

where $E_{\mathrm{low}}$ \cite{MBD14} and $E_{\mathrm{high}}$ define the range of the new experimental data \cite{LCD11}. At lower and higher energies the evaluated cross sections $\sigma_{lib}$ have been adopted from the current data libraries with an optional normalization factor  

\begin{equation}
  \eta=\frac{\int_{E_{\mathrm{low}}}^{E_{\mathrm{high}}} \sigma_{\mathrm{exp}}(E)\mbox{d}E }{\int_{E_{\mathrm{low}}}^{E_{\mathrm{high}}} \sigma_{\mathrm{lib}}(E)\mbox{d}E }
\end{equation}

For the update of the $^{197}$Au(n,$\gamma$) cross section we adopted $\eta = 1$ and the following uncertainties for the data sets used:
  \begin{itemize}
    \item{$\pm5$\% in region $I_{\mathrm{low}}$ ($E_\mathrm{n}\leq 3.5$ keV) (ENDF/B-VII.1 \cite{CHO11})}
    \item{$\pm1$\% in region $I_{\mathrm{high}}$ ($E_\mathrm{n}\geq 400$ keV), where the cross section is an established standard (ENDF/B-VII.1 \cite{CHO11}), and}
    \item{$\pm1$\% systematic uncertainty in region $I_{\mathrm{exp}}$ ($3.5$ keV $\leq E_\mathrm{n}  \leq 400$ keV) together with the statistics quoted in Refs. \cite{LCD11,MBD14}.}
  \end{itemize} 
 
In Table~\ref{Tab:macs_30} the corresponding $MACS$ results for $kT=30$~keV are compared for the combination of three experimental data sets and three different libraries.

\begin{table}[htb]
 \caption{$MACS$ (mb) for $kT=30$~keV. See Fig. \ref{fig:197au_ng_overview} for the differential cross sections.}
   \label{Tab:macs_30}
   \renewcommand{\arraystretch}{1.5} % enlarge line spacing
   \begin{tabular}{cccc}
    Data set      				& ENDF  		& JEFF			& JENDL 	\\
    \hline
    Macklin \cite{Mac82a}	& 	586.2				& 588.3				& 587.1			\\
    Lederer \cite{LCD11}  &   610.6      			& 618.4				& 616.2 		\\
    Massimi \cite{MBD14}  &   610.5      			& 614.6				& 610.6 		\\
    ENDF/B-VII.1 				& 	616.5				& ***				& ***			\\
    JEFF-3.2 					& 	***					& 626.0				& ***			\\
    JENDL-4.0 					& 	***					& ***				& 634.0			\\
   \end{tabular}
\end{table}

The comparison shows remarkable agreement within 1\% between the results based on the new TOF data \cite{LCD11,MBD14} and the ENDF/B-VII.1 evaluation, all very well compatible with the direct quotes of Lederer {\it et al.} \cite{LCD11} and Massimi {\it et al.} \cite{MBD14} of $611\pm22$ mb and $613.3\pm6$ mb, respectively. Somewhat larger differences are obtained with the evaluated cross sections from the other libraries. In view of this situation, an improved set of $MACS$ data for $^{197}$Au has been determined by combining the new TOF results \cite{LCD11,MBD14} above 3.5~keV with the ENDF/B-VII.1 evaluation below that energy as summarized in Table \ref{Tab:goldmacs}.\\

\begin{table}[htb]
 \caption{Improved $MACS$ values (mb) of $^{197}$Au for the range of thermal energies of relevance for s-process nucleosynthesis. }
   \label{Tab:goldmacs}
   \renewcommand{\arraystretch}{1.5} % enlarge line spacing
   \begin{tabular}{cc}
    $kT$ (keV)	& $MACS_\mathrm{197Au}$ (mb) 	\\
    \hline
    5			& 	2056 $\pm$ 37\\
    10			& 1241 $\pm$ 14	\\
    15			& 940 $\pm$ 10			\\
    20			& 781 $\pm$ 8					\\
    25			& 	681 $\pm$ 7			\\
    30			& 	612 $\pm$ 6			\\
    40			& 	521 $\pm$ 5	\\
    50			& 	463 $\pm$ 5			\\
    60			& 	423 $\pm$ 4			\\
    80			& 	367 $\pm$ 4		\\
    100			& 	329 $\pm$ 3			\\
   \end{tabular}
\end{table}

The excellent agreement of the new TOF data motivated the Standard Commission of the IAEA to consider a timely revision of the gold cross section and to envisage an extension of the gold standard into the keV region \cite{Cap18}. 

We will discuss the impact of the new cross section on a number of TOF-measurements in the next section. The change of the differential neutron capture cross section of $^{197}$Au has also implication for past and future activation experiments, which used or will use gold as a reference. Ratynski \& K\"appeler recommended $SACS=586$~mb  for the spectrum described in Figs. \ref{fig:25kev_spectra} and \ref{fig:activation_setup} for a spherical sample covering the entire neutron cone. Based on the new cross section, we recommend:
\begin{equation}
SACS_\mathrm{197Au, sphere} = 622.7 \pm 6.2 ~\mathrm{mb}
\end{equation}
For flat samples covering the entire neutron cone, which will be used in most experiments, we recommend
\begin{equation}
SACS_\mathrm{197Au, flat} = 651.6 \pm 6.5 ~\mathrm{mb}
\end{equation}
The difference of 5\% is a result of the fact that low-energy neutrons are on average emitted at larger angles and will therefore pass through more sample material than high-energy neutrons, compare Figs. \ref{fig:activation_setup} and \ref{fig:ratynski_setup}. The cross section differential cross section at higher energies, however, is smaller.

\section{Renormalization of TOF measurements} 
\label{sec:renormalization}

A consequence of the revised MACS-data of $^{197}$Au is the re-evaluation of all cross sections obtained in previous TOF measurements, which were using the gold cross section recommended in \cite{RaK88} as a reference. This concerns, for example, the data listed in the compilation of the Karlsruhe Astrophysical Database of Nucleosynthesis in Stars (KADoNiS) \cite{DHK06}, which need to be corrected as they were consistently normalized to that work. In particular, this holds true for all TOF measurements carried out at the Karlsruhe Van-de-Graaff accelerator. 

To this end, the $MACS$ are calculated according to Eq.~(\ref{eq:MACS_exp_lib}), where $E_{low}$ and $E_{high}$ define the range of experimental data, $\sigma_{exp}$ denotes the experimental results- if possible directly the cross section ratio to $^{197}$Au(n,$\gamma$), and $\eta$ is the respective normalization factor over the range of the experimental data, which falls in the range between 0.8 and 1.2. The experimental data are complemented with evaluated cross sections, $\sigma_{lib}$, which were taken from ENDFB-VII.1 except for a few cases.

The adopted procedure is using the following input:

\begin{itemize}
  \item{$^{197}$Au}
  \begin{itemize}
    \item{$I_{\mathrm{low}}$} - 5\%, ENDF/B-VII.1, \cite{CHO11}
    \item{$I_{\mathrm{high}}$} - 1\%, ENDF/B-VII.1, \cite{CHO11}
    \item{$I_{\mathrm{exp}}$} - 1\% systematics, statistics as quoted \cite{LCD11,MBD14}
  \end{itemize}  
  \item{$^{170, 171, 172,173, 174, 176}$Yb}
  \begin{itemize}
    \item{$I_{\mathrm{low}}$} - 20\%, JENDL-4.0 - \cite{KSO11}
    \item{$I_{\mathrm{high}}$} - 10\%, JENDL-4.0 - \cite{KSO11}
    \item{$I_{\mathrm{exp}}$} - as quoted \cite{WVA00a}
  \end{itemize}  
  \item{$^{180}$Ta}
  \begin{itemize}
    \item{$I_{\mathrm{low}}$} - 20\%, JEFF-3.2 \cite{JEFF30,JEFF31}
    \item{$I_{\mathrm{high}}$} - 10\%, JEFF-3.2 \cite{JEFF30,JEFF31}
    \item{$I_{\mathrm{exp}}$} - as quoted \cite{WVA04}
  \end{itemize}  
  \item{All other isotopes}
  \begin{itemize}
    \item{$I_{\mathrm{low}}$} - 20\%, ENDF/B-VII.1, \cite{CHO11}
    \item{$I_{\mathrm{high}}$} - 10\%, ENDF/B-VII.1, \cite{CHO11}
    \item{$I_{\mathrm{exp}}$} - as quoted 
  \end{itemize}  
\end{itemize}  

\begin{figure}
\begin{center}
 \renewcommand{\baselinestretch}{1}
 \includegraphics{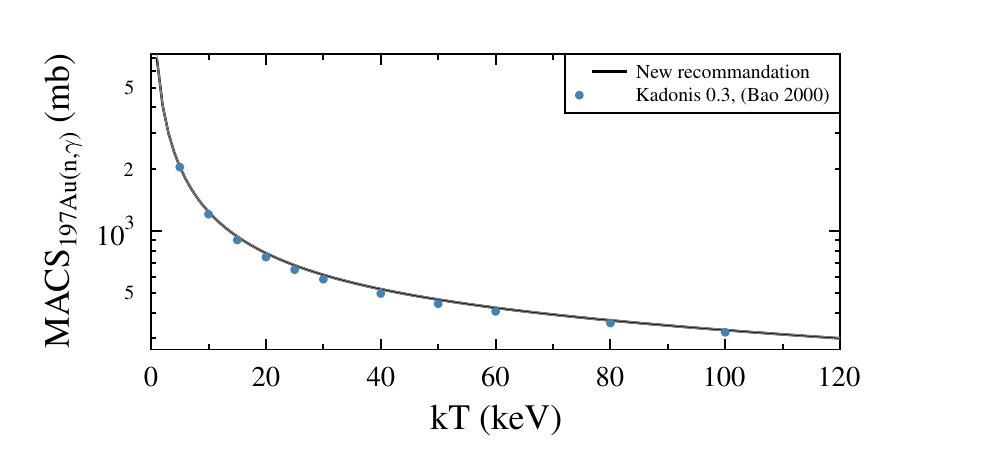}
 \caption{\label{fig:197au_comparison0} New (based on \cite{LCD11,MBD14,CHO11})and old \cite{RaK88,BBK00,DHK06}  recommendation for the $MACS$ of $^{197}$Au.}
\end{center}
\end{figure}

\begin{figure}
\begin{center}
 \renewcommand{\baselinestretch}{1}
 \includegraphics{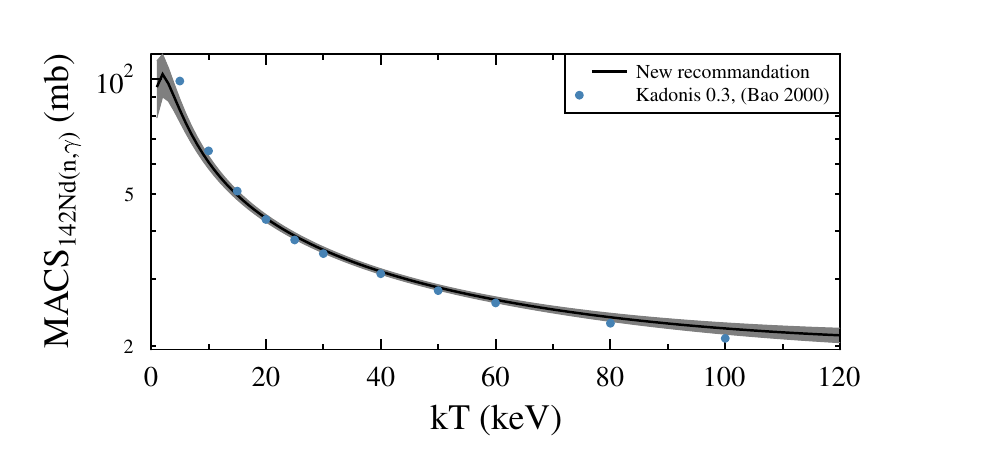}
 \caption{\label{fig:142_comparison0} New and old \cite{WVK98a,WVK98b,BBK00,DHK06} recommendation for the $MACS$ of $^{142}$Nd.}
\end{center}
\end{figure}

\begin{figure}
\begin{center}
 \renewcommand{\baselinestretch}{1}
 \includegraphics{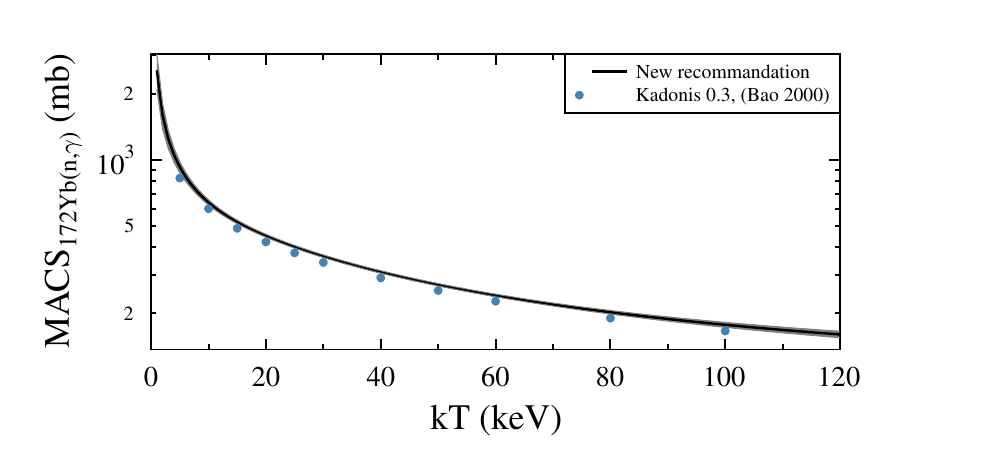}
 \caption{\label{fig:172yb_comparison0} New and old \cite{WVA00a,BBK00,DHK06} recommendation for the $MACS$ of $^{172}$Yb.}
\end{center}
\end{figure}

\begin{figure}
\begin{center}
 \renewcommand{\baselinestretch}{1}
 \includegraphics{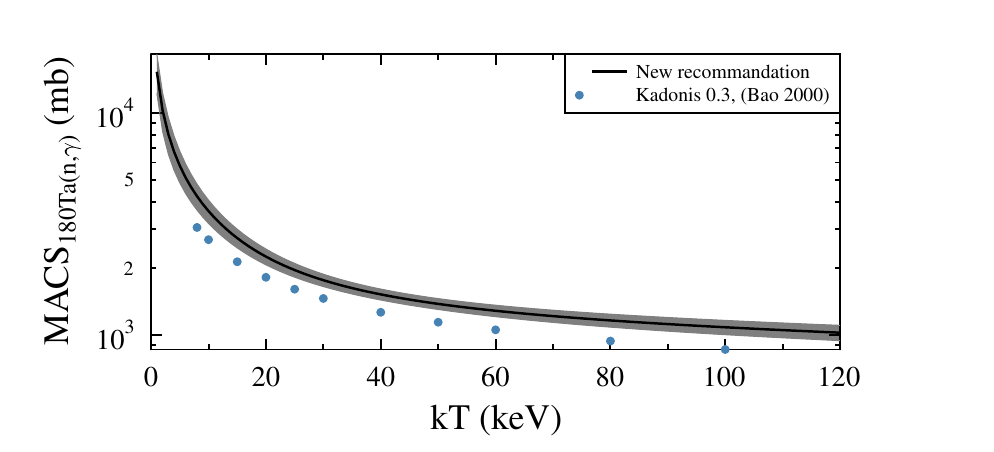}
 \caption{\label{fig:180ta_comparison0} New and old \cite{WVA04,BBK00,DHK06}  recommendation for the $MACS$ of $^{180}$Ta.}
\end{center}
\end{figure}

The essential features of the updated $MACS$ tables are illustrated at a few characteristic examples. The gold $MACS$ in Fig.~\ref{fig:197au_comparison0} represents a case, which could be based on accurate energy-differential data in all three regions. Accordingly, the only difference to the previous KADoNiS versions is due to the 5\% renormalization described in Sec. \ref{sec:au}. The $MACS$ values for $^{172}$Yb are typical for most cases derived from the accurate TOF data obtained with the Karlsruhe 4$\pi$ BaF$_2$ detector, which are spanning a neutron energy range from a few to about 220~keV. Accordingly, the larger uncertainties from the evaluated data adopted in the low and high energy regions are affecting the $MACS$ data at $kT\geq70$ keV (Fig.\ref{fig:172yb_comparison0}). Somewhat stronger effects in the low energy part are observed for isotopes with a pronounced resonance structure, e.g. for $^{142}$Nd (Fig.\ref{fig:142_comparison0}). An illustrative example is the case of $^{180}$Ta (Fig~\ref{fig:180ta_comparison0}). The extremely rare isotope was only available with an enrichment of 5.5\% \cite{WVA01,WVA04}. The remainder of the sample was $^{181}$Ta. This resulted in a very limited range of experimental data. The lowest energy bin was 10-12.5~keV. In particular the $MACS$ at low temperatures is therefore basically not constrained by experimental data. The evaluations used by Wisshak $et~al.$ \cite{NKR92,RaT00,BeM82} to supplement the experimental data of $^{180}$Ta(n,$\gamma$) had a different energy dependence than the currently used evaluation. 

In total 64 sets of MACS data have been updated and will be included in the next version of KADoNiS (1.0). The corresponding $MACS$ can be found in the appendix.

\section{Indirect approaches}
\label{sec:indirect}

Direct measurements of neutron-induced cross sections are particularly difficult. Indirect methods are therefore often the only possibility to improve our knowledge. Well established indirect approaches are replacing the (n,$\gamma$) reaction with a surrogate reaction or measuring the time-reversed  ($\gamma$,n) reaction. So far not done at all is the inverse kinematics approach, which is in fact also a direct measurement.

\subsection{Surrogate}

\begin{figure}
\begin{center}
 \renewcommand{\baselinestretch}{1}
 \includegraphics[width=0.49\textwidth]{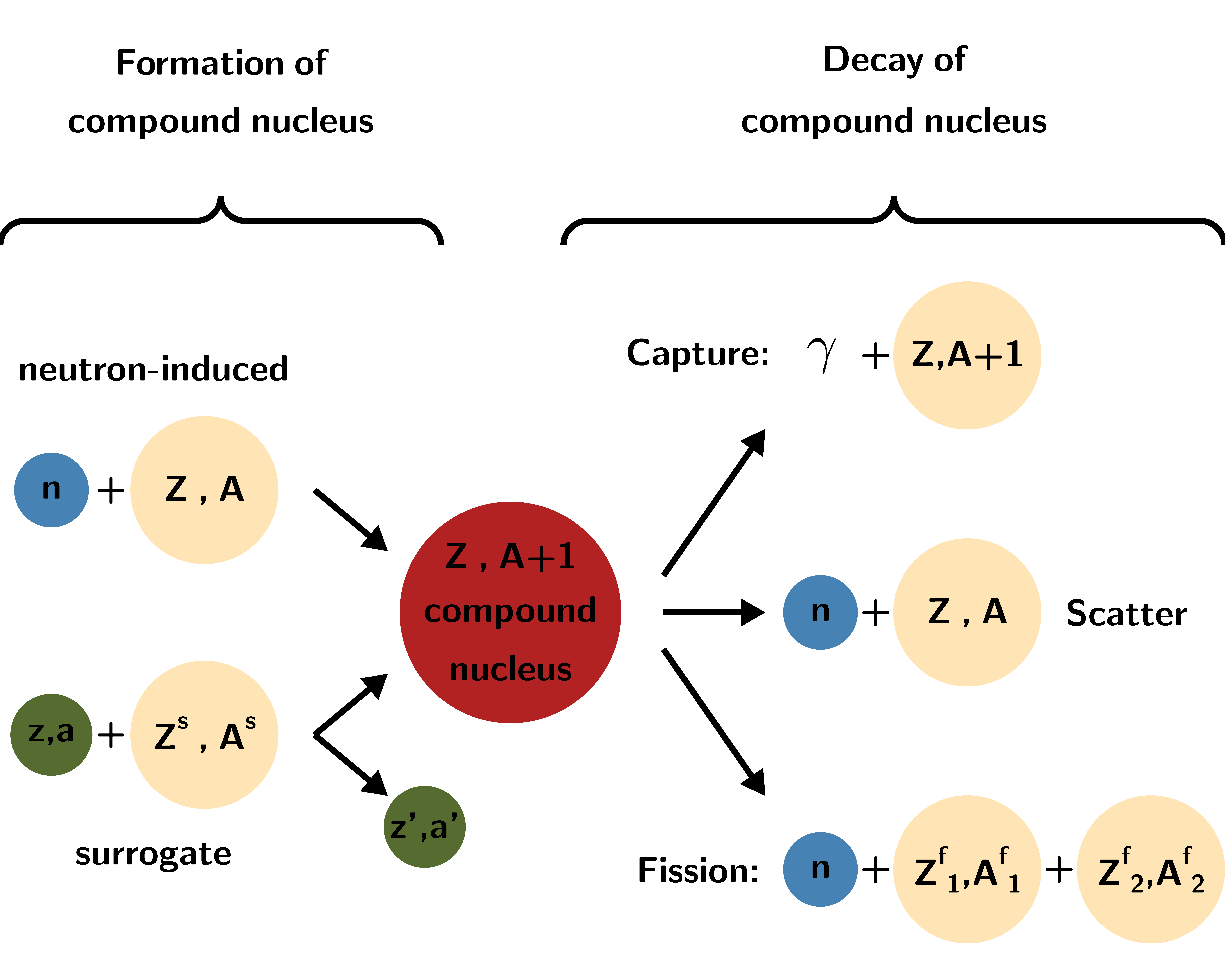}
 \caption{\label{fig:n_induced_fission_ng_surrogate} A neutron-induced reaction can be described as a multi-step
process. The nucleus will first absorb the neutron forming a compound
nucleus in an excited state. Afterwards it can either de-excite via $\gamma$-emission
(capture), neutron emission (inelastic scattering) or split into two fragments
(fission). The fission fragments in-turn are highly excited and will typically
emit neutrons before they decay towards the valley of stability. In a surrogate
reaction, the compound nucleus is produced by a different reaction.}
\end{center}
\end{figure}

Surrogate reactions have been successfully used for obtaining neutron-induced fission cross sections
\cite{EAB05}. This approach is using a charged particle reaction for producing the same compound
system as in the neutron reaction of interest, Fig.~\ref{fig:n_induced_fission_ng_surrogate}. In this way, a short-lived target isotope can be replaced by 
a stable or longer-lived target. For neutron capture reactions, however, the method is challenged 
because the compound nucleus that is produced in the surrogate reaction is characterized by a
spin-parity distribution that can be very different from the spin-parity distribution of the compound 
nucleus occurring in the direct (n,$\gamma$) reaction \cite{FDE07,EBD12,EBC17}.

\subsection{Time-reversed}

\begin{figure}
\begin{center}
 \renewcommand{\baselinestretch}{1}
 \includegraphics[width=0.49\textwidth]{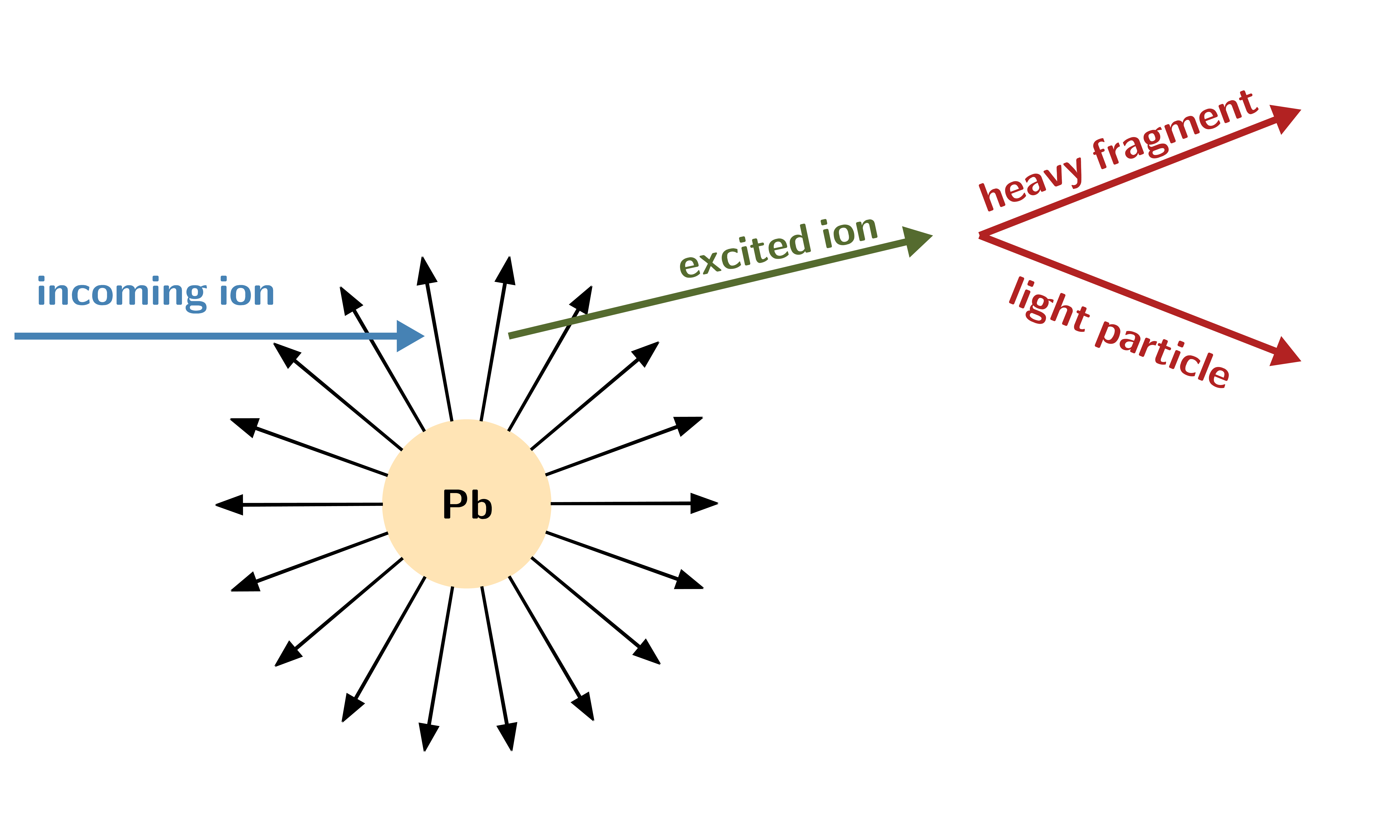}
 \caption{\label{fig:coulomb_breakup} Schematic drawing of the Coulomb dissociation.
If an ion traveling close to the speed of light passes close
by a high-Z nucleus, it can get excited via interaction with
the strong Coulomb field. If the excitation is higher than the
binding energy of the constituents of the nucleus, they can be
emitted, hence dissociated. The emitted particles are typically protons or neutrons, 
but sometimes also $\alpha$-particles.}
\end{center}
\end{figure}

The Coulomb dissociation (CD) method can be used to determine
the desired cross section  of the reaction A(n,$\gamma$)B via the inverse reaction B($\gamma$,n)A
by applying the detailed balance theorem, Fig.~\ref{fig:coulomb_breakup} \cite{BBR86}.  It has been shown that this method can be successfully
applied if the structure of the involved nuclei is not too complicated, as in the case of the reaction 
$^{14}$C(n,$\gamma$)$^{15}$C \cite{RHF08,NFA09}. In case of heavier nuclei, this approach is usually less conclusive, since the CD cross section only 
constraints the direct decay to the ground state of the compound nucleus \cite{UAA14}.
If the reaction product B is short-lived, the 
CD method can be applied at radioactive beam facilities \cite{HTW17}.

Limitations of the CD method are (i) the applicability of this method to heavier nuclei close to the 
valley of stability due to the high level density in the compound nucleus, and (ii) because the 
resolution of current facilities of $\geq100$~keV is not sufficient to constrain the astrophysically 
relevant cross section.

Restriction (i) is alleviated in r-process studies, because the level density is rapidly decreasing
as the Q-values drops towards the neutron drip line. This implies that fewer levels are important,
and the part of the capture cross section, which can be constrained via the inverse reaction, increases. 
Restriction (ii) motivated the 
development of improved experimental approaches such as NeuLAND@FAIR, which aims for an 
energy resolution of better than 50~keV \cite{BAA11}.

If the product is stable or very long-lived, also real photons can be used to study B($\gamma$, n)A
reactions \cite{SMV03,Wel08}. In principle the same restrictions apply as for the Coulomb dissociation method.

\subsection{Inverse kinematics}

\begin{figure}
\begin{center}
 \renewcommand{\baselinestretch}{1}
 \includegraphics[width=0.49\textwidth]{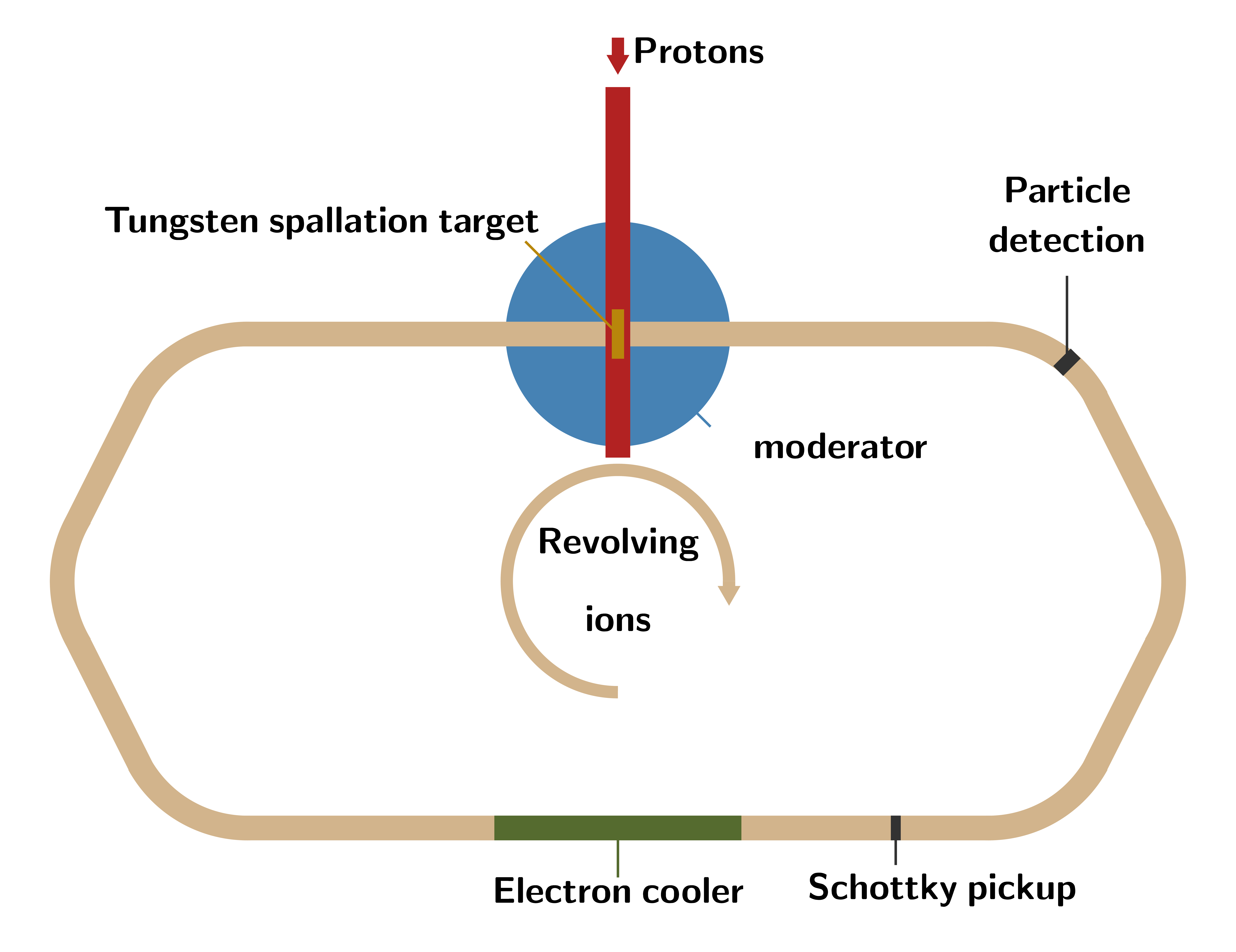}
 \caption{\label{fig:ring_spallation} Neutrons are
produced by protons impinging on a tungsten spallation target
(brown). The proton beam pipe (red) is orientated perpendicular
to the ion beam pipe (light brown). The beam pipes do not
intersect. The neutrons produced in the spallation process are
moderated in the surrounding moderator (blue). They penetrate
the ion beam pipe and act as a neutron target for the ions. The ion
beam pipe is part of a storage ring outside the moderator. The
storage ring may contain additional equipment like an electron
cooler (green), Schottky pickups and particle detectors (gray).}
\end{center}
\end{figure}

A completely different approach is to investigate neutron-induced reactions in inverse kinematics. 
This requires a beam of radioactive ions cycling in a storage ring with 100~$A$keV or 
less and a neutron target. Radioactive ions close to stability can be produced with high intensities 
using ISOL-techniques and storage rings for low beam energies, which require extremely high 
vacuum, are under construction, e.g. the CRYRING at GSI/FAIR \cite{HHS13} or the CSR at 
MPK/Heidelberg \cite{HBB11}. The neutron target could be either a reactor coupled with the 
storage ring to obtain an interaction zone near the core \cite{ReL14} or the moderator surrounding 
a spallation target \cite{RGH17}. Different materials with low neutron-absorbing cross sections like D$_{2}$O, Be or C are suited
\cite{RBD18}. The scheme of such a setup is sketched in Fig.~\ref{fig:ring_spallation}.

\section{Conclusion}
Most of the experiments determining neutron-induced cross sections in the astrophysically important energy regime between 1~keV and 1~MeV are either based on the activation or the time-of-flight method. Even after decades of application, both techniques have lots of potential for improvements. Very often nuclear data, which are used during the analysis of the experiments, will get improved later. This includes decay properties but also reference cross sections. A careful re-evaluation of published results is only possible, if all the necessary raw data are provided. The new evaluation of the $^{197}$Au(n,$\gamma$) cross section implied the re-evaluation of 63 other isotopes with experimental information from TOF experiments. We recommend a new spectrum-averaged cross section for the widely used $^7$Li(p,n) activation setup with neutron energies around $kT\approx25$~keV. This will affect many isotopes and will be published separately. 

\begin{acknowledgement}
This research has received funding from the European Research Council under the European Unions's Seventh Framework Programme (FP/2007-2013) / ERC Grant Agreement n. 615126, the DFG (RE 3461/4-1) and HIC for FAIR.
\end{acknowledgement}

\section*{Appendix} 
\label{sec:appendix}
 
\begin{figure}
\begin{center}
 \renewcommand{\baselinestretch}{1}
 \includegraphics{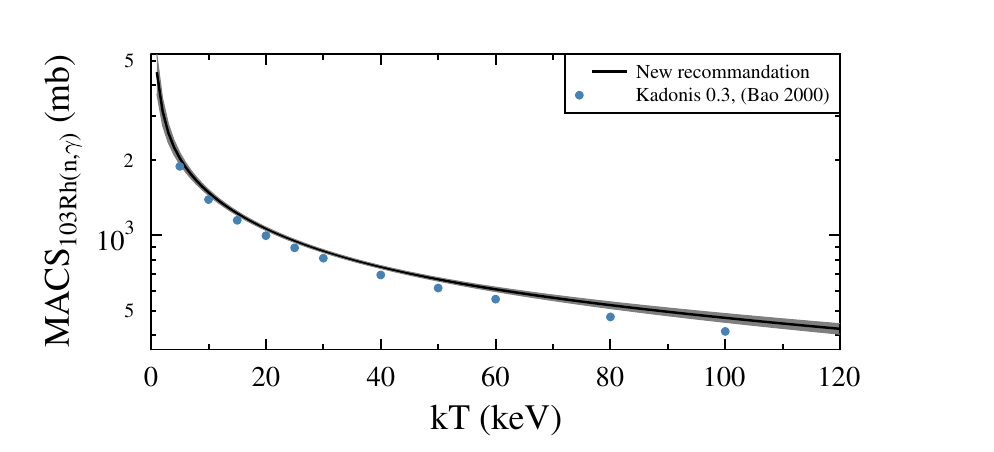}
 \includegraphics{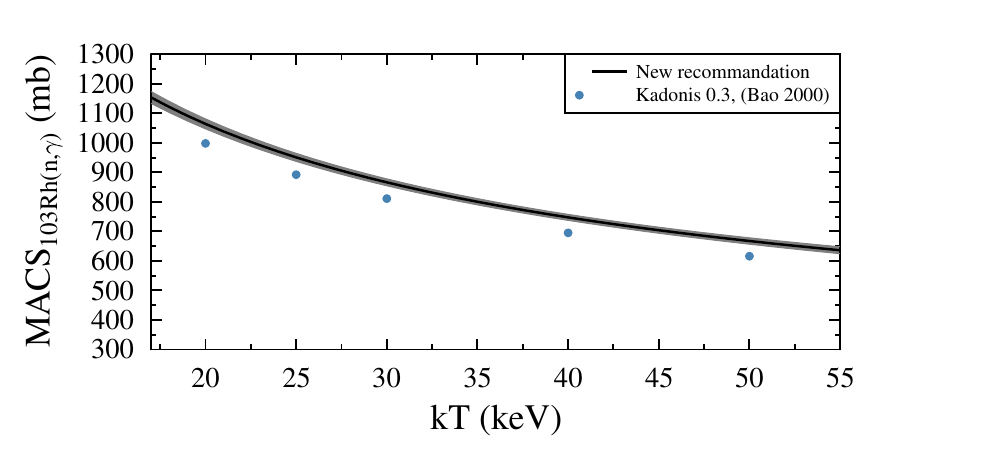}
 \caption{\label{fig:103rh_comparison} New and old \cite{WVK90,BBK00,DHK06} recommendation for the $MACS$ of $^{103}$Rh.}
\end{center}
\end{figure}

\begin{figure}
\begin{center}
 \renewcommand{\baselinestretch}{1}
 \includegraphics{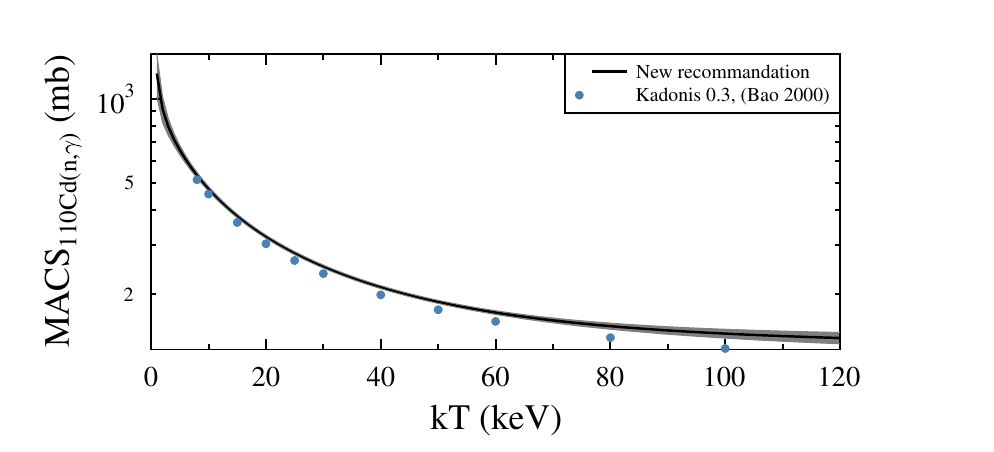}
 \includegraphics{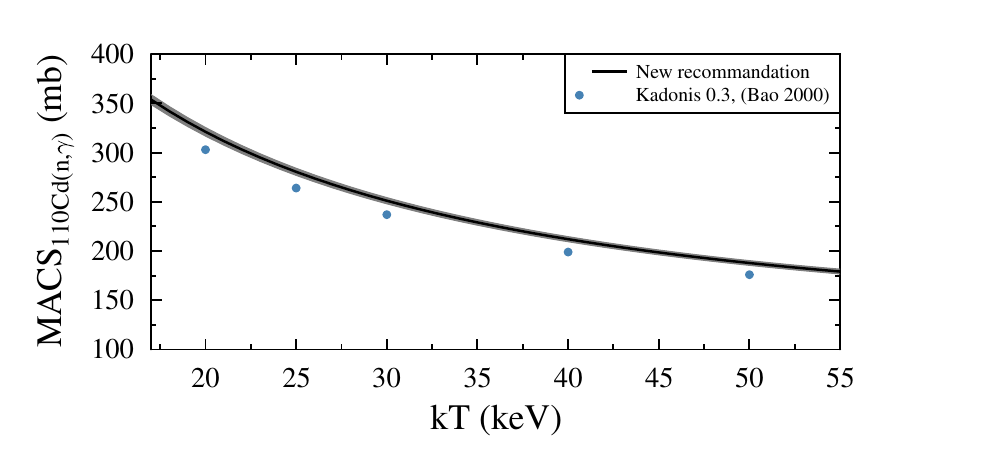}
 \caption{\label{fig:110cd_comparison} New and old \cite{WVK02,BBK00,DHK06}  recommendation for the $MACS$ of $^{110}$Cd.}
\end{center}
\end{figure}

\begin{figure}
\begin{center}
 \renewcommand{\baselinestretch}{1}
 \includegraphics{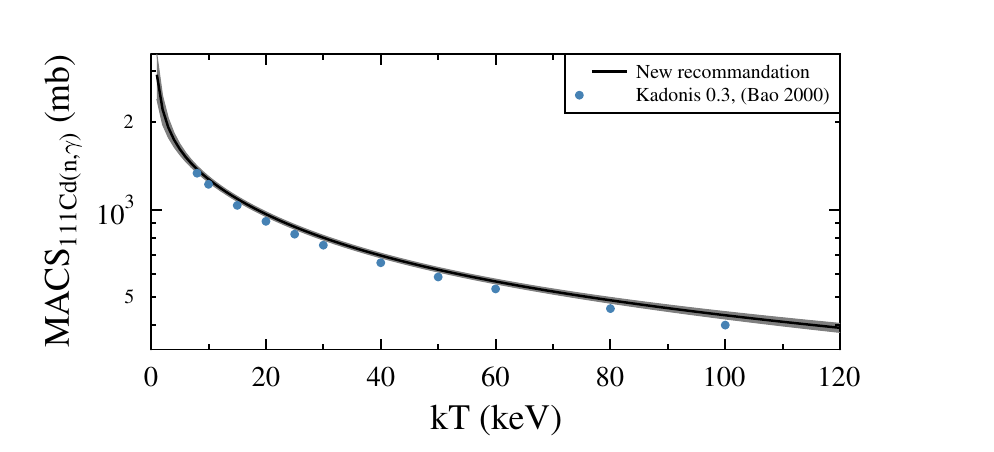}
 \includegraphics{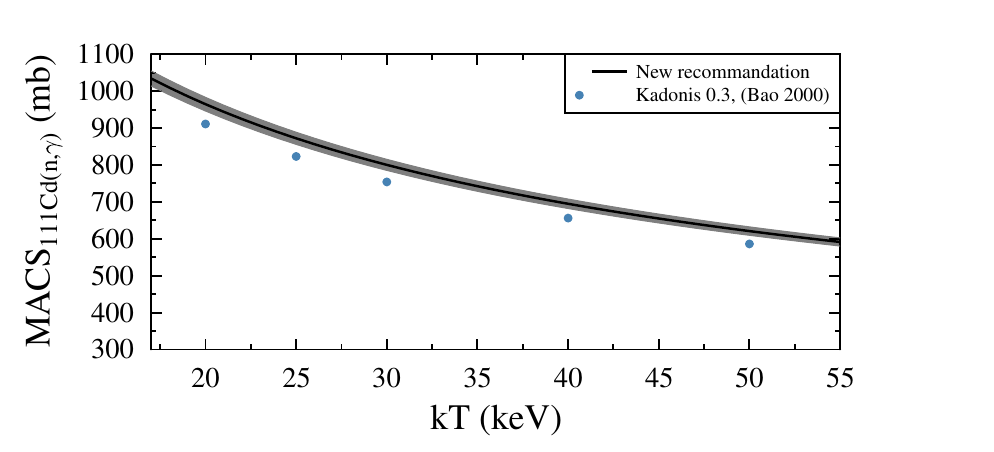}
 \caption{\label{fig:111cd_comparison} New and old \cite{WVK02,BBK00,DHK06}  recommendation for the $MACS$ of $^{111}$Cd.}
\end{center}
\end{figure}

\begin{figure}
\begin{center}
 \renewcommand{\baselinestretch}{1}
 \includegraphics{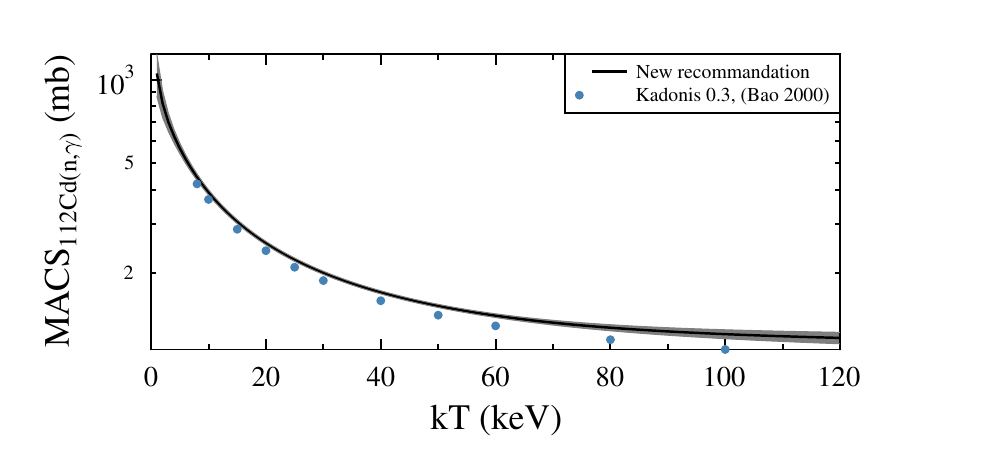}
 \includegraphics{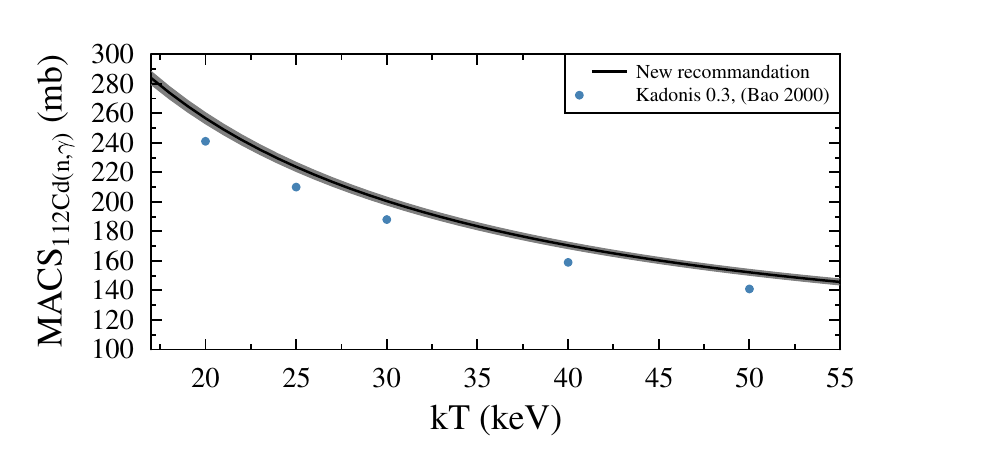}
 \caption{\label{fig:112cd_comparison} New and old \cite{WVK02,BBK00,DHK06}  recommendation for the $MACS$ of $^{112}$Cd.}
\end{center}
\end{figure}

\begin{figure}
\begin{center}
 \renewcommand{\baselinestretch}{1}
 \includegraphics{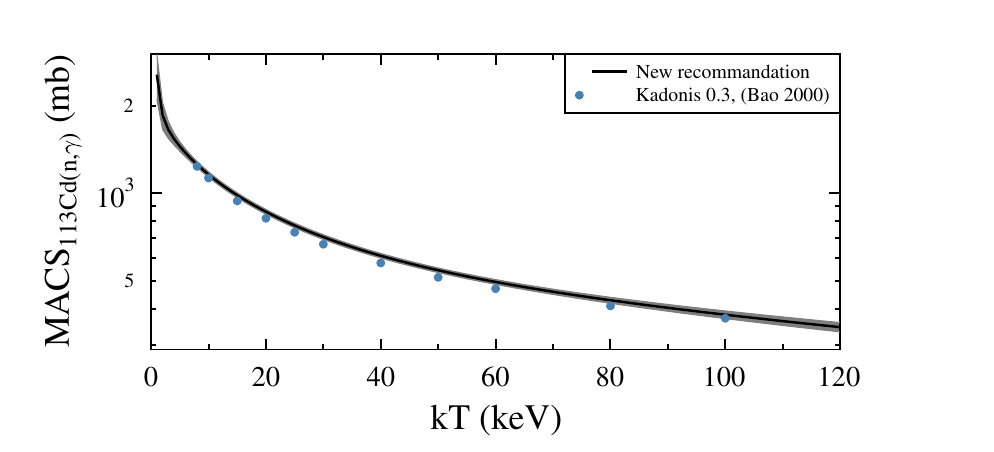}
 \includegraphics{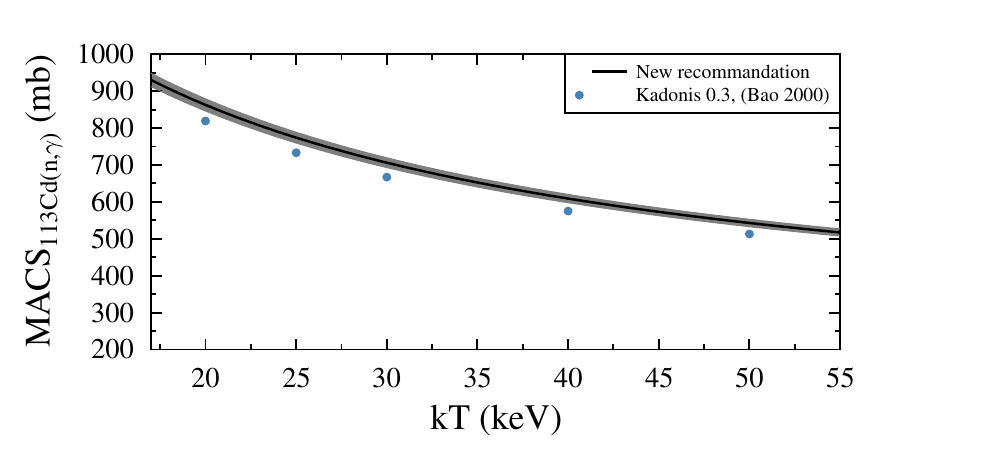}
 \caption{\label{fig:113cd_comparison} New and old \cite{WVK02,BBK00,DHK06}  recommendation for the $MACS$ of $^{113}$Cd.}
\end{center}
\end{figure}

\begin{figure}
\begin{center}
 \renewcommand{\baselinestretch}{1}
 \includegraphics{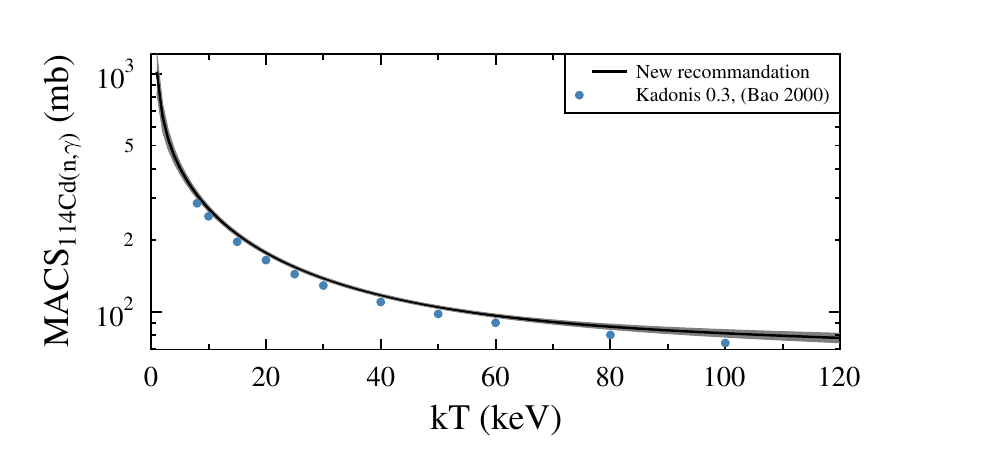}
 \includegraphics{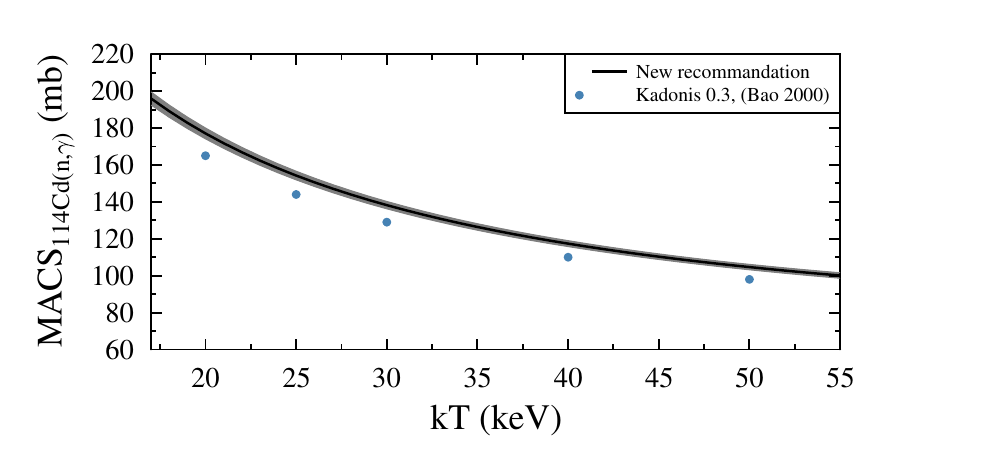}
 \caption{\label{fig:114cd_comparison} New and old \cite{WVK02,BBK00,DHK06}  recommendation for the $MACS$ of $^{114}$Cd.}
\end{center}
\end{figure}

\begin{figure}
\begin{center}
 \renewcommand{\baselinestretch}{1}
 \includegraphics{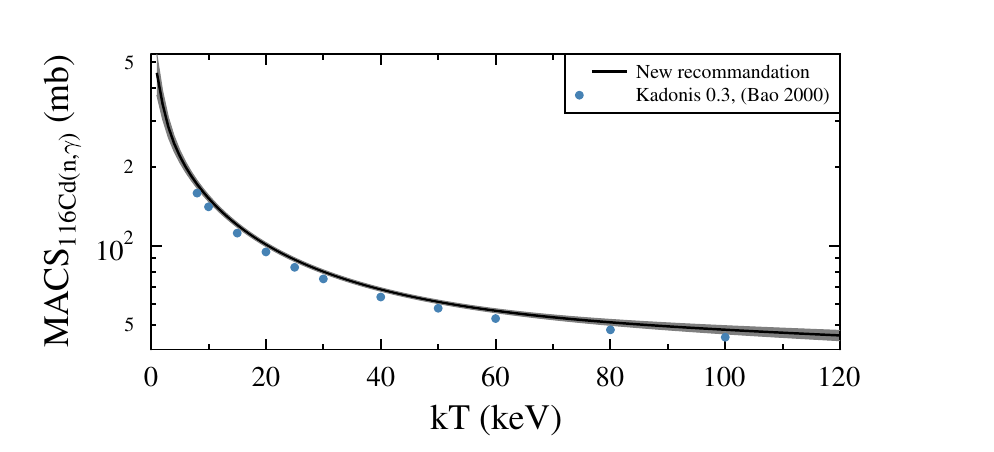}
 \includegraphics{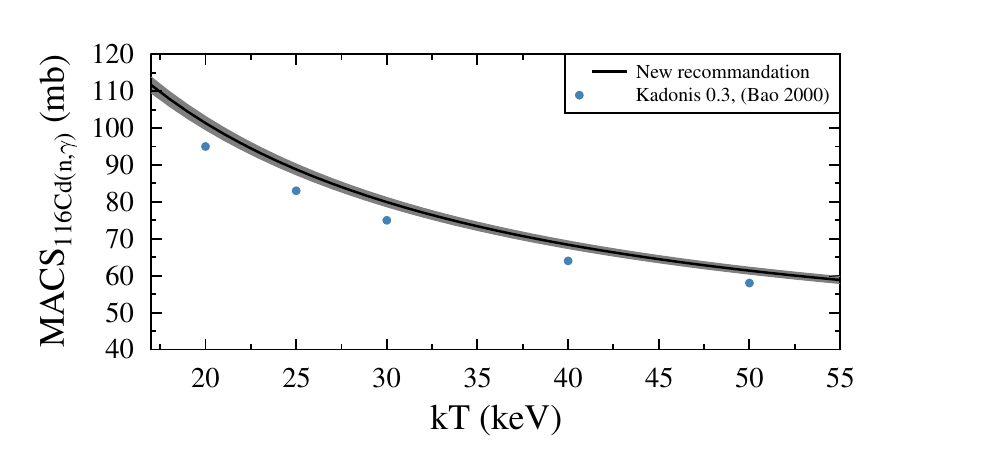}
 \caption{\label{fig:116cd_comparison} New and old \cite{WVK02,BBK00,DHK06}  recommendation for the $MACS$ of $^{116}$Cd.}
\end{center}
\end{figure}

\begin{figure}
\begin{center}
 \renewcommand{\baselinestretch}{1}
 \includegraphics{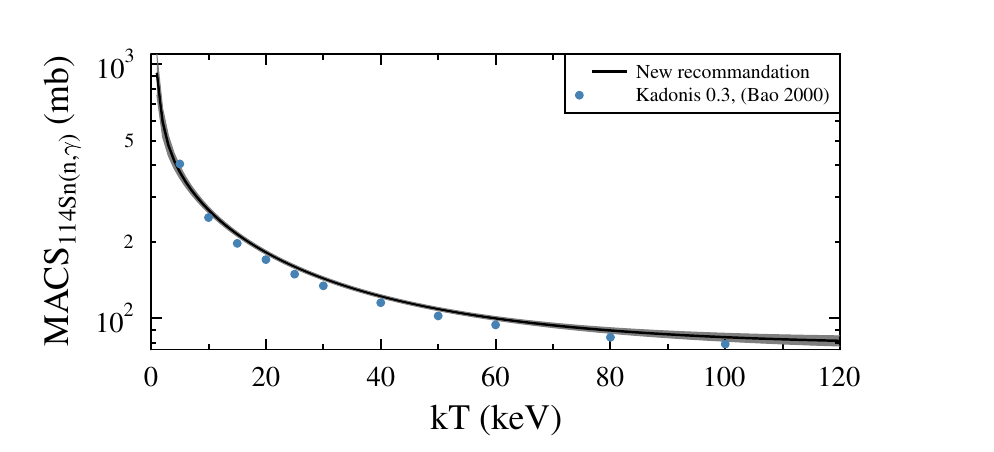}
 \includegraphics{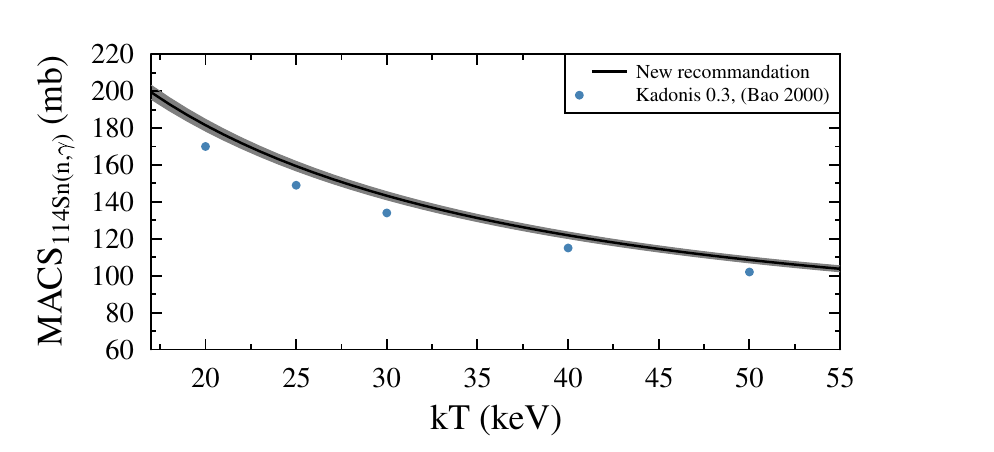}
 \caption{\label{fig:114sn_comparison} New and old \cite{WVT96a,BBK00,DHK06}  recommendation for the $MACS$ of $^{114}$Sn.}
\end{center}
\end{figure}

\begin{figure}
\begin{center}
 \renewcommand{\baselinestretch}{1}
 \includegraphics{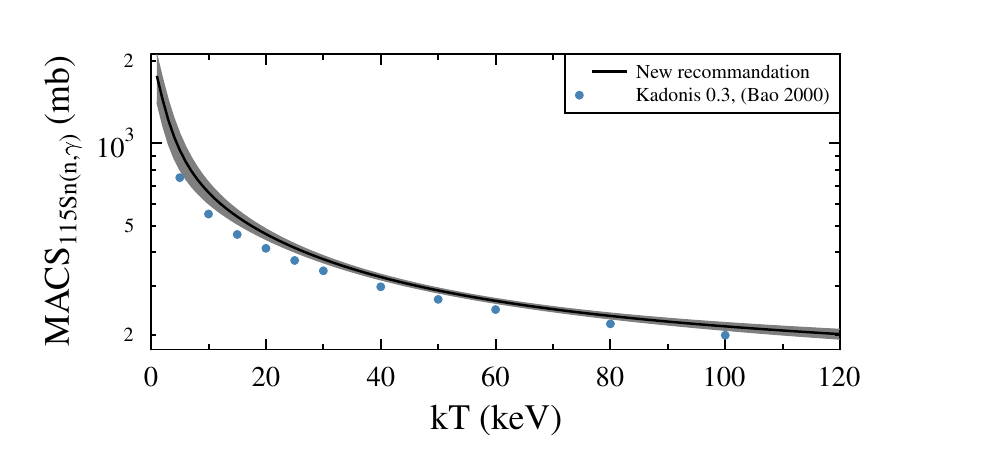}
 \includegraphics{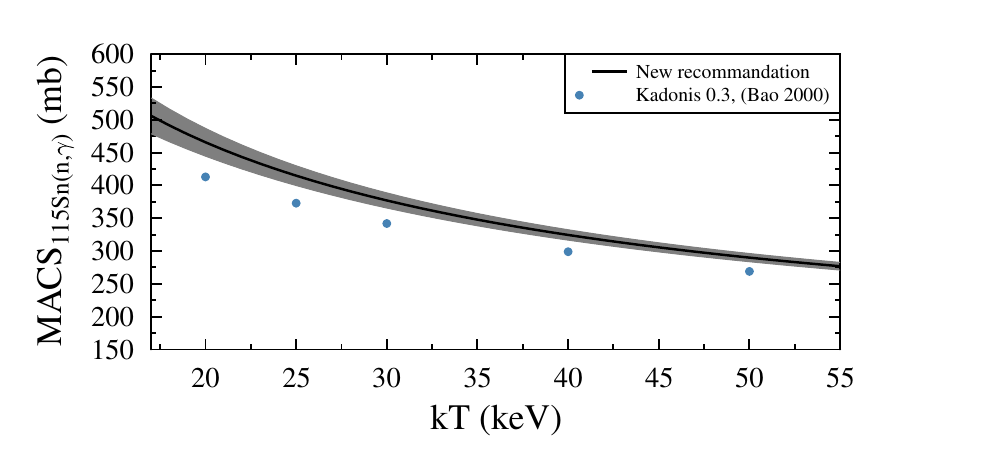}
 \caption{\label{fig:115sn_comparison} New and old \cite{WVT96a,BBK00,DHK06}  recommendation for the $MACS$ of $^{115}$Sn.}
\end{center}
\end{figure}

\begin{figure}
\begin{center}
 \renewcommand{\baselinestretch}{1}
 \includegraphics{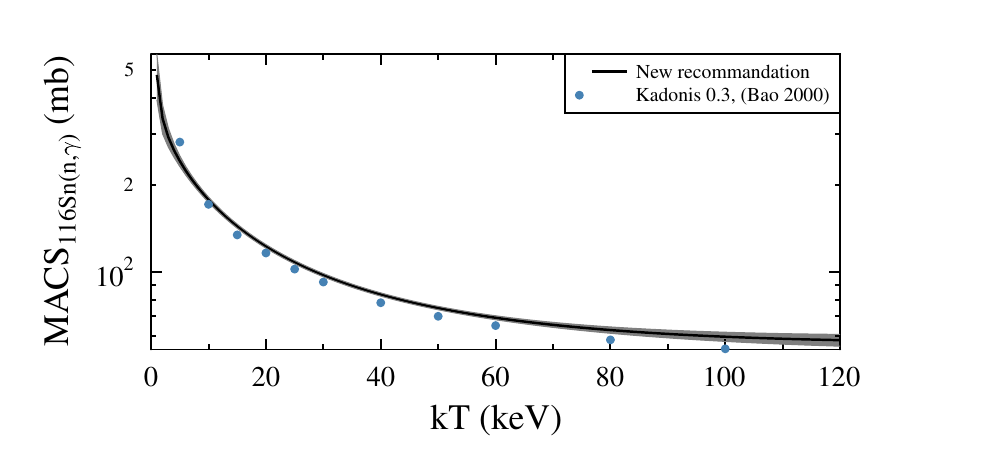}
 \includegraphics{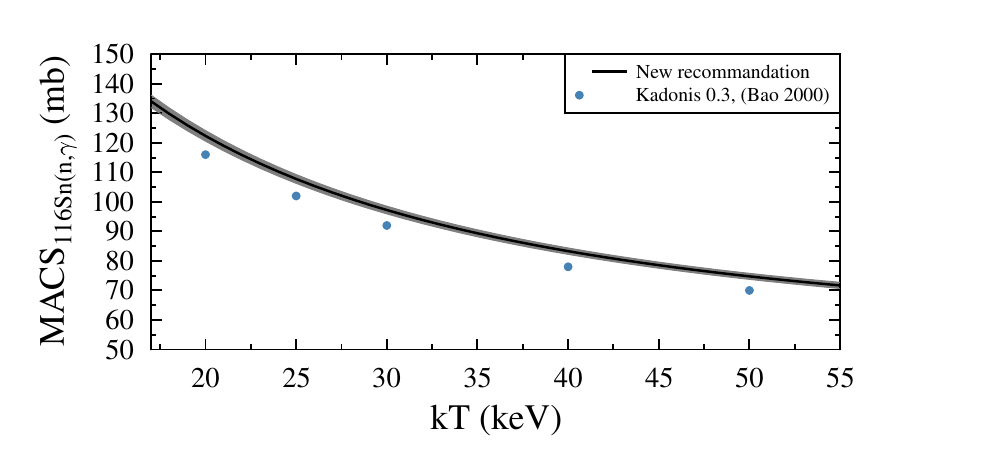}
 \caption{\label{fig:116sn_comparison} New and old \cite{WVT96a,BBK00,DHK06}  recommendation for the $MACS$ of $^{116}$Sn.}
\end{center}
\end{figure}

\begin{figure}
\begin{center}
 \renewcommand{\baselinestretch}{1}
 \includegraphics{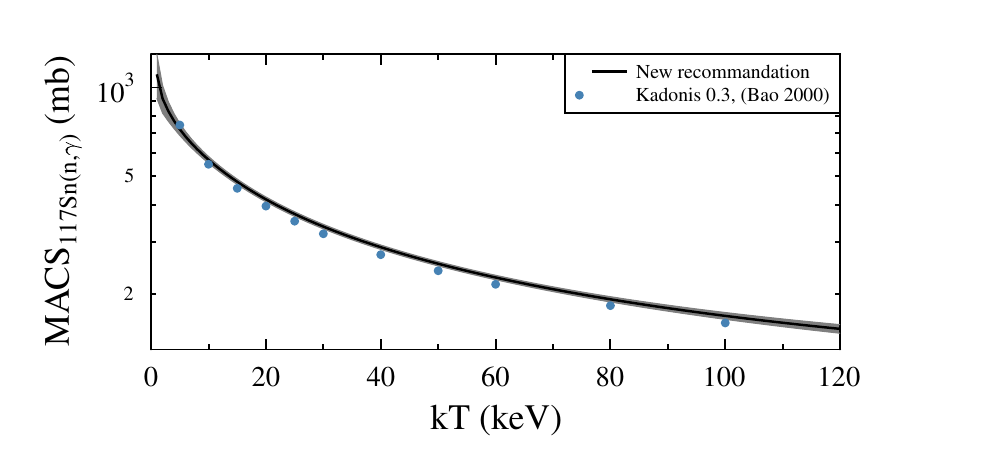}
 \includegraphics{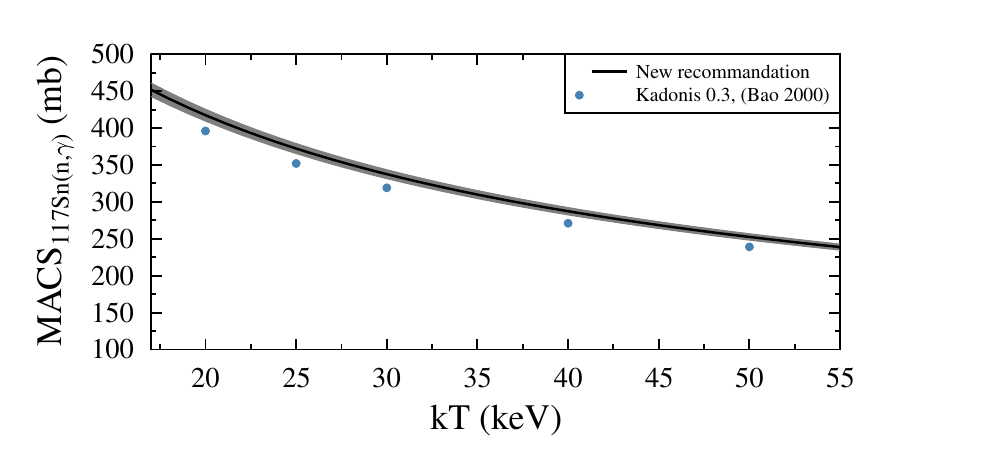}
 \caption{\label{fig:117sn_comparison} New and old \cite{WVT96a,BBK00,DHK06}  recommendation for the $MACS$ of $^{117}$Sn.}
\end{center}
\end{figure}

\begin{figure}
\begin{center}
 \renewcommand{\baselinestretch}{1}
 \includegraphics{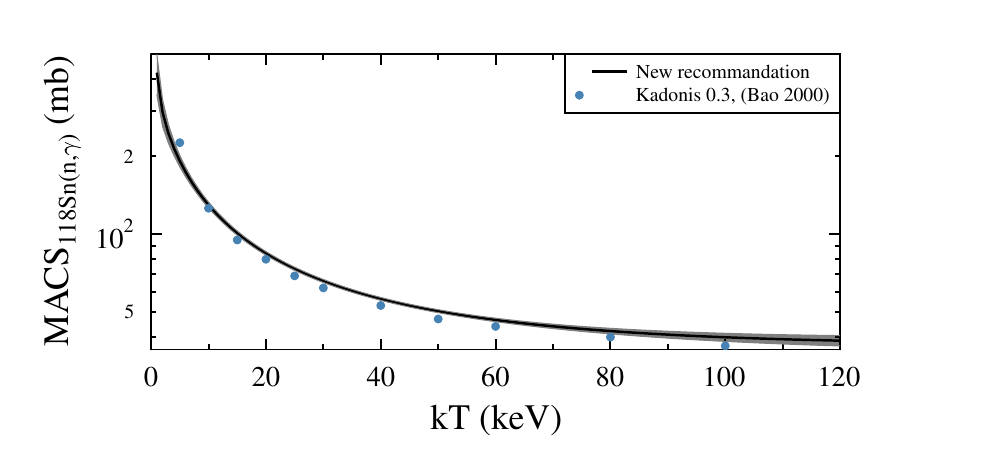}
 \includegraphics{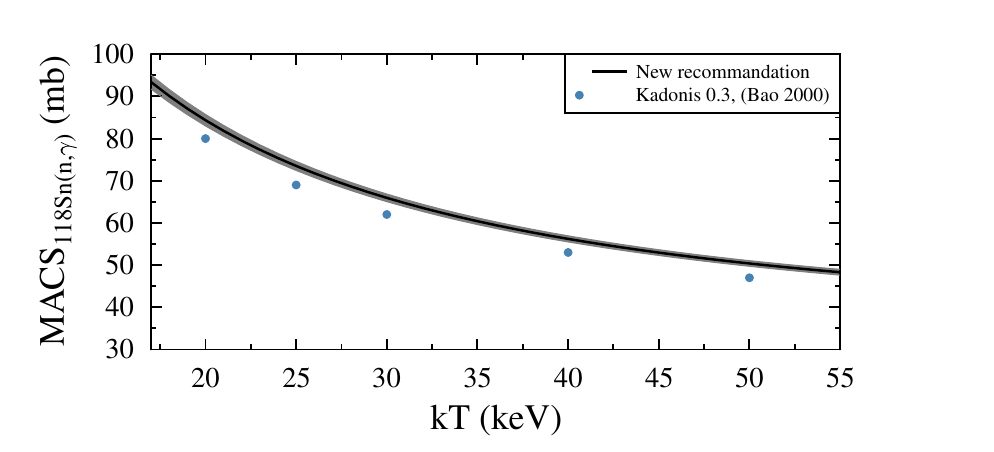}
 \caption{\label{fig:118sn_comparison} New and old \cite{WVT96a,BBK00,DHK06}  recommendation for the $MACS$ of $^{118}$Sn.}
\end{center}
\end{figure}

\begin{figure}
\begin{center}
 \renewcommand{\baselinestretch}{1}
 \includegraphics{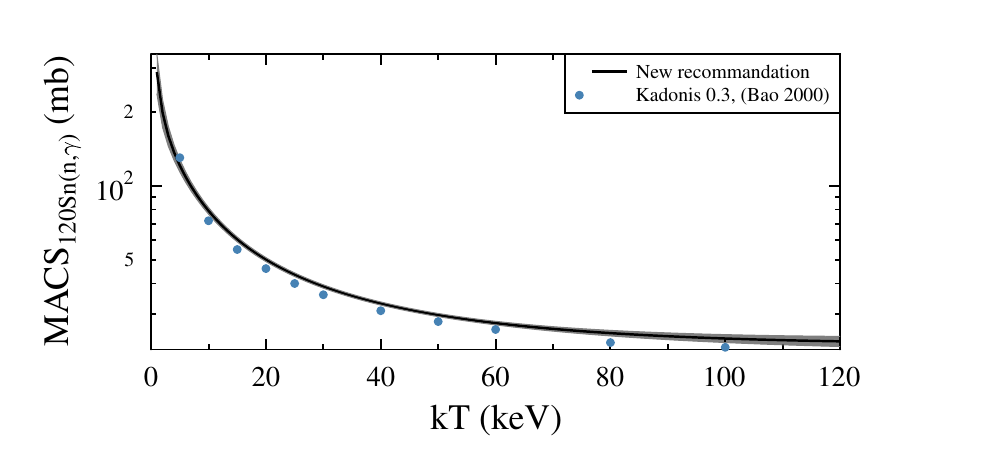}
 \includegraphics{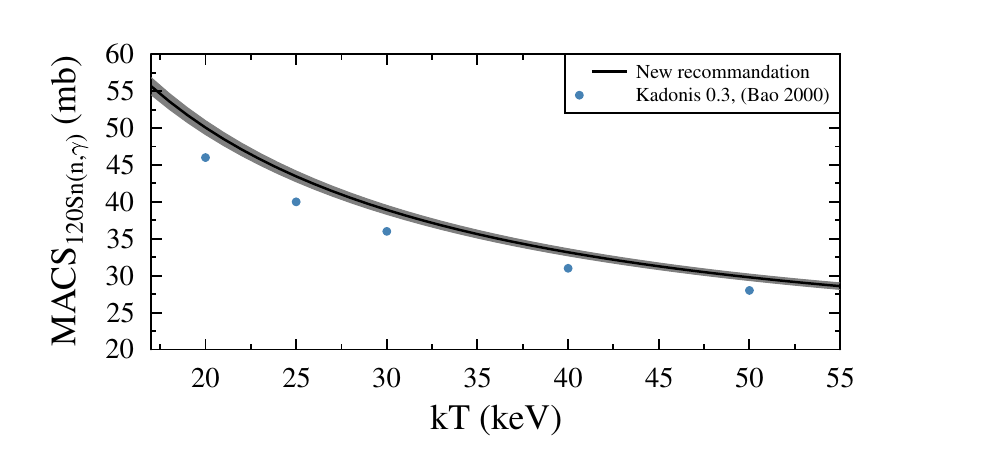}
 \caption{\label{fig:120sn_comparison} New and old \cite{WVT96a,BBK00,DHK06}  recommendation for the $MACS$ of $^{120}$Sn.}
\end{center}
\end{figure}

\begin{figure}
\begin{center}
 \renewcommand{\baselinestretch}{1}
 \includegraphics{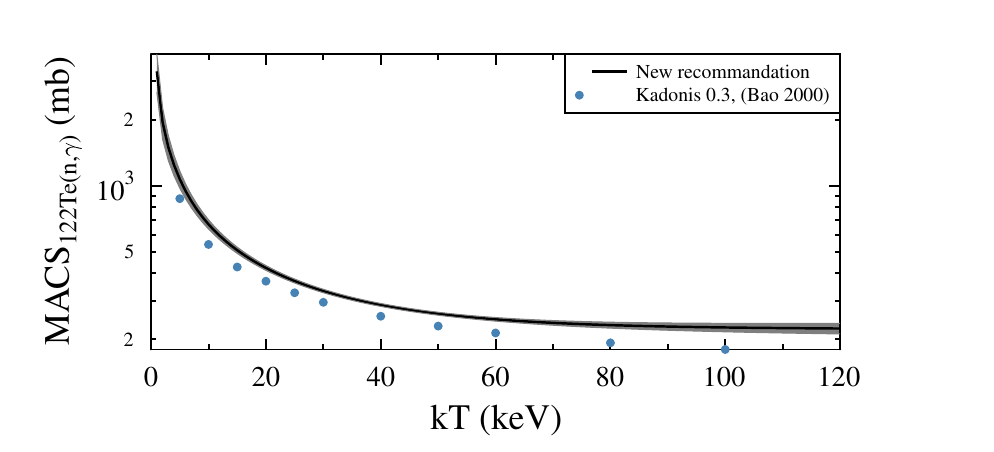}
 \includegraphics{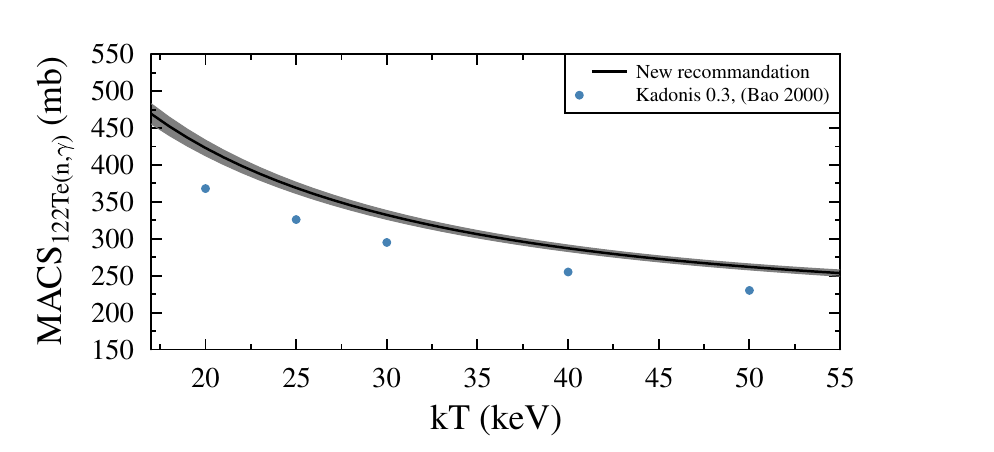}
 \caption{\label{fig:122te_comparison} New and old \cite{WVK92,BBK00,DHK06}  recommendation for the $MACS$ of $^{122}$Te.}
\end{center}
\end{figure}

\begin{figure}
\begin{center}
 \renewcommand{\baselinestretch}{1}
 \includegraphics{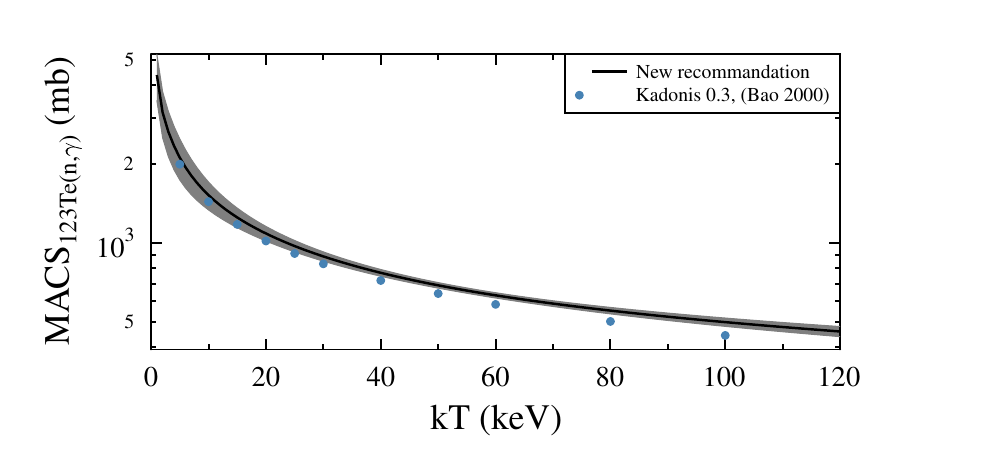}
 \includegraphics{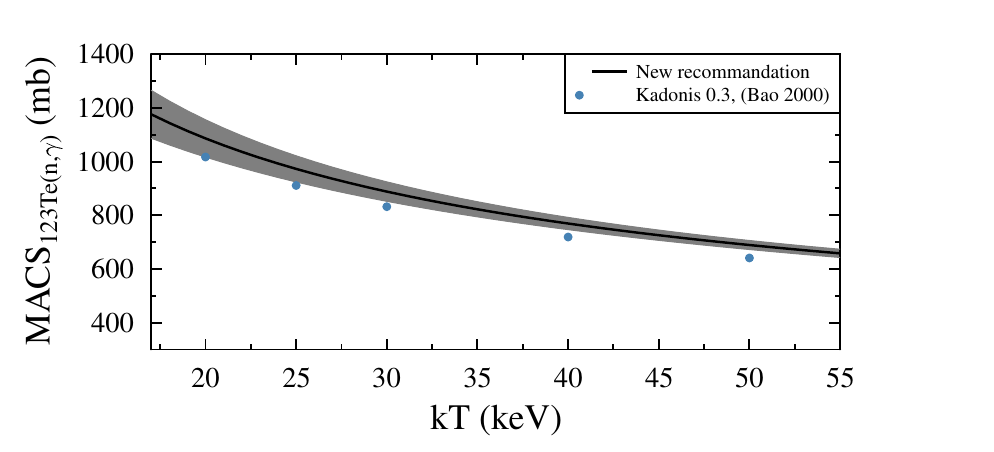}
 \caption{\label{fig:123te_comparison} New and old \cite{WVK92,BBK00,DHK06}  recommendation for the $MACS$ of $^{123}$Te.}
\end{center}
\end{figure}

\begin{figure}
\begin{center}
 \renewcommand{\baselinestretch}{1}
 \includegraphics{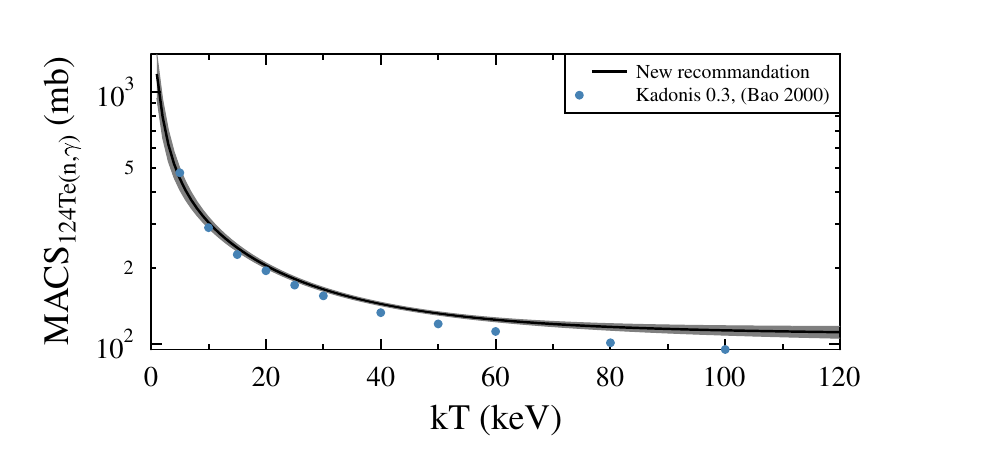}
 \includegraphics{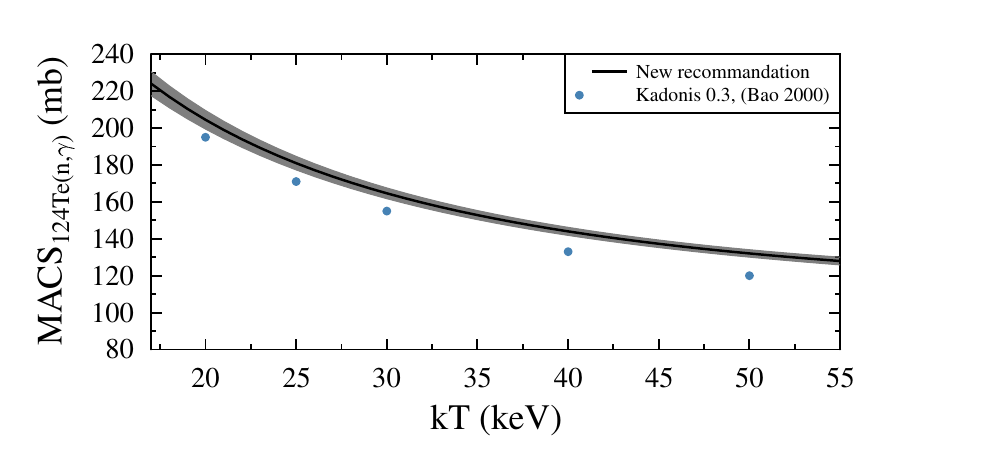}
 \caption{\label{fig:124te_comparison} New and old \cite{WVK92,BBK00,DHK06}  recommendation for the $MACS$ of $^{124}$Te.}
\end{center}
\end{figure}

\begin{figure}
\begin{center}
 \renewcommand{\baselinestretch}{1}
 \includegraphics{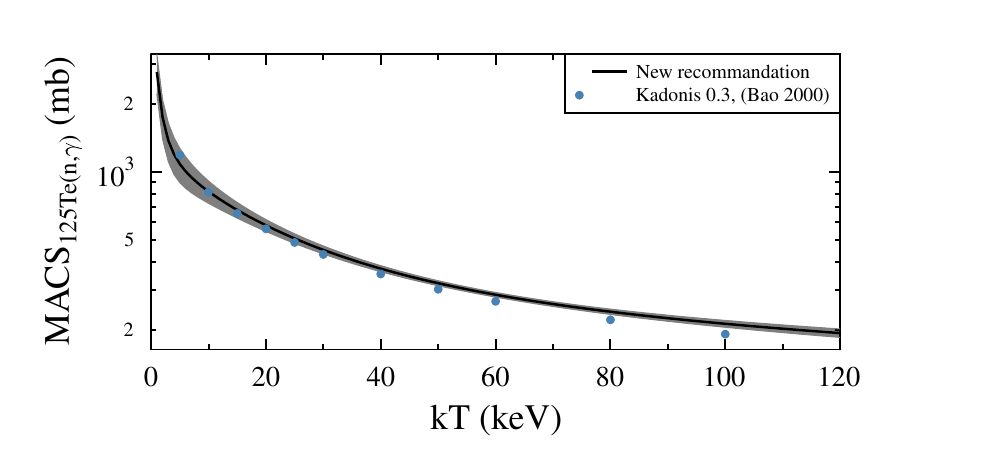}
 \includegraphics{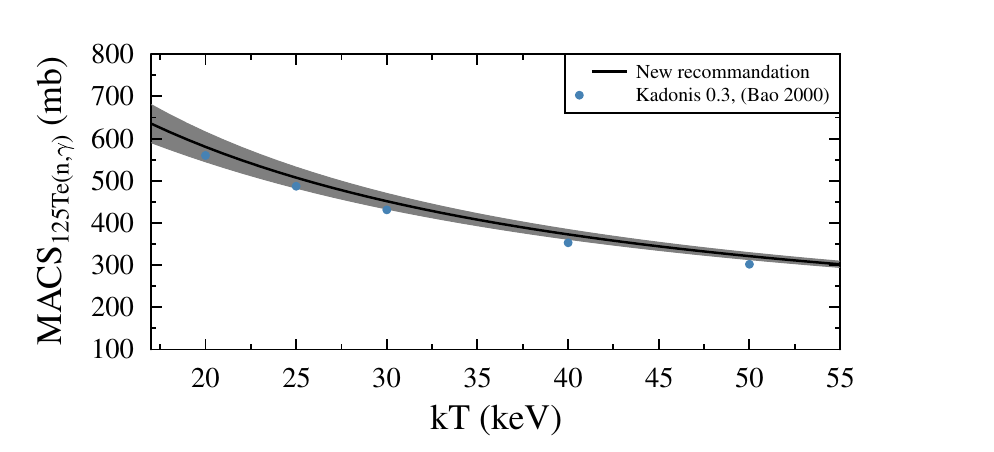}
 \caption{\label{fig:125te_comparison} New and old \cite{WVK92,BBK00,DHK06}  recommendation for the $MACS$ of $^{125}$Te.}
\end{center}
\end{figure}

\begin{figure}
\begin{center}
 \renewcommand{\baselinestretch}{1}
 \includegraphics{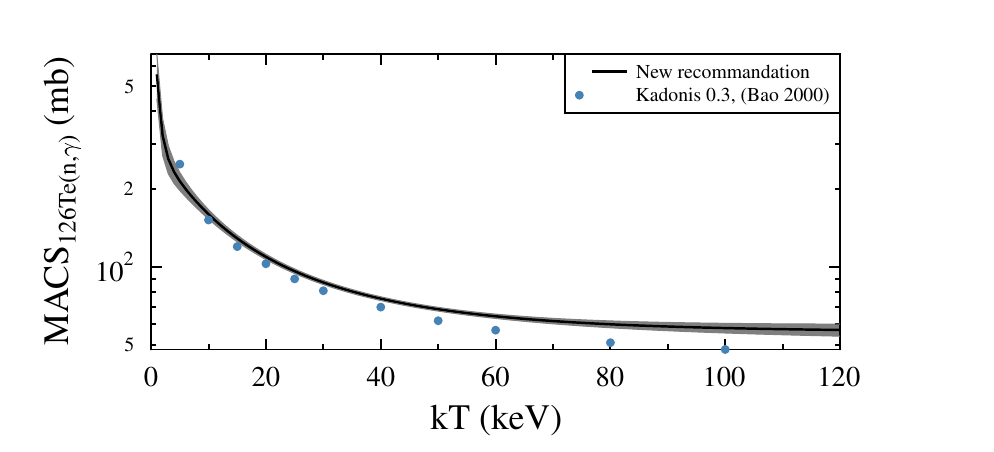}
 \includegraphics{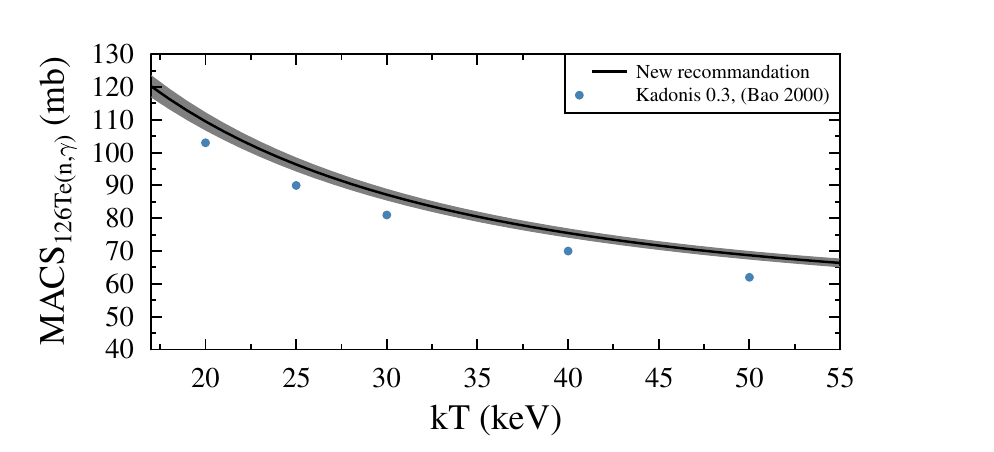}
 \caption{\label{fig:126te_comparison} New and old \cite{WVK92,BBK00,DHK06}  recommendation for the $MACS$ of $^{126}$Te.}
\end{center}
\end{figure}

\begin{figure}
\begin{center}
 \renewcommand{\baselinestretch}{1}
 \includegraphics{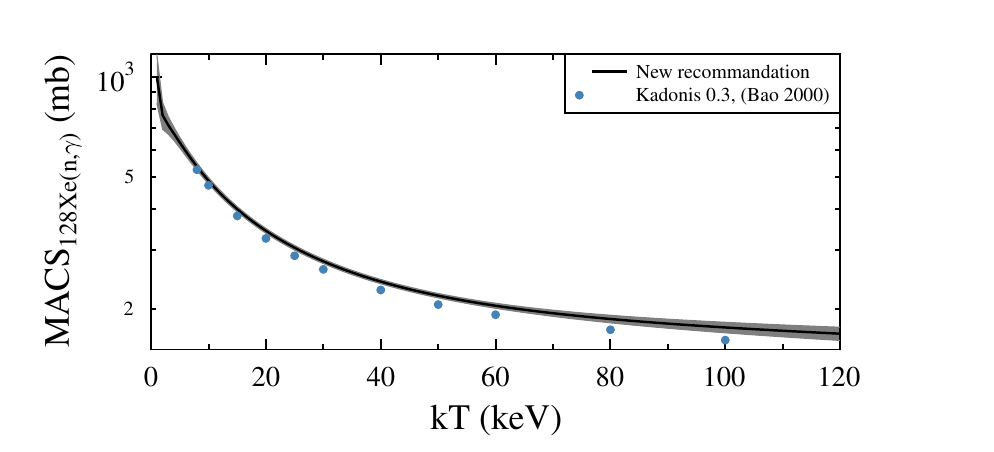}
 \includegraphics{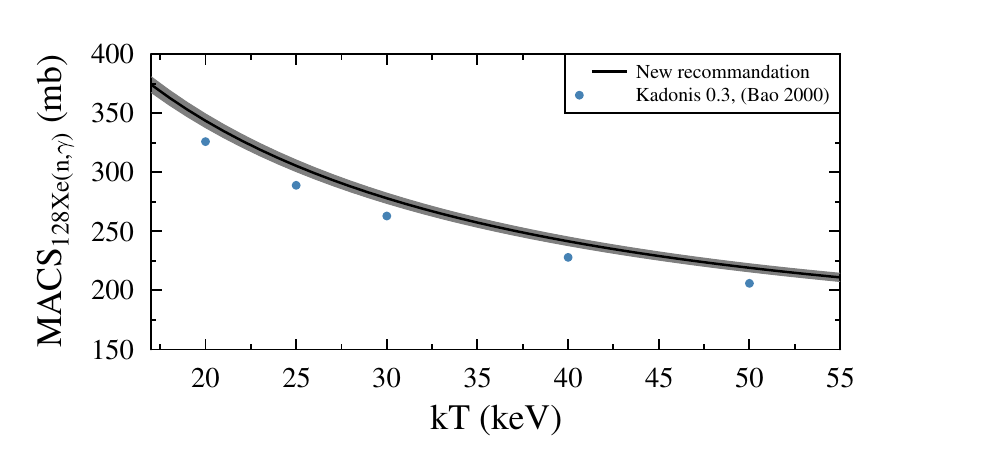}
 \caption{\label{fig:128xe_comparison} New and old \cite{RHK02,BBK00,DHK06}  recommendation for the $MACS$ of $^{128}$Xe.}
\end{center}
\end{figure}

\begin{figure}
\begin{center}
 \renewcommand{\baselinestretch}{1}
 \includegraphics{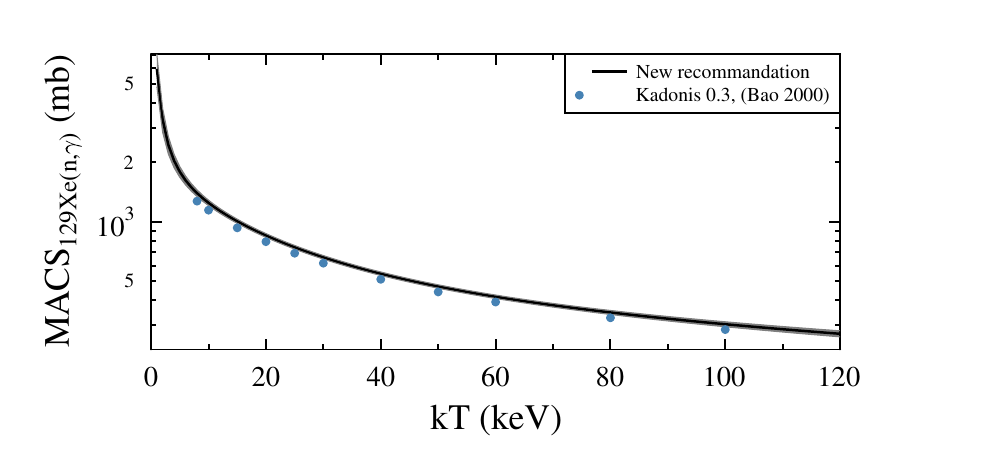}
 \includegraphics{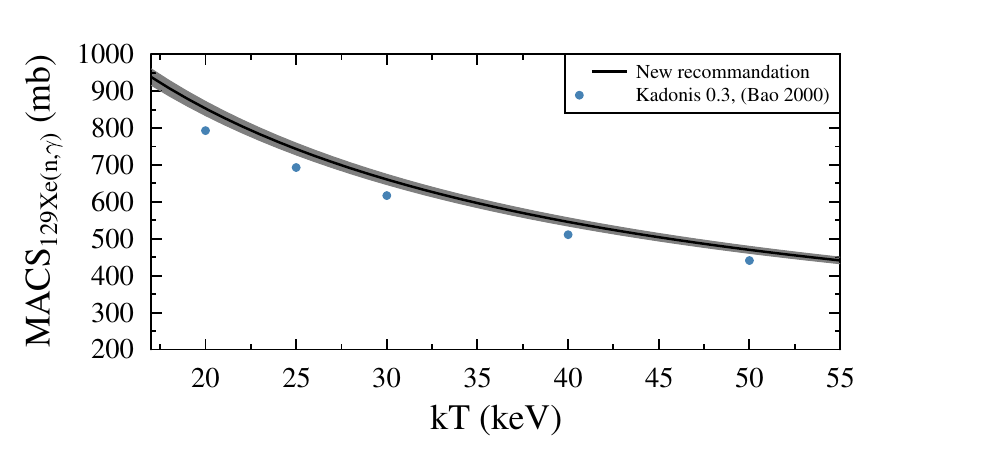}
 \caption{\label{fig:129xe_comparison} New and old \cite{RHK02,BBK00,DHK06}  recommendation for the $MACS$ of $^{129}$Xe.}
\end{center}
\end{figure}

\begin{figure}
\begin{center}
 \renewcommand{\baselinestretch}{1}
 \includegraphics{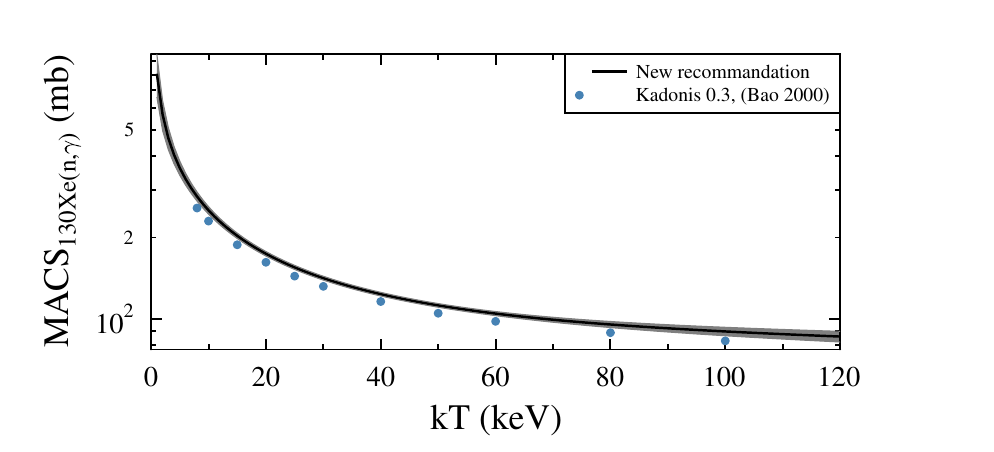}
 \includegraphics{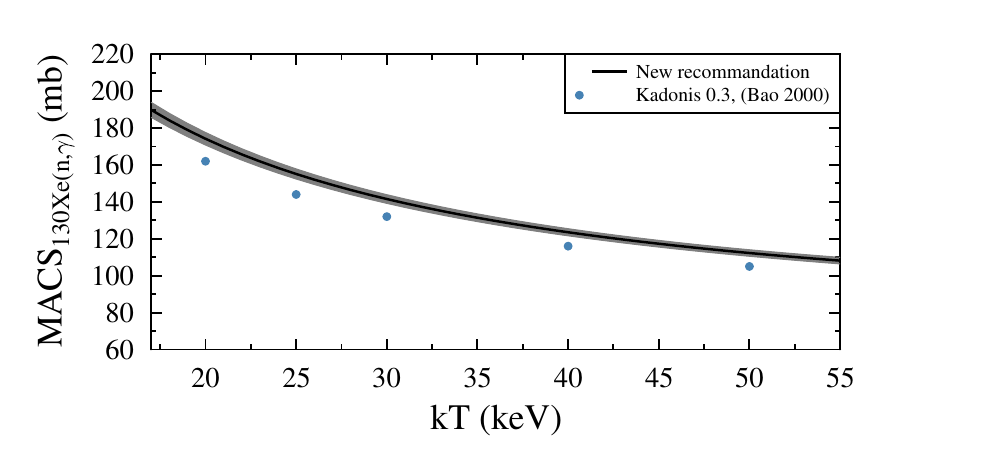}
 \caption{\label{fig:130xe_comparison} New and old \cite{RHK02,BBK00,DHK06}  recommendation for the $MACS$ of $^{130}$Xe.}
\end{center}
\end{figure}

\begin{figure}
\begin{center}
 \renewcommand{\baselinestretch}{1}
 \includegraphics{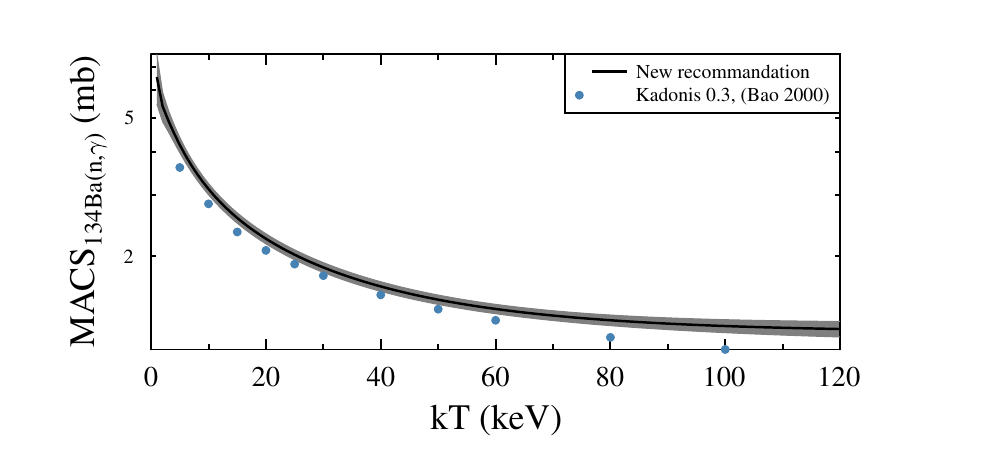}
 \includegraphics{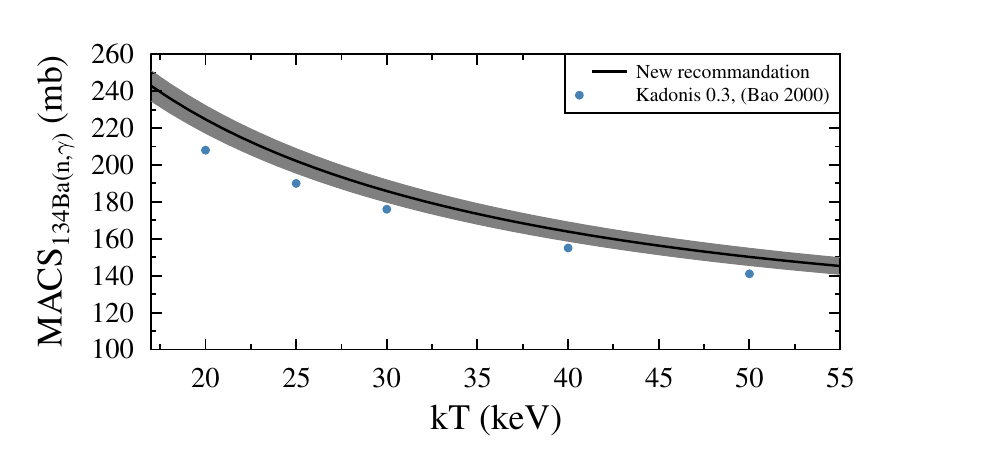}
 \caption{\label{fig:134ba_comparison} New and old \cite{VWG94b,BBK00,DHK06}  recommendation for the $MACS$ of $^{134}$Ba.}
\end{center}
\end{figure}

\begin{figure}
\begin{center}
 \renewcommand{\baselinestretch}{1}
 \includegraphics{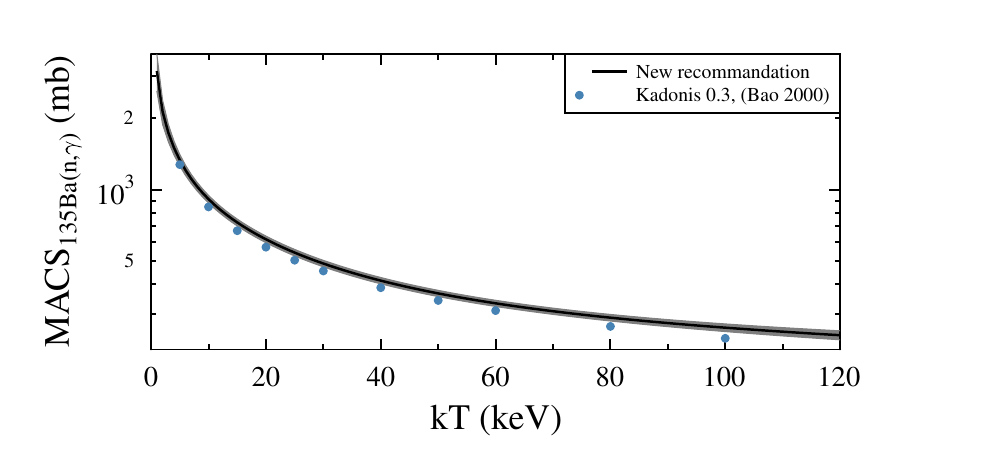}
 \includegraphics{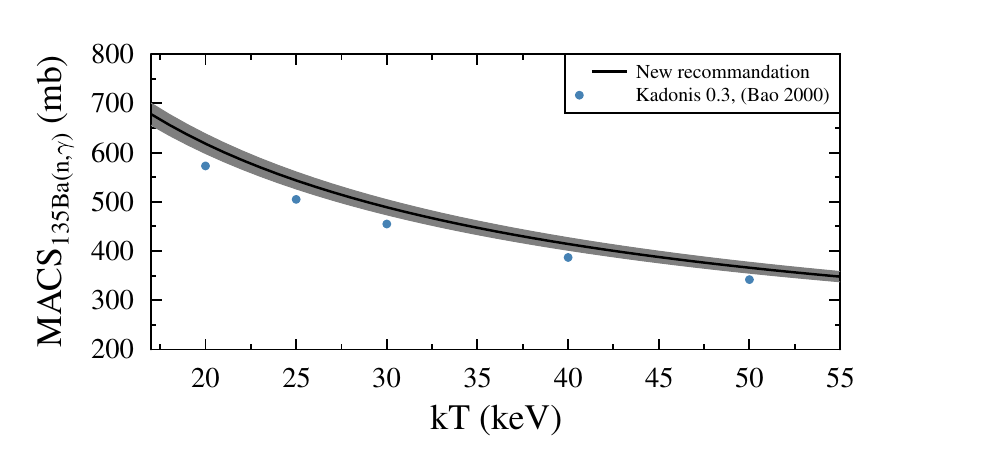}
 \caption{\label{fig:135ba_comparison} New and old \cite{VWG94b,BBK00,DHK06}  recommendation for the $MACS$ of $^{135}$Ba.}
\end{center}
\end{figure}

\begin{figure}
\begin{center}
 \renewcommand{\baselinestretch}{1}
 \includegraphics{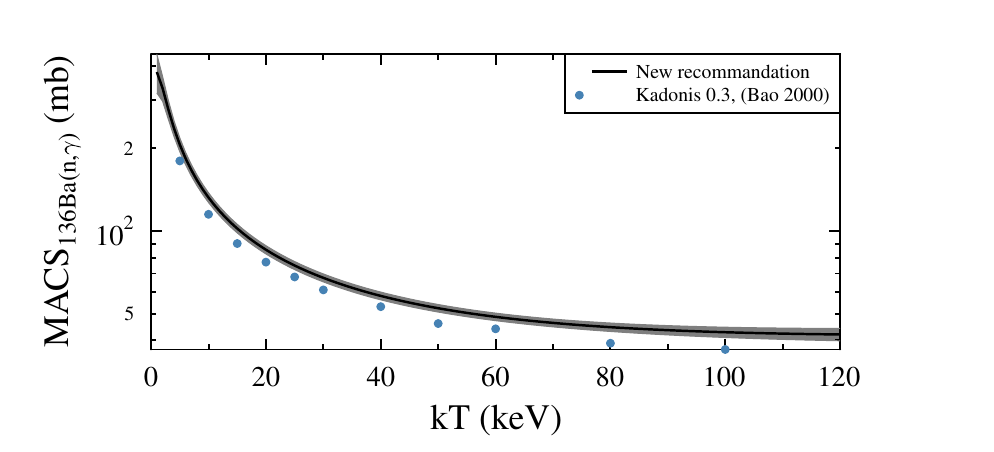}
 \includegraphics{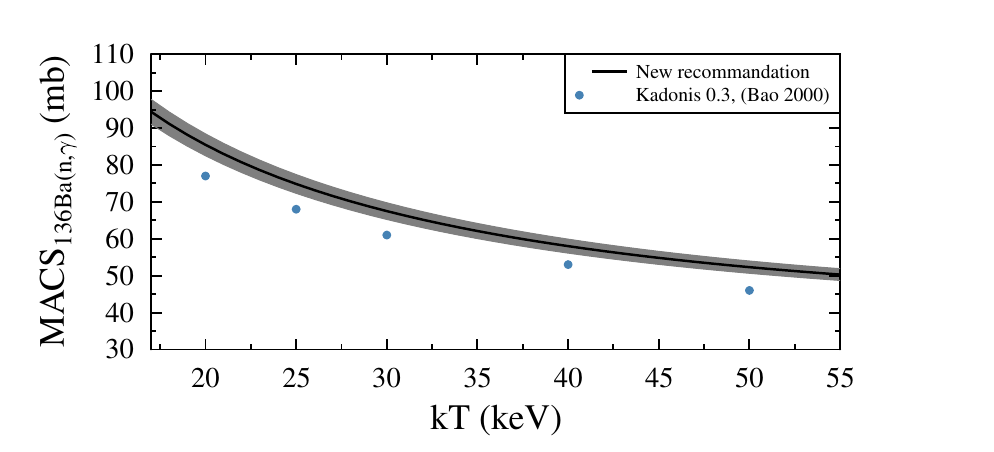}
 \caption{\label{fig:136ba_comparison} New and old \cite{VWG94b,BBK00,DHK06}  recommendation for the $MACS$ of $^{136}$Ba.}
\end{center}
\end{figure}

\begin{figure}
\begin{center}
 \renewcommand{\baselinestretch}{1}
 \includegraphics{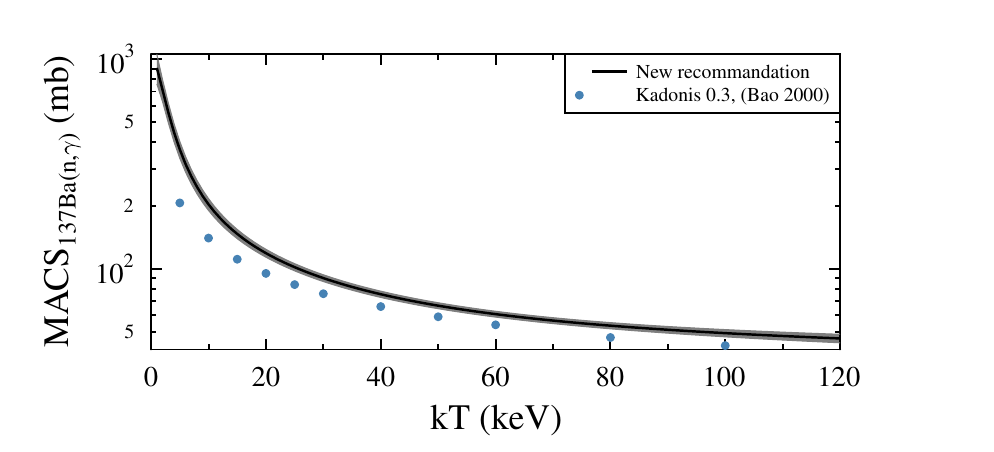}
 \includegraphics{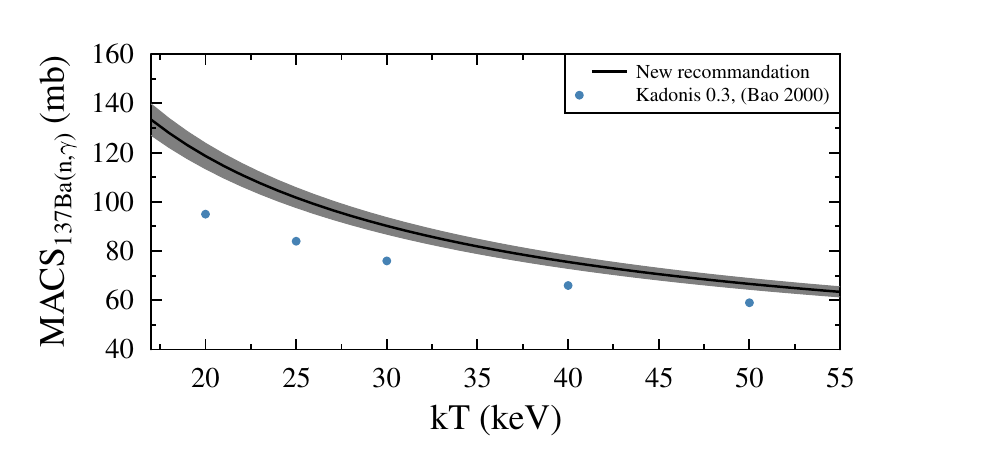}
 \caption{\label{fig:137ba_comparison} New and old \cite{VWG94b,BBK00,DHK06}  recommendation for the $MACS$ of $^{137}$Ba.}
\end{center}
\end{figure}

\begin{figure}
\begin{center}
 \renewcommand{\baselinestretch}{1}
 \includegraphics{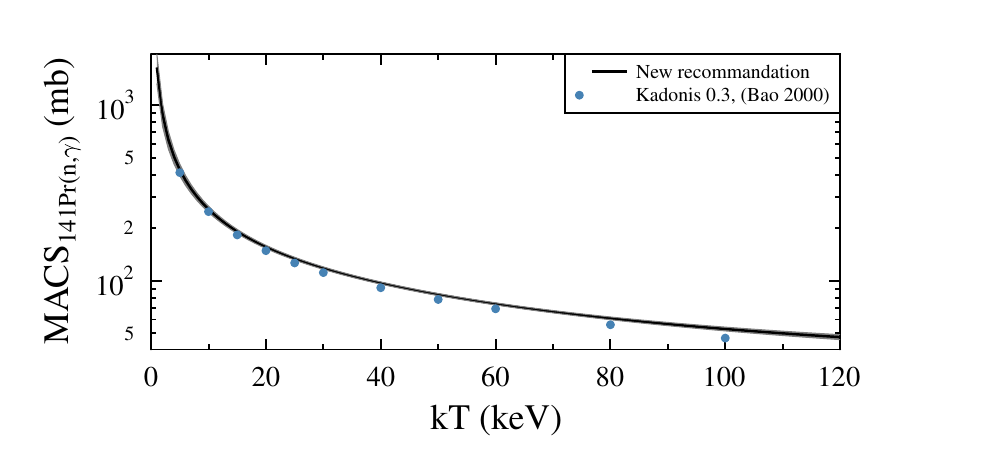}
 \includegraphics{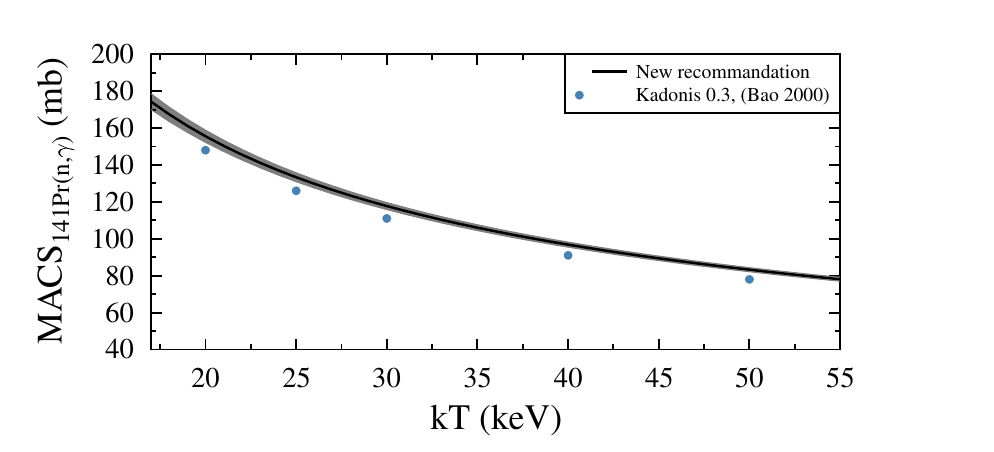}
 \caption{\label{fig:141pr_comparison} New and old \cite{VWA99,BBK00,DHK06}  recommendation for the $MACS$ of $^{141}$Pr.}
\end{center}
\end{figure}

\begin{figure}
\begin{center}
 \renewcommand{\baselinestretch}{1}
 \includegraphics{142nd_MACS_log_auto}
 \includegraphics{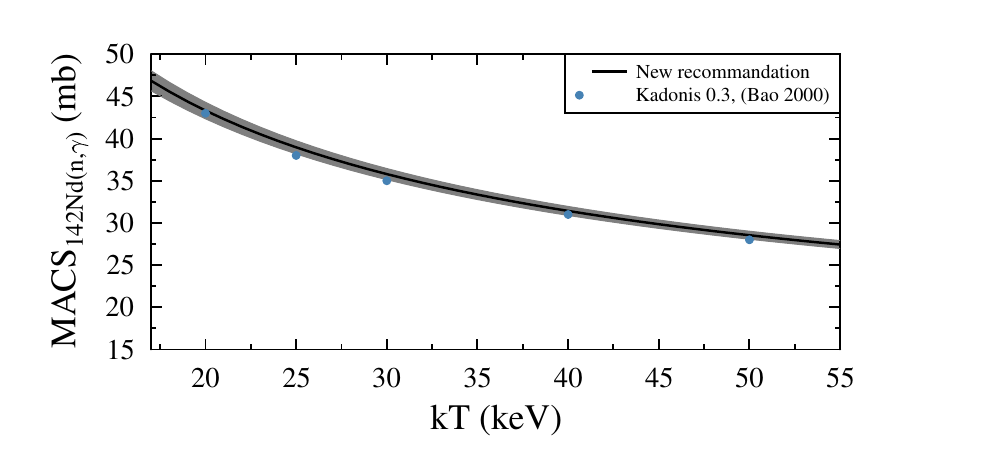}
 \caption{\label{fig:142nd_comparison} New and old \cite{WVK98a,BBK00,DHK06}  recommendation for the $MACS$ of $^{142}$Nd.}
\end{center}
\end{figure}

\begin{figure}
\begin{center}
 \renewcommand{\baselinestretch}{1}
 \includegraphics{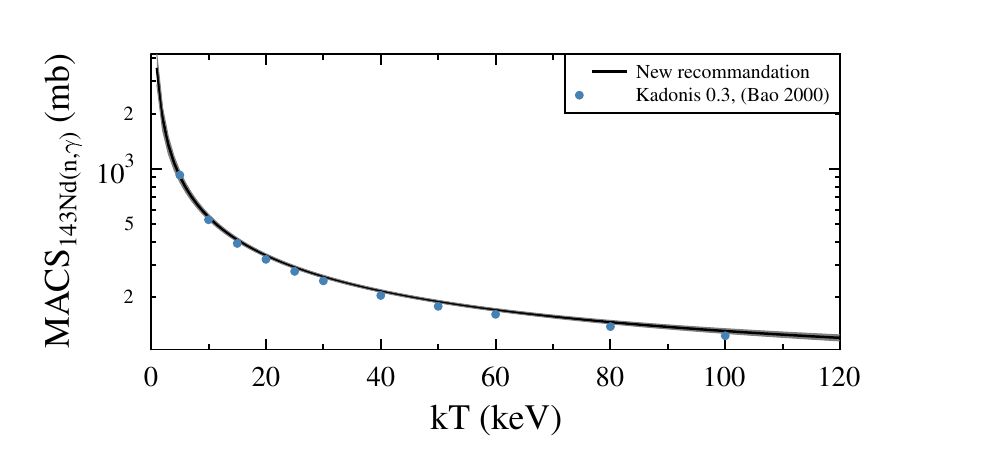}
 \includegraphics{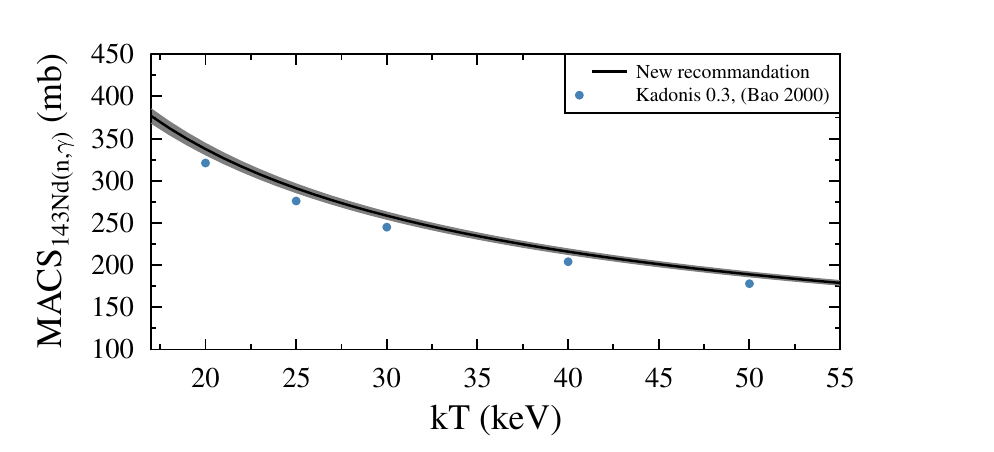}
 \caption{\label{fig:143nd_comparison} New and old \cite{WVK98a,BBK00,DHK06}  recommendation for the $MACS$ of $^{143}$Nd.}
\end{center}
\end{figure}

\begin{figure}
\begin{center}
 \renewcommand{\baselinestretch}{1}
 \includegraphics{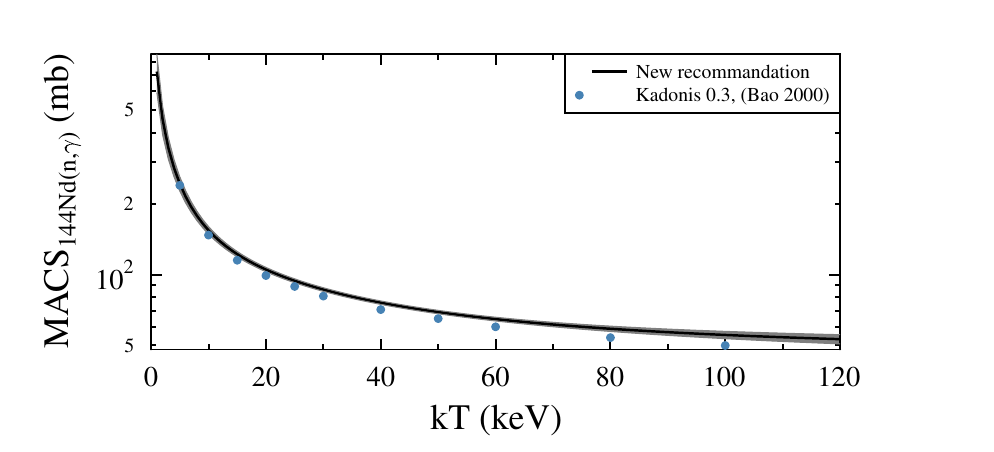}
 \includegraphics{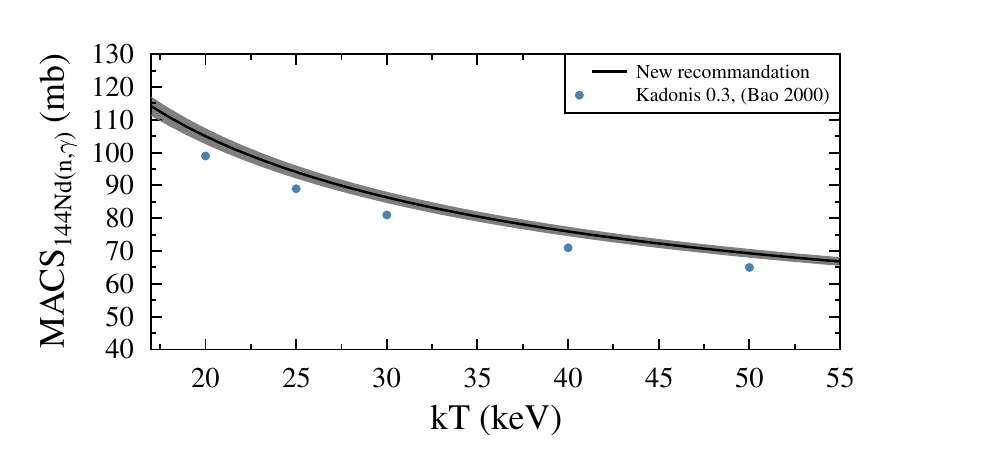}
 \caption{\label{fig:144nd_comparison} New and old \cite{WVK98a,BBK00,DHK06}  recommendation for the $MACS$ of $^{144}$Nd.}
\end{center}
\end{figure}

\begin{figure}
\begin{center}
 \renewcommand{\baselinestretch}{1}
 \includegraphics{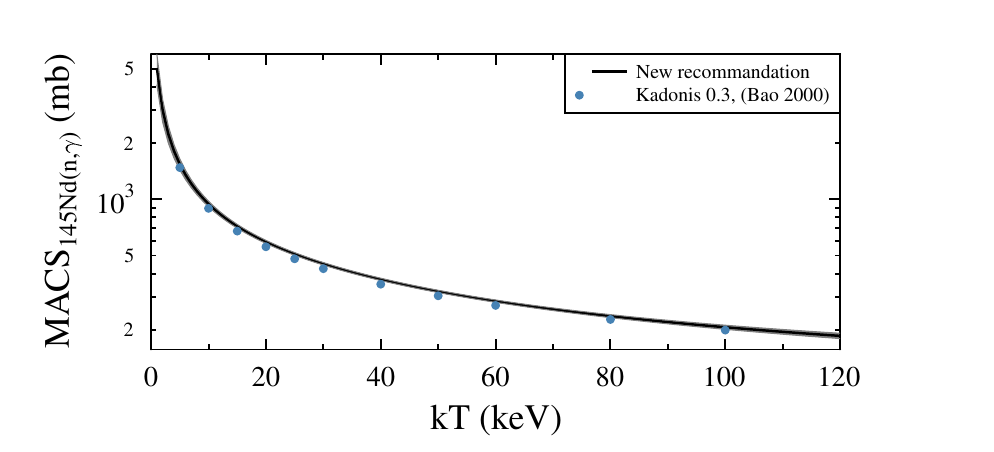}
 \includegraphics{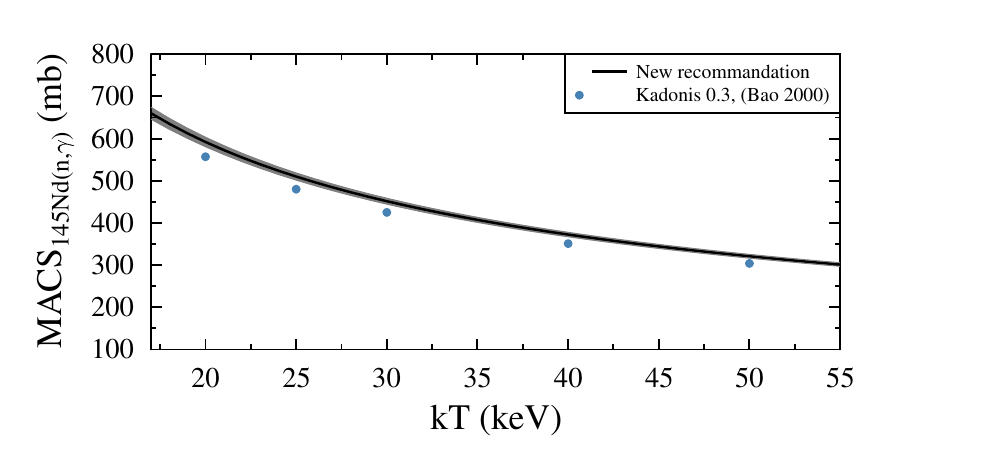}
 \caption{\label{fig:145nd_comparison} New and old \cite{WVK98a,BBK00,DHK06}  recommendation for the $MACS$ of $^{145}$Nd.}
\end{center}
\end{figure}

\begin{figure}
\begin{center}
 \renewcommand{\baselinestretch}{1}
 \includegraphics{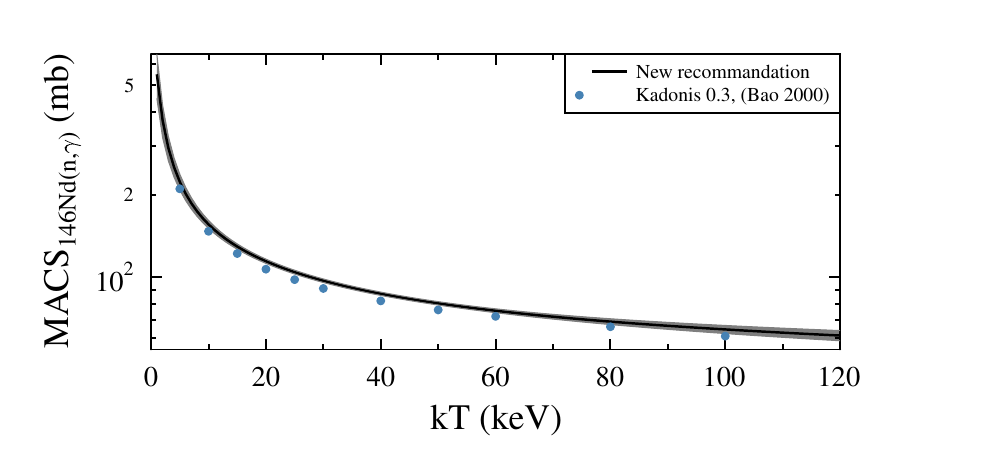}
 \includegraphics{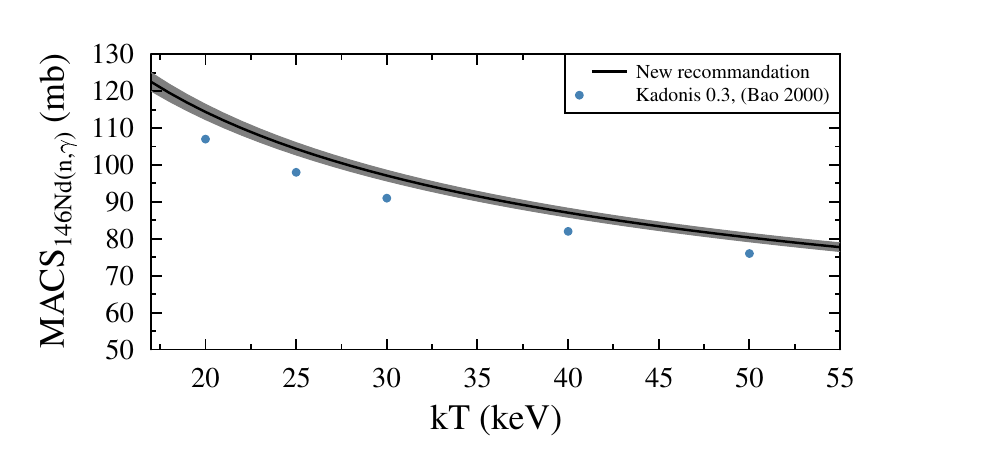}
 \caption{\label{fig:146nd_comparison} New and old \cite{WVK98a,BBK00,DHK06}  recommendation for the $MACS$ of $^{146}$Nd.}
\end{center}
\end{figure}

\begin{figure}
\begin{center}
 \renewcommand{\baselinestretch}{1}
 \includegraphics{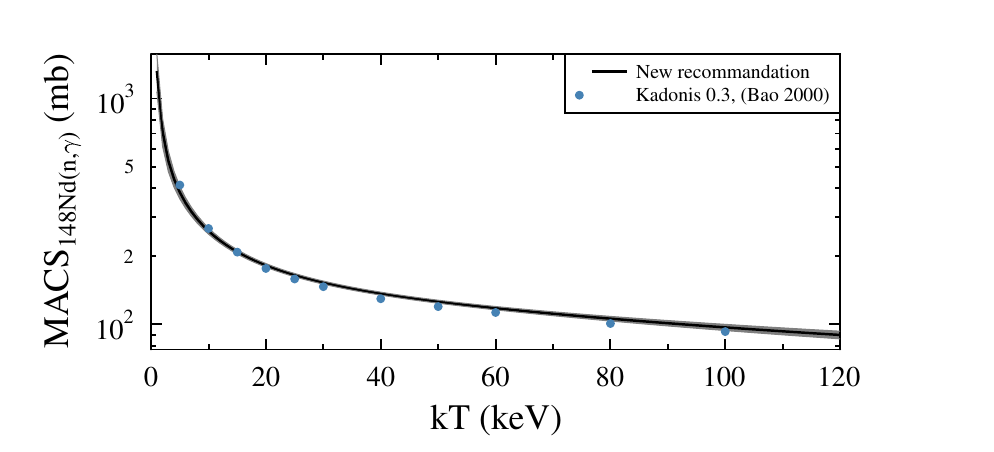}
 \includegraphics{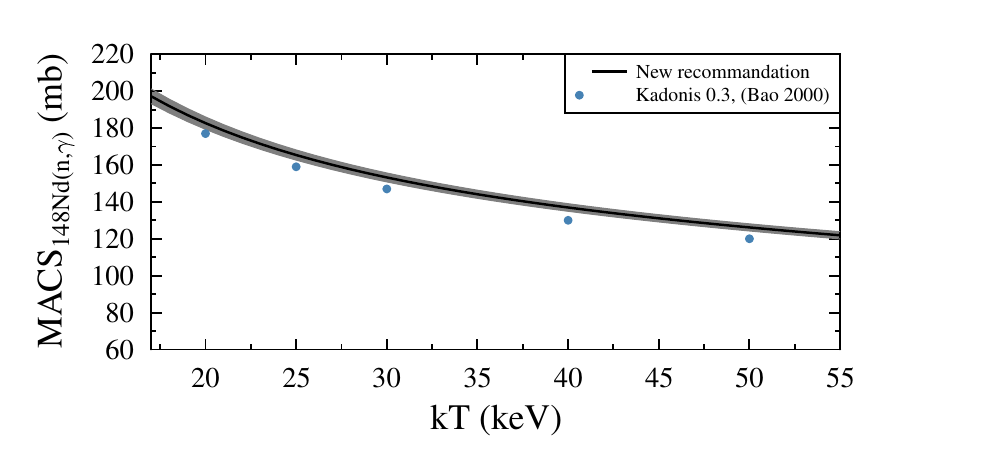}
 \caption{\label{fig:148nd_comparison} New and old \cite{WVK98a,BBK00,DHK06}  recommendation for the $MACS$ of $^{148}$Nd.}
\end{center}
\end{figure}

\begin{figure}
\begin{center}
 \renewcommand{\baselinestretch}{1}
 \includegraphics{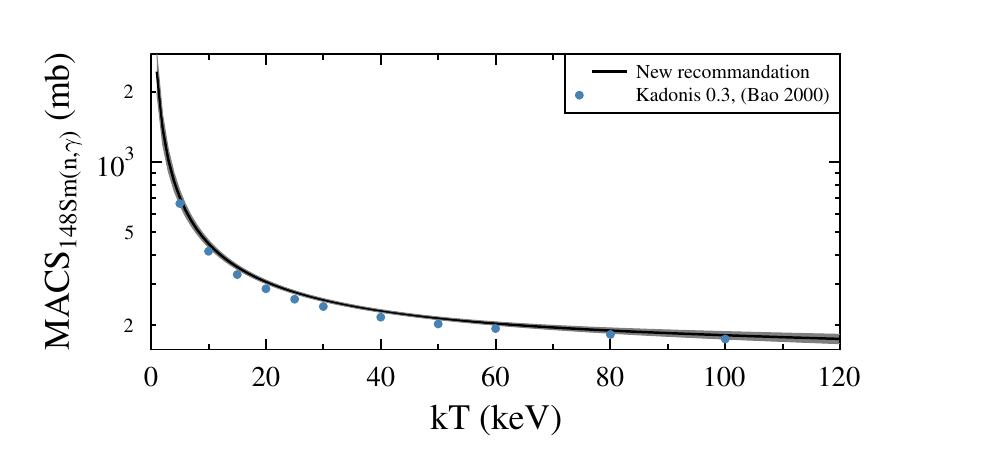}
 \includegraphics{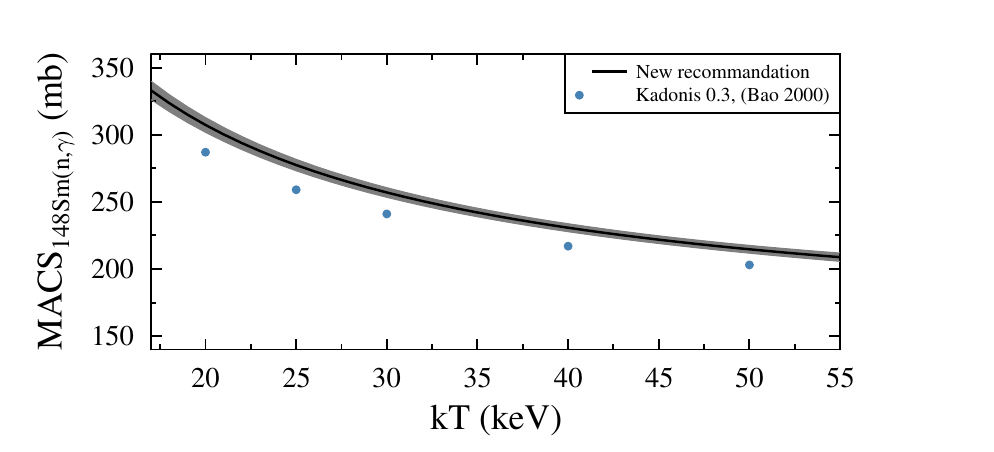}
 \caption{\label{fig:148sm_comparison} New and old \cite{WGV93,BBK00,DHK06}  recommendation for the $MACS$ of $^{148}$Sm.}
\end{center}
\end{figure}

\begin{figure}
\begin{center}
 \renewcommand{\baselinestretch}{1}
 \includegraphics{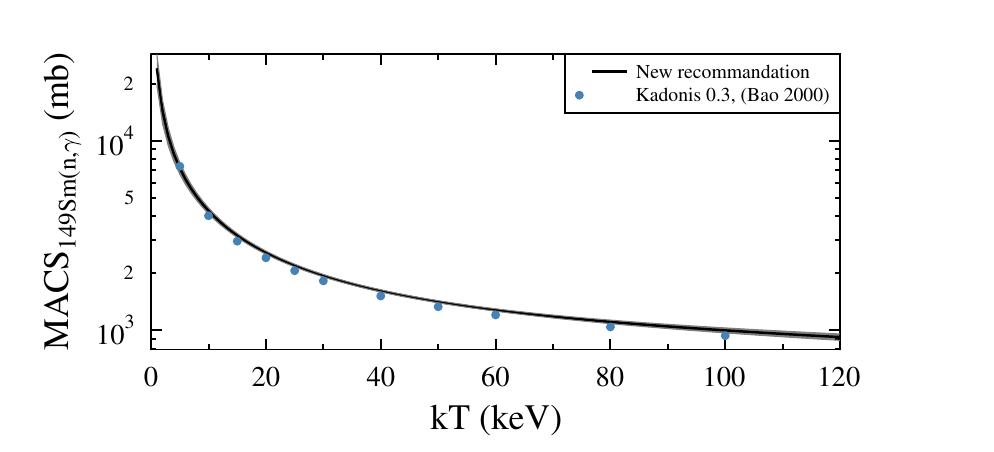}
 \includegraphics{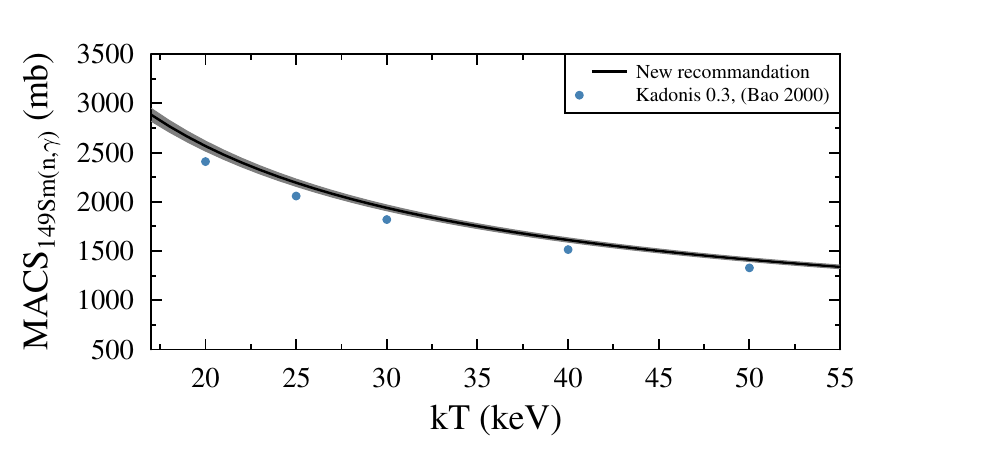}
 \caption{\label{fig:149sm_comparison} New and old \cite{WGV93,BBK00,DHK06}  recommendation for the $MACS$ of $^{149}$Sm.}
\end{center}
\end{figure}

\begin{figure}
\begin{center}
 \renewcommand{\baselinestretch}{1}
 \includegraphics{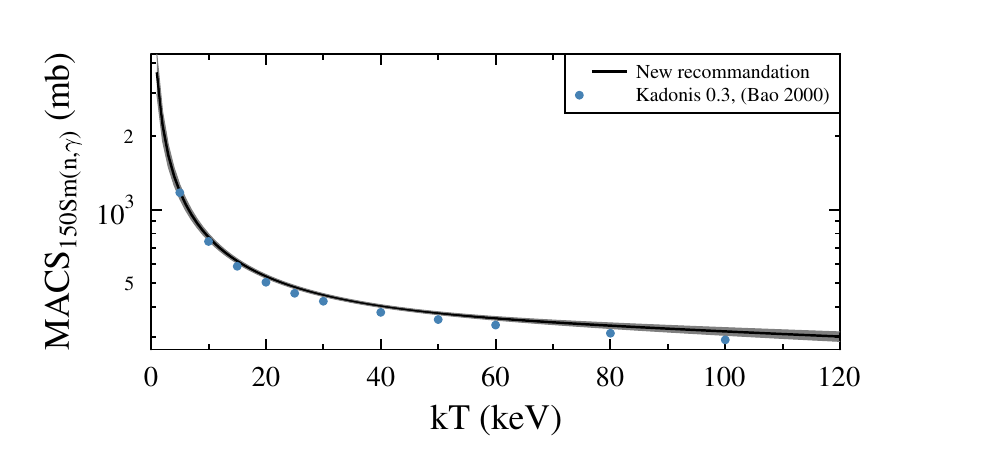}
 \includegraphics{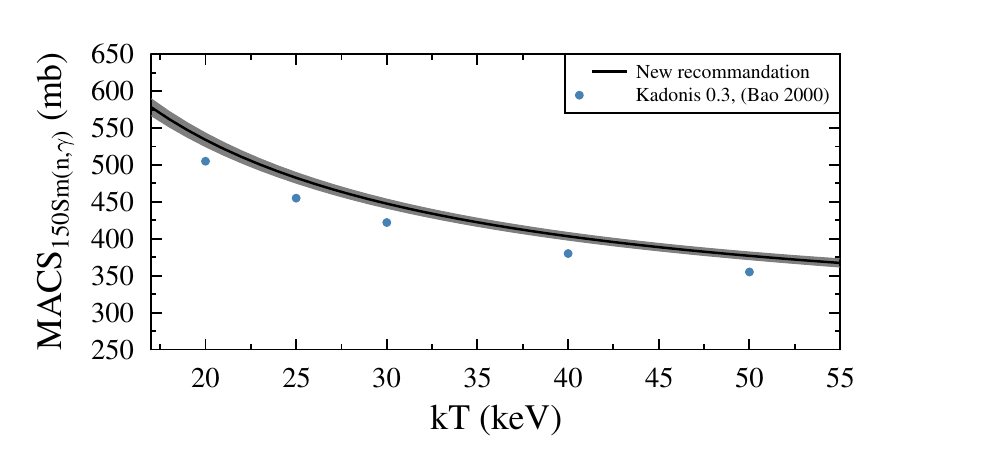}
 \caption{\label{fig:150sm_comparison} New and old \cite{WGV93,BBK00,DHK06}  recommendation for the $MACS$ of $^{150}$Sm.}
\end{center}
\end{figure}

\begin{figure}
\begin{center}
 \renewcommand{\baselinestretch}{1}
 \includegraphics{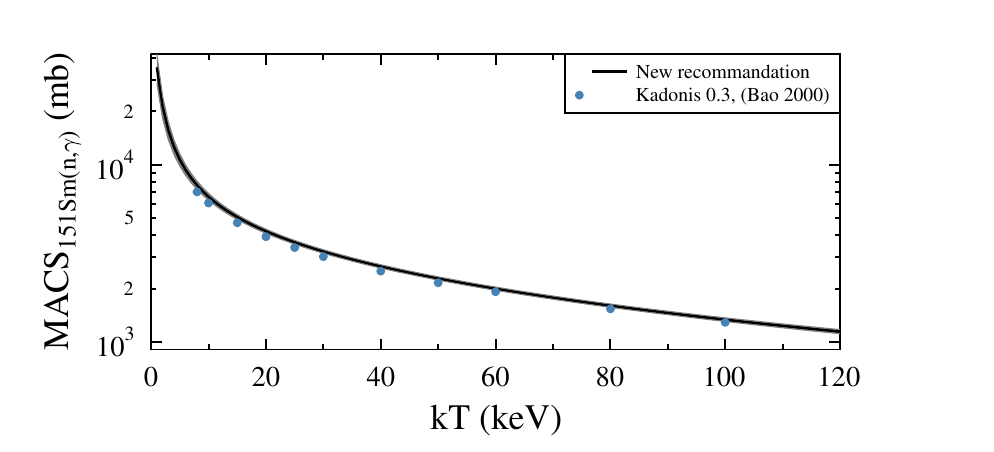}
 \includegraphics{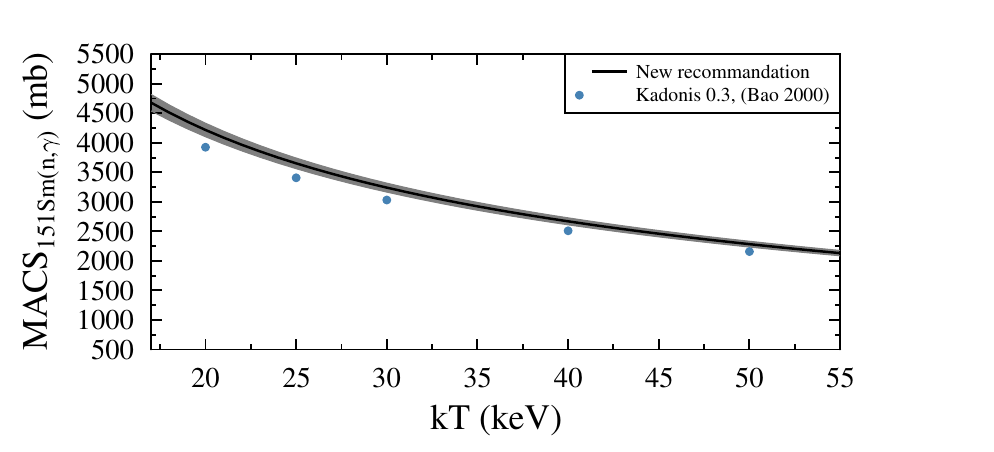}
 \caption{\label{fig:151sm_comparison} New and old \cite{WVK06,BBK00,DHK06}  recommendation for the $MACS$ of $^{151}$Sm. The re-normalized values are in very good agreement with the TOF measurement performed at n\_TOF \cite{AAA04}. }
\end{center}
\end{figure}

\begin{figure}
\begin{center}
 \renewcommand{\baselinestretch}{1}
 \includegraphics{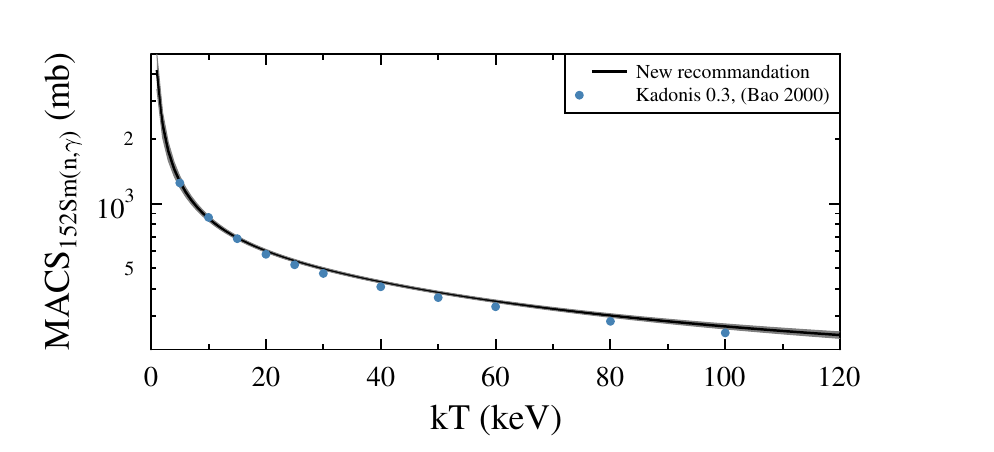}
 \includegraphics{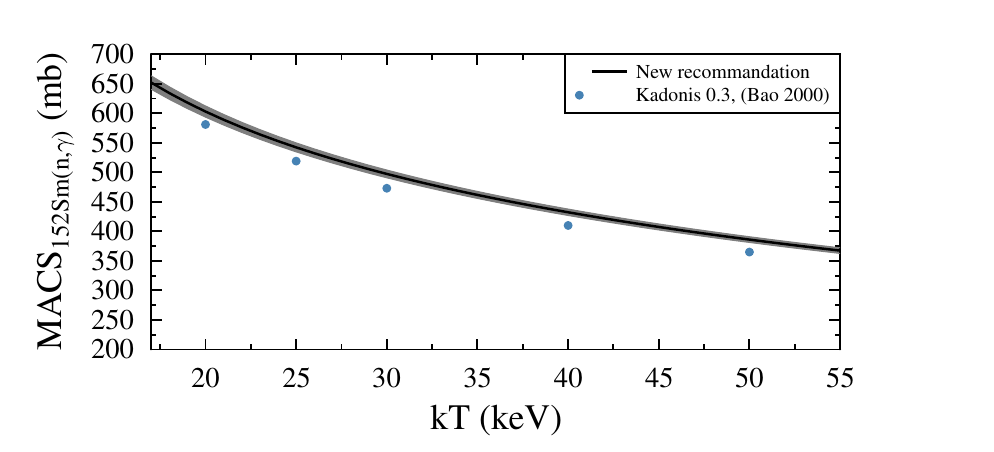}
 \caption{\label{fig:152sm_comparison} New and old \cite{WGV93,BBK00,DHK06}  recommendation for the $MACS$ of $^{152}$Sm.}
\end{center}
\end{figure}

\begin{figure}
\begin{center}
 \renewcommand{\baselinestretch}{1}
 \includegraphics{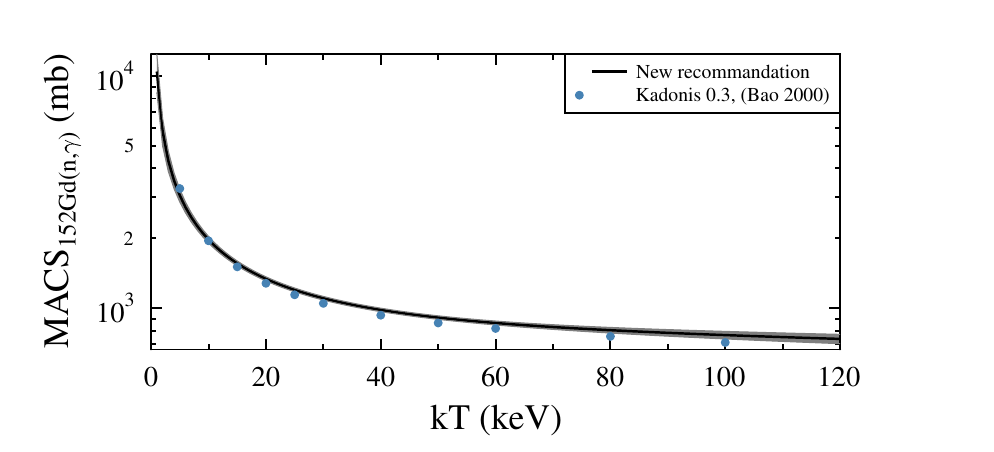}
 \includegraphics{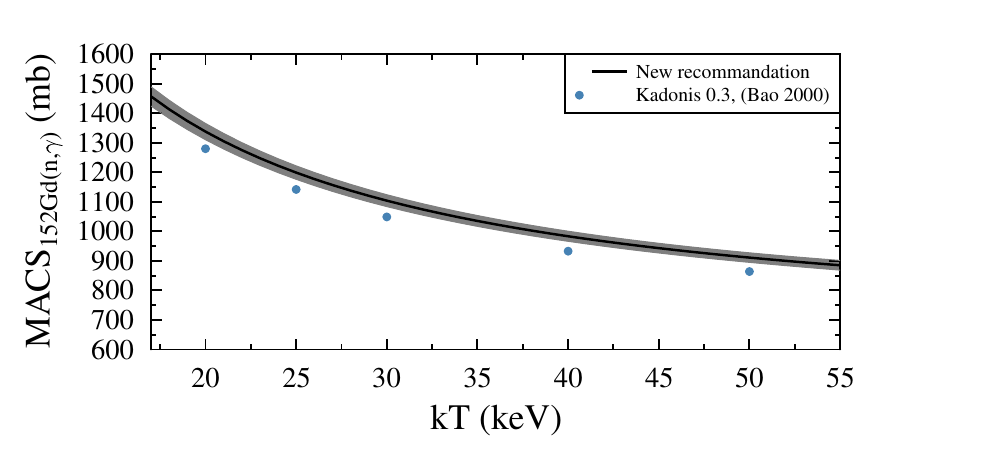}
 \caption{\label{fig:152gd_comparison} New and old \cite{WVK95b,BBK00,DHK06}  recommendation for the $MACS$ of $^{152}$Gd.}
\end{center}
\end{figure}

\begin{figure}
\begin{center}
 \renewcommand{\baselinestretch}{1}
 \includegraphics{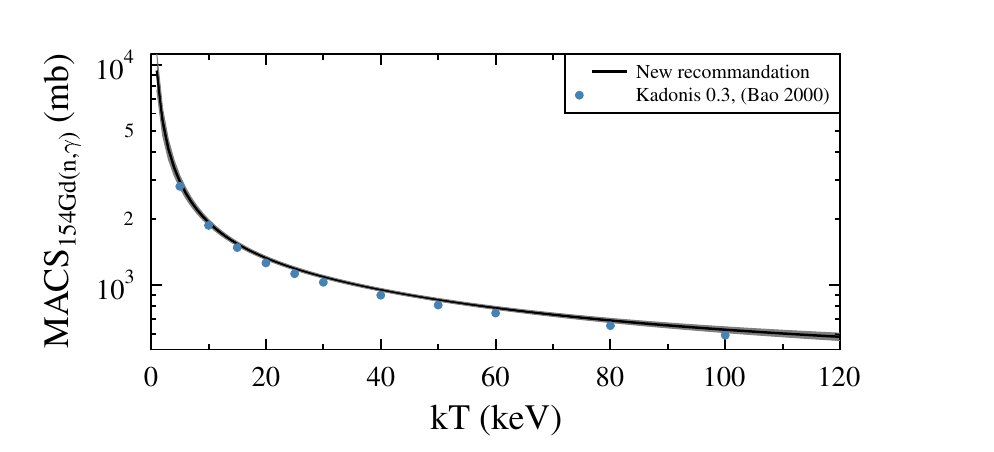}
 \includegraphics{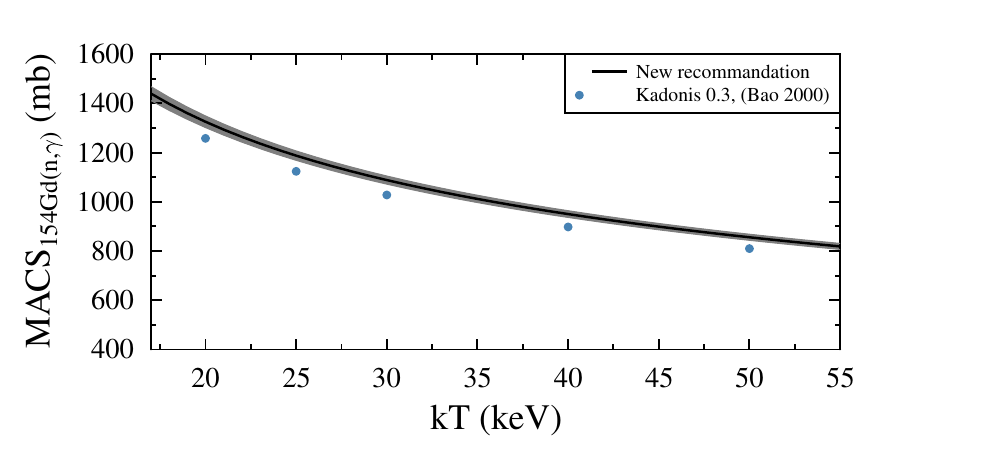}
 \caption{\label{fig:154gd_comparison} New and old \cite{WVK95b,BBK00,DHK06}  recommendation for the $MACS$ of $^{154}$Gd.}
\end{center}
\end{figure}

\begin{figure}
\begin{center}
 \renewcommand{\baselinestretch}{1}
 \includegraphics{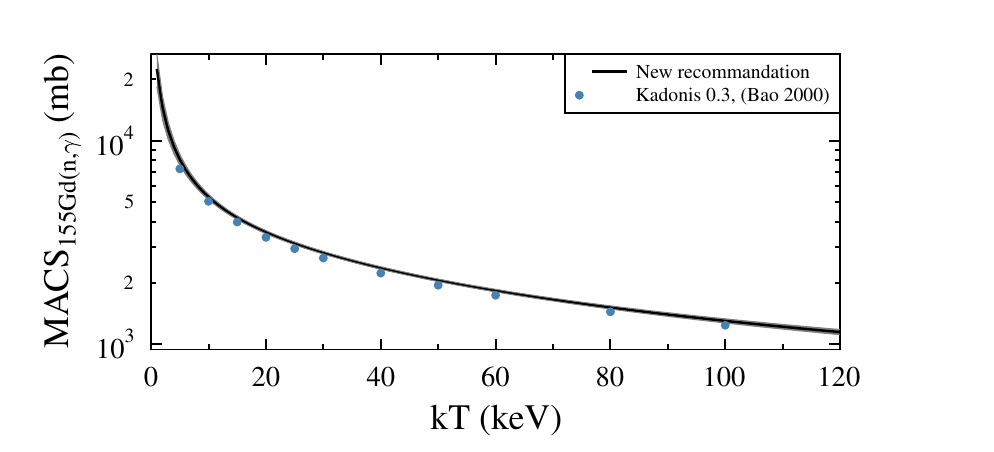}
 \includegraphics{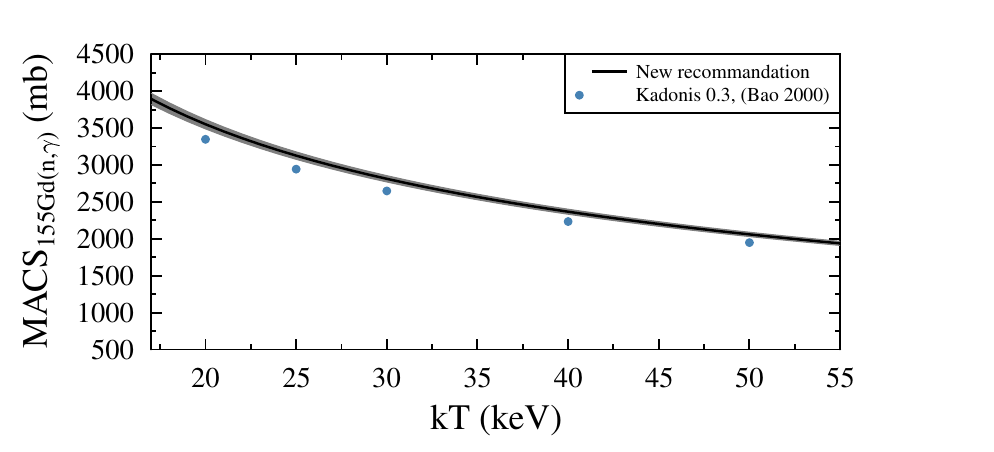}
 \caption{\label{fig:155gd_comparison} New and old \cite{WVK95b,BBK00,DHK06}  recommendation for the $MACS$ of $^{155}$Gd.}
\end{center}
\end{figure}

\begin{figure}
\begin{center}
 \renewcommand{\baselinestretch}{1}
 \includegraphics{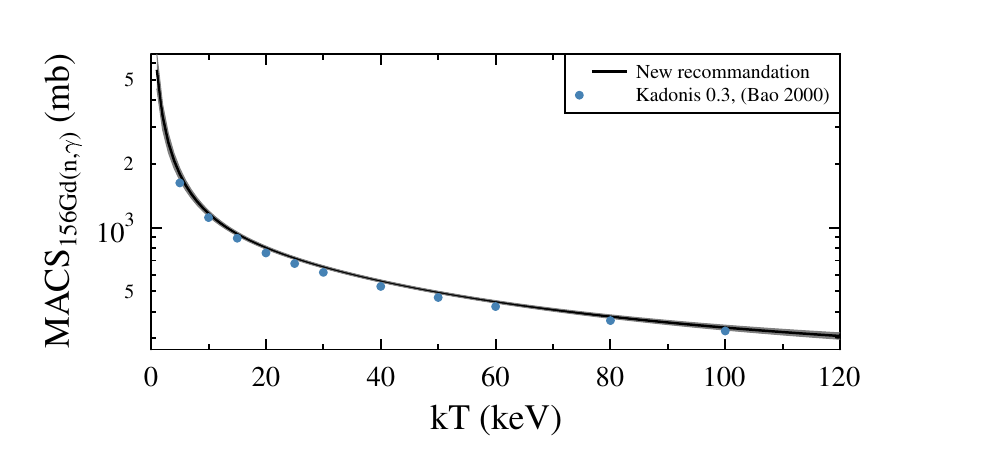}
 \includegraphics{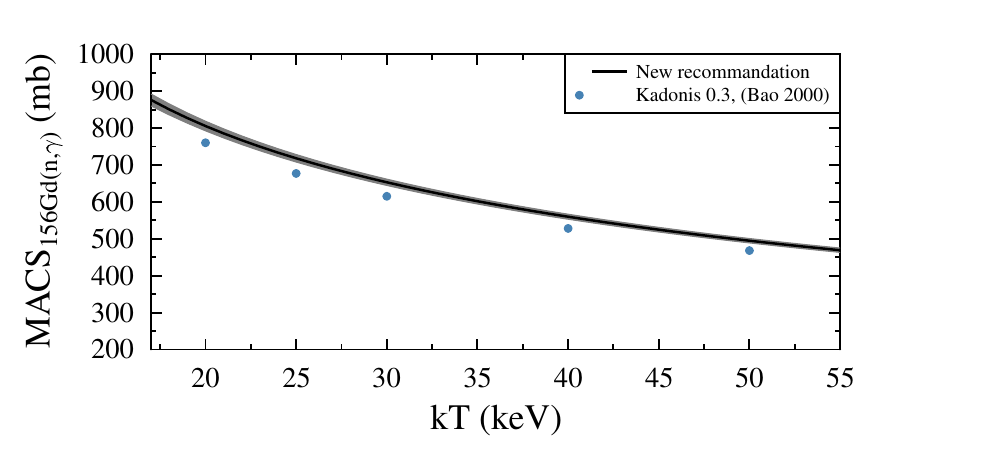}
 \caption{\label{fig:156gd_comparison} New and old \cite{WVK95b,BBK00,DHK06}  recommendation for the $MACS$ of $^{156}$Gd.}
\end{center}
\end{figure}

\begin{figure}
\begin{center}
 \renewcommand{\baselinestretch}{1}
 \includegraphics{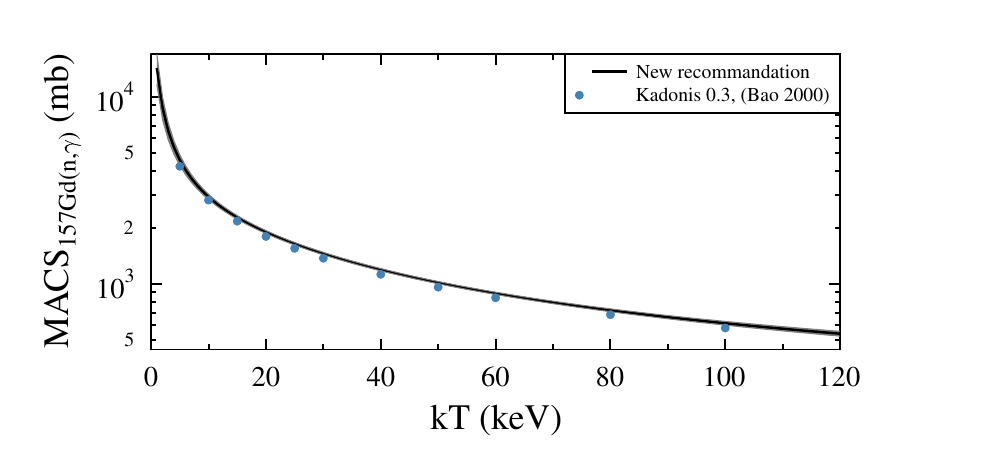}
 \includegraphics{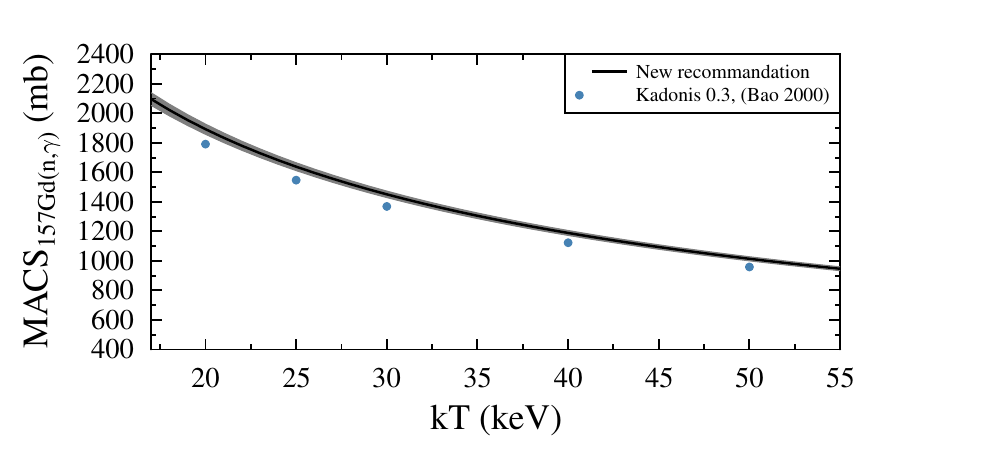}
 \caption{\label{fig:157gd_comparison} New and old \cite{WVK95b,BBK00,DHK06}  recommendation for the $MACS$ of $^{157}$Gd.}
\end{center}
\end{figure}

\begin{figure}
\begin{center}
 \renewcommand{\baselinestretch}{1}
 \includegraphics{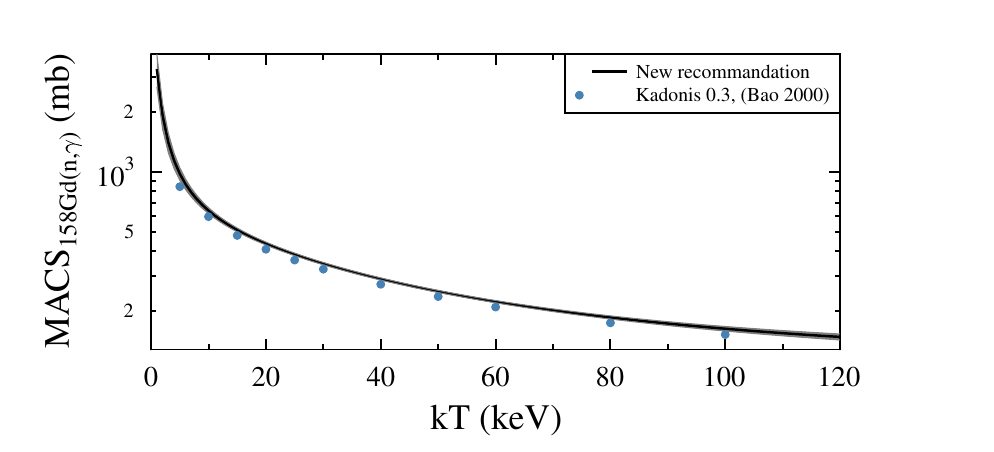}
 \includegraphics{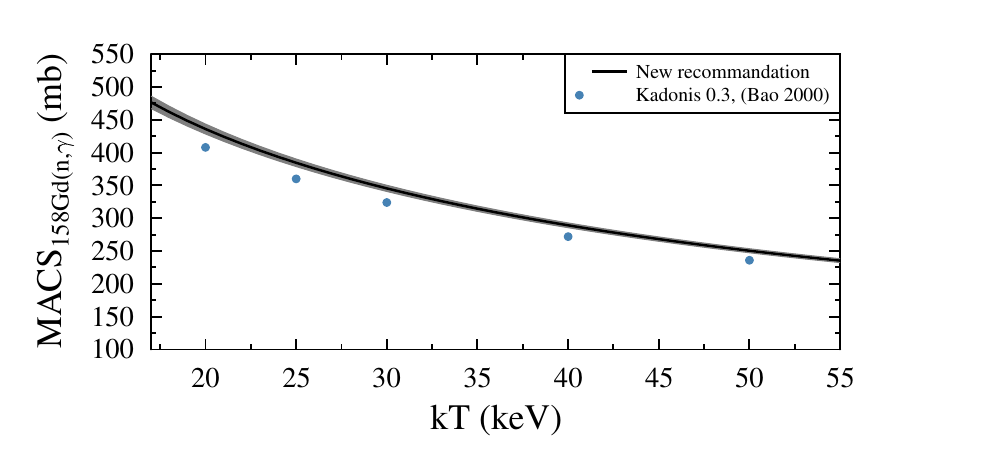}
 \caption{\label{fig:158gd_comparison} New and old \cite{WVK95b,BBK00,DHK06}  recommendation for the $MACS$ of $^{158}$Gd.}
\end{center}
\end{figure}

\begin{figure}
\begin{center}
 \renewcommand{\baselinestretch}{1}
 \includegraphics{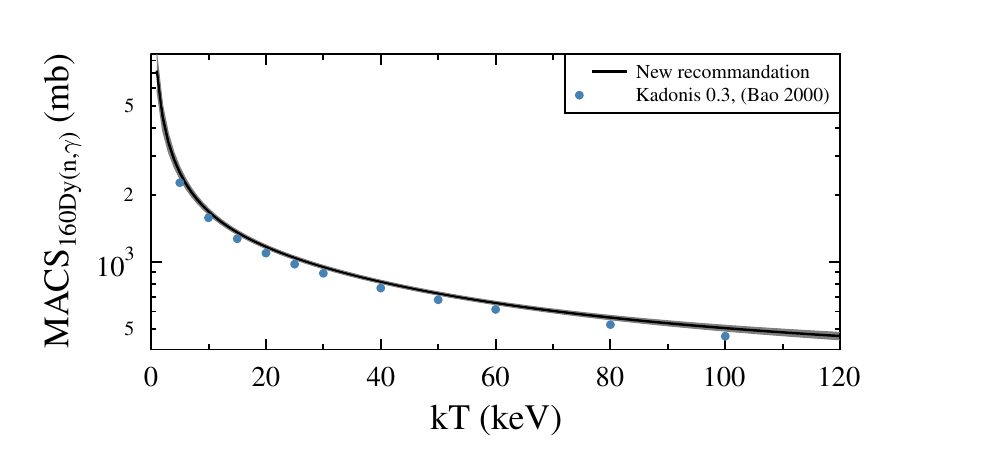}
 \includegraphics{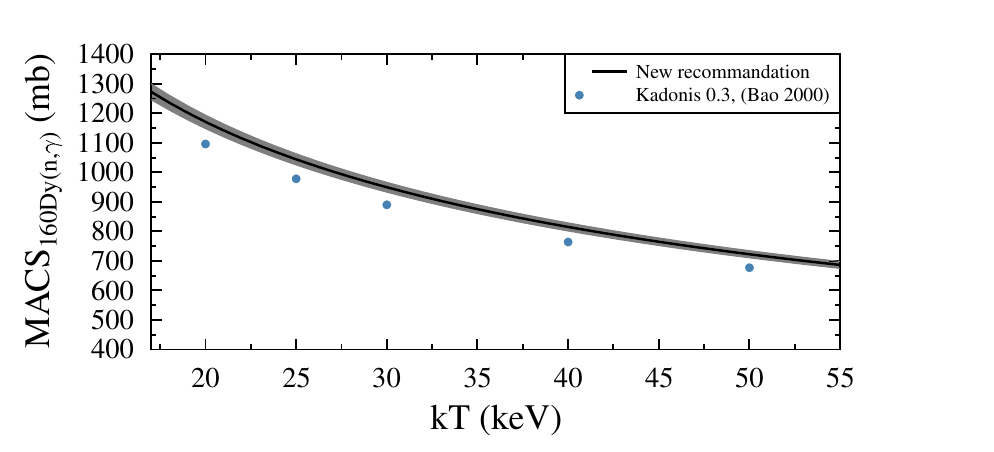}
 \caption{\label{fig:160dy_comparison} New and old \cite{VWA99,BBK00,DHK06}  recommendation for the $MACS$ of $^{160}$Dy.}
\end{center}
\end{figure}

\begin{figure}
\begin{center}
 \renewcommand{\baselinestretch}{1}
 \includegraphics{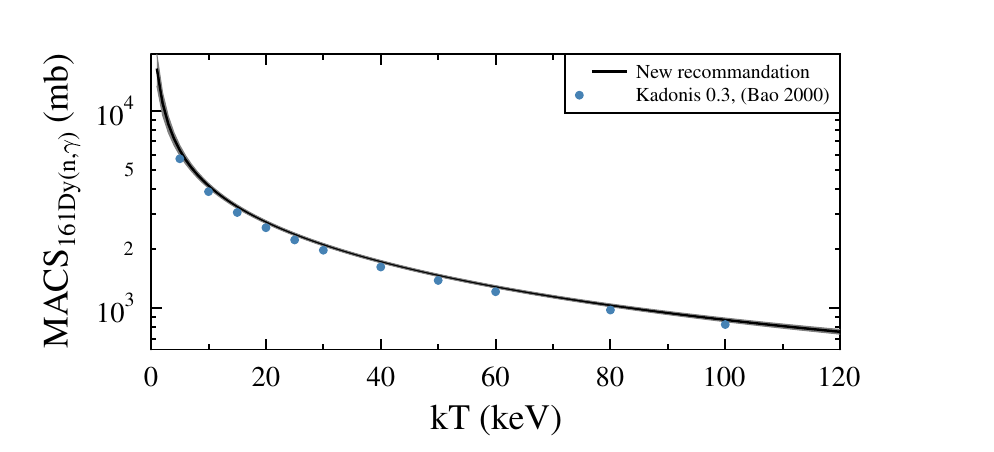}
 \includegraphics{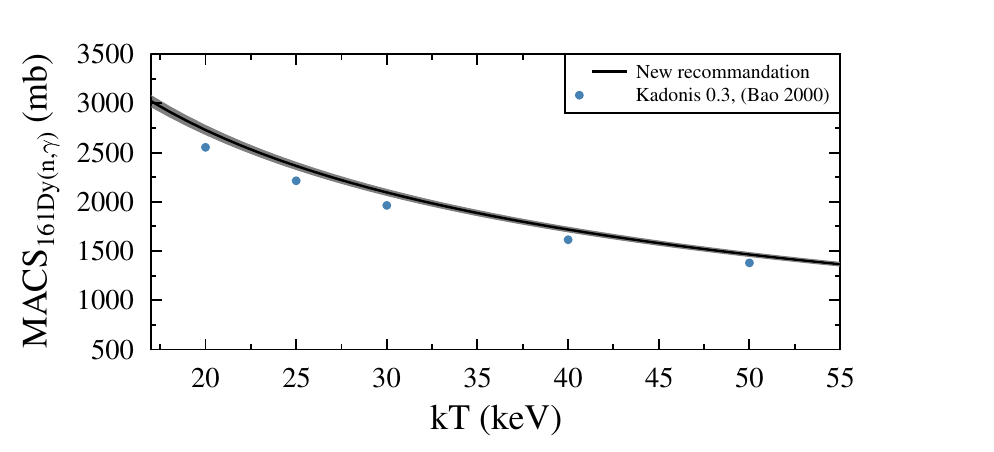}
 \caption{\label{fig:161dy_comparison} New and old \cite{VWA99,BBK00,DHK06}  recommendation for the $MACS$ of $^{161}$Dy.}
\end{center}
\end{figure}

\begin{figure}
\begin{center}
 \renewcommand{\baselinestretch}{1}
 \includegraphics{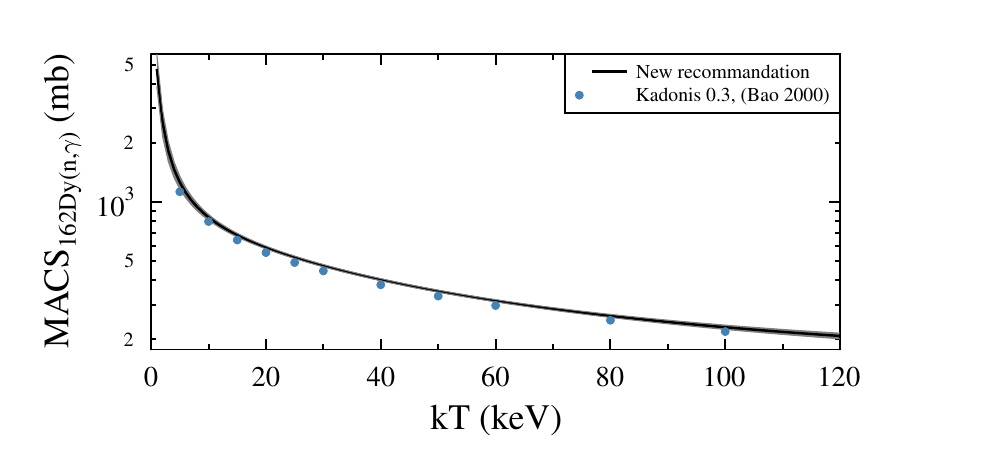}
 \includegraphics{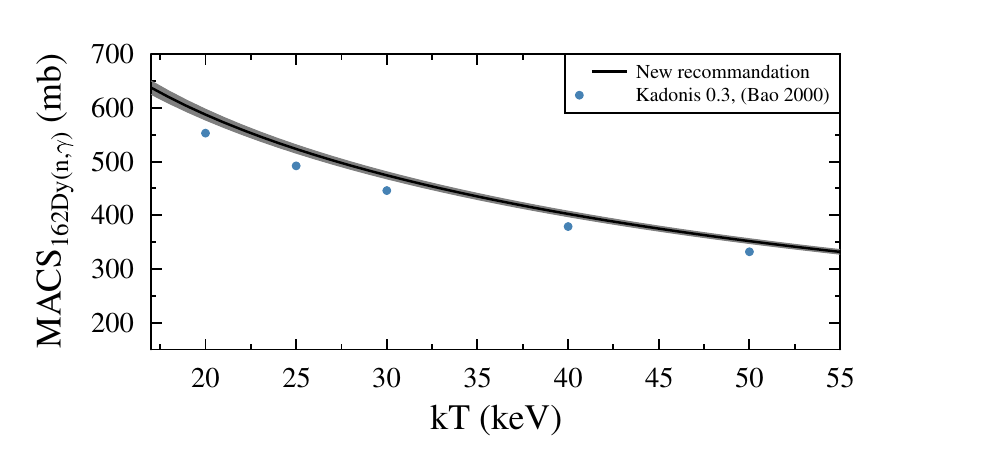}
 \caption{\label{fig:162dy_comparison} New and old \cite{VWA99,BBK00,DHK06}  recommendation for the $MACS$ of $^{162}$Dy.}
\end{center}
\end{figure}

\begin{figure}
\begin{center}
 \renewcommand{\baselinestretch}{1}
 \includegraphics{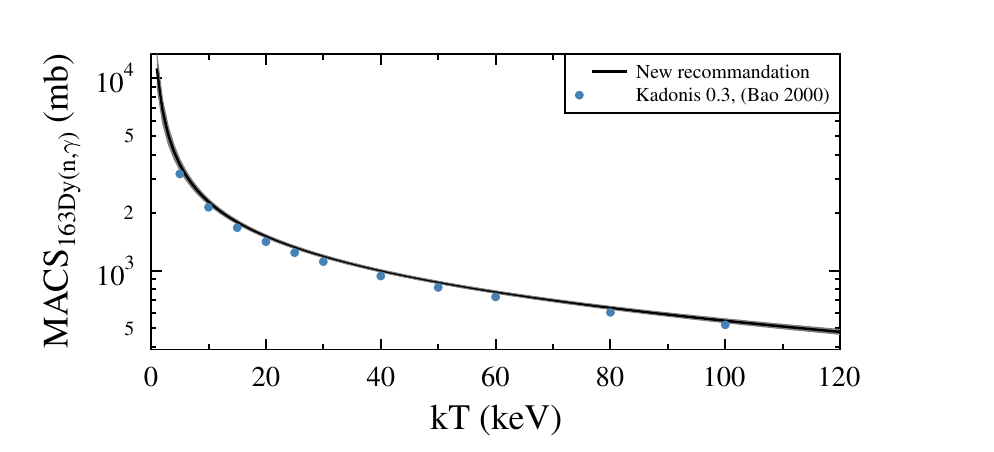}
 \includegraphics{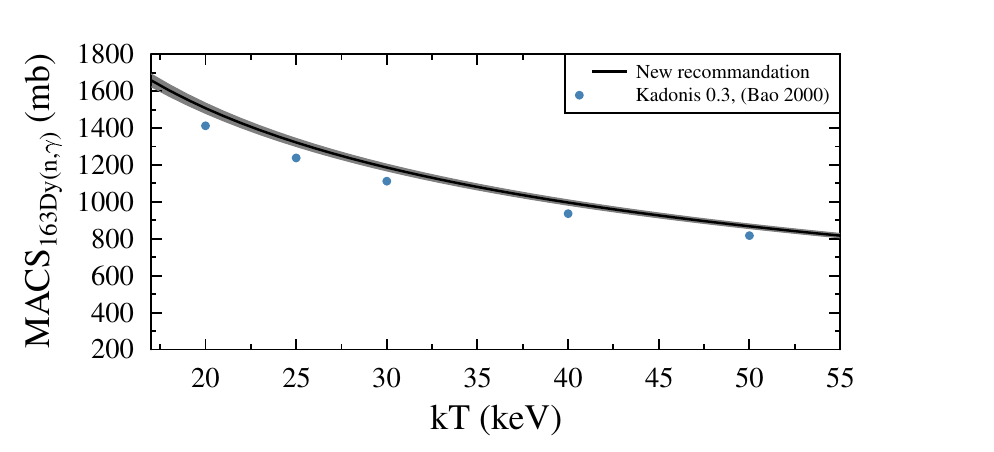}
 \caption{\label{fig:163dy_comparison} New and old \cite{VWA99,BBK00,DHK06}  recommendation for the $MACS$ of $^{163}$Dy.}
\end{center}
\end{figure}

\begin{figure}
\begin{center}
 \renewcommand{\baselinestretch}{1}
 \includegraphics{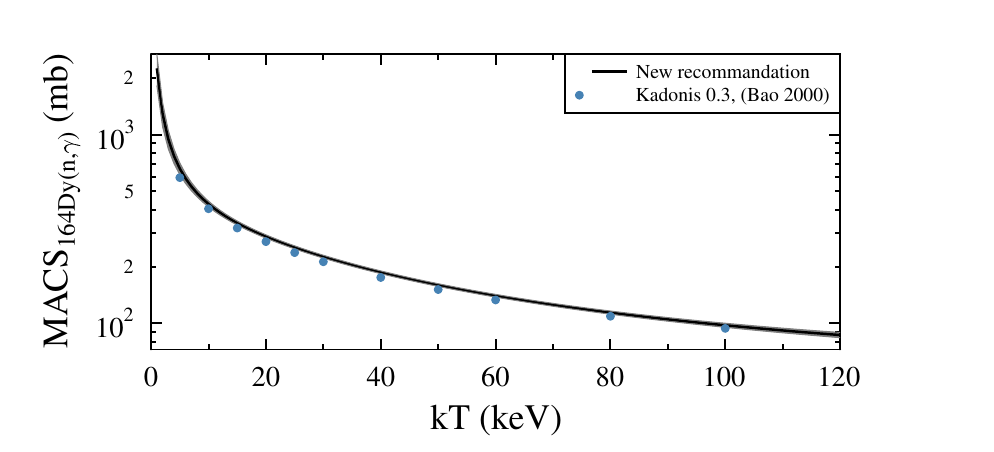}
 \includegraphics{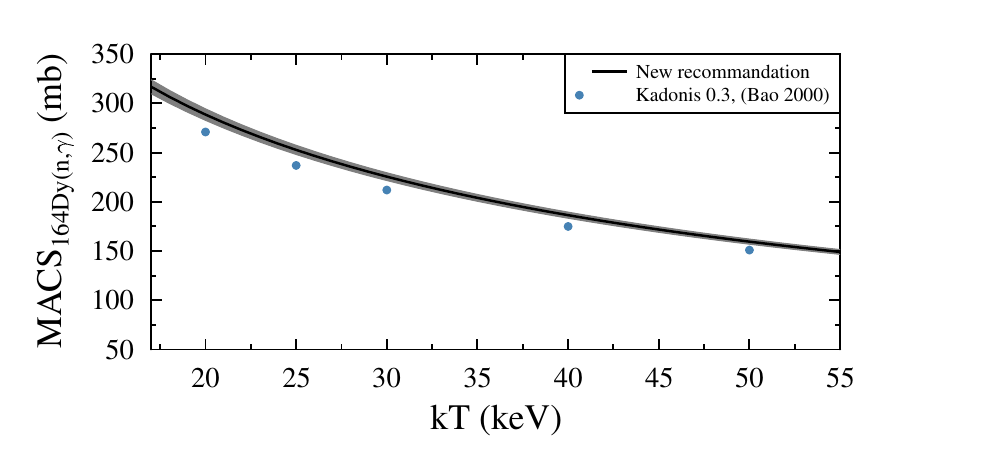}
 \caption{\label{fig:164dy_comparison} New and old \cite{VWA99,BBK00,DHK06}  recommendation for the $MACS$ of $^{164}$Dy.}
\end{center}
\end{figure}

\begin{figure}
\begin{center}
 \renewcommand{\baselinestretch}{1}
 \includegraphics{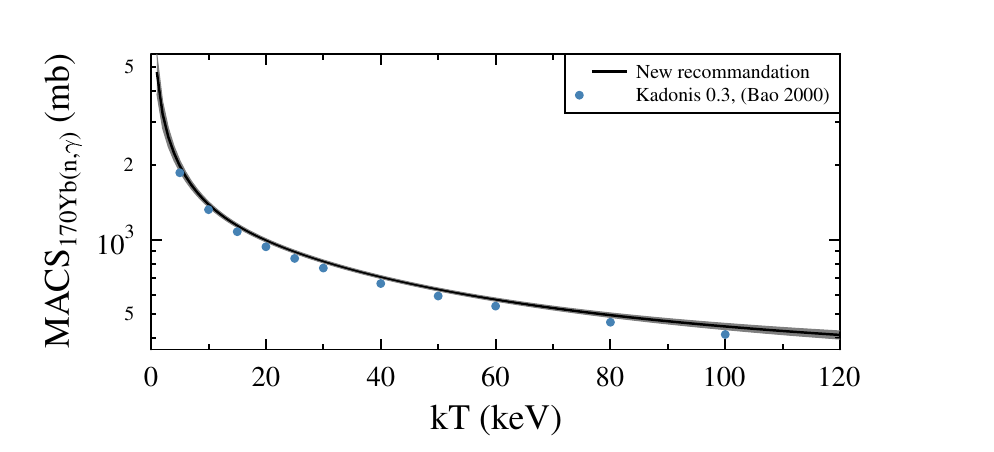}
 \includegraphics{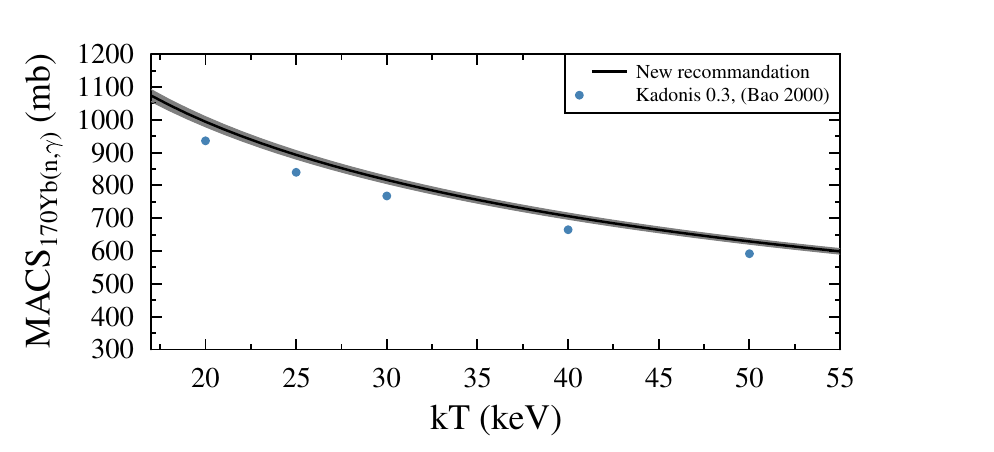}
 \caption{\label{fig:170yb_comparison} New and old \cite{WVA00a,BBK00,DHK06}  recommendation for the $MACS$ of $^{170}$Yb.}
\end{center}
\end{figure}

\begin{figure}
\begin{center}
 \renewcommand{\baselinestretch}{1}
 \includegraphics{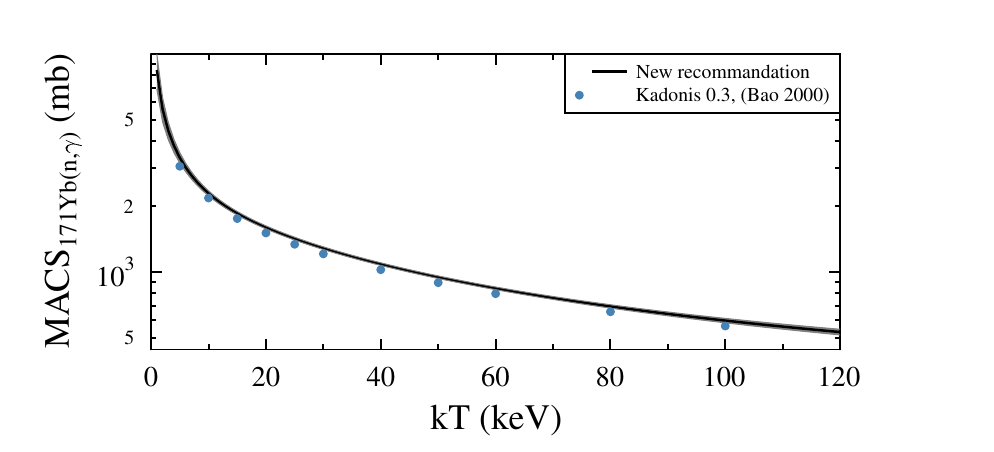}
 \includegraphics{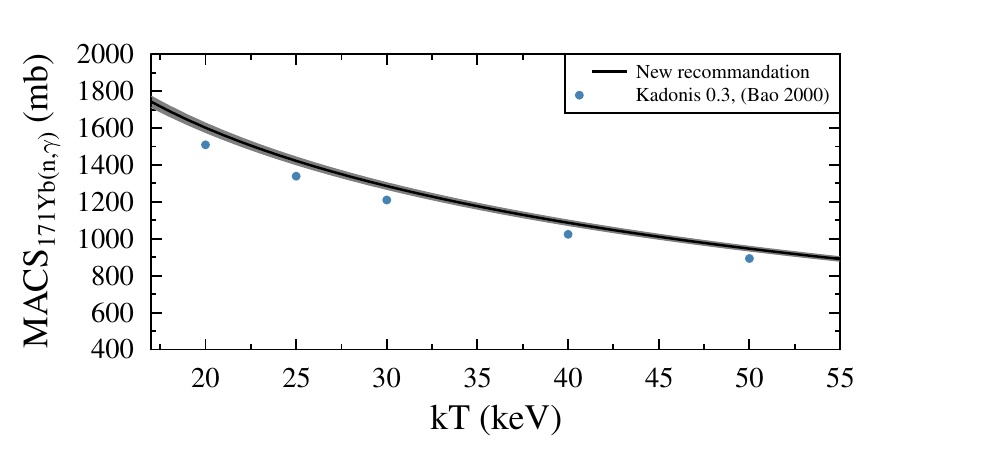}
 \caption{\label{fig:171yb_comparison} New and old \cite{WVA00a,BBK00,DHK06}  recommendation for the $MACS$ of $^{171}$Yb.}
\end{center}
\end{figure}

\begin{figure}
\begin{center}
 \renewcommand{\baselinestretch}{1}
 \includegraphics{172yb_MACS_log_auto}
 \includegraphics{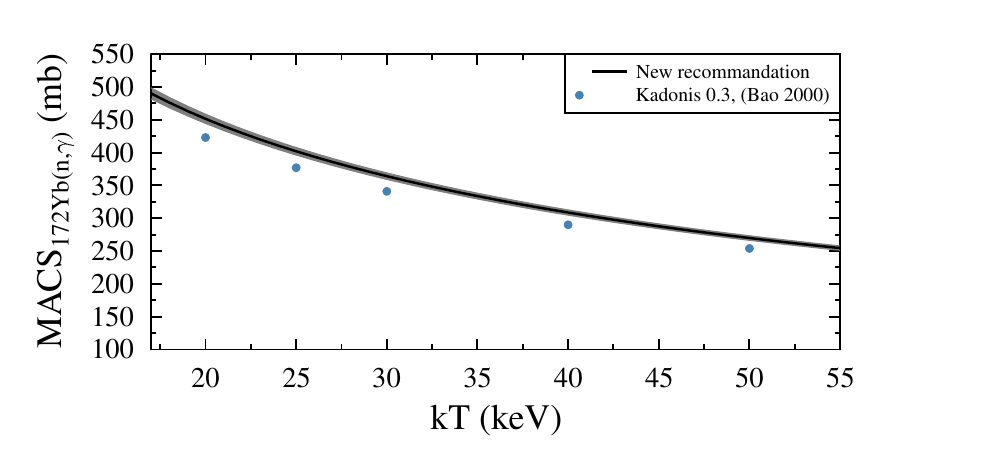}
 \caption{\label{fig:172yb_comparison} New and old \cite{WVA00a,BBK00,DHK06}  recommendation for the $MACS$ of $^{172}$Yb.}
\end{center}
\end{figure}

\begin{figure}
\begin{center}
 \renewcommand{\baselinestretch}{1}
 \includegraphics{197au_MACS_log_auto}
 \includegraphics{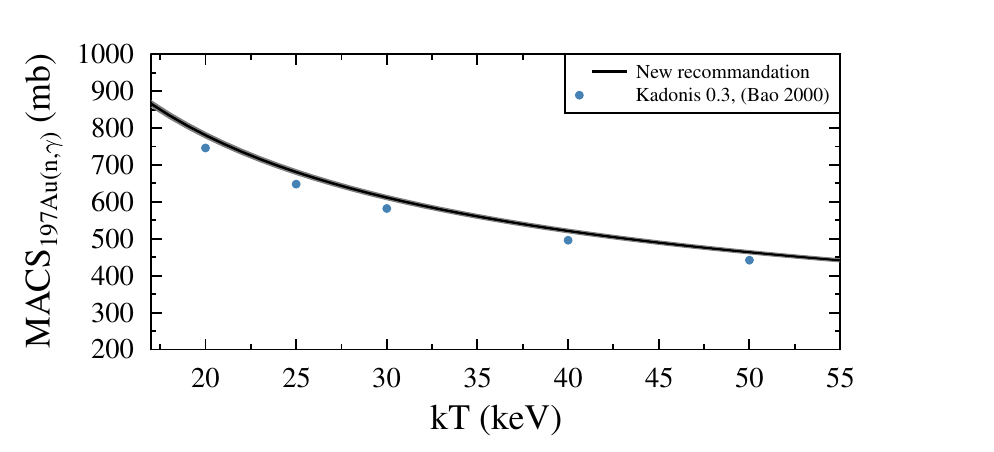}
 \caption{\label{fig:197au_comparison} New (based on \cite{LCD11,MBD14,CHO11}) and old \cite{RaK88,BBK00,DHK06}  recommendation for the $MACS$ of $^{197}$Au.}
\end{center}
\end{figure}

Figs.~\ref{fig:103rh_comparison}-\ref{fig:197au_comparison} show the results of the re-normalization for 64 isotopes separately.

\clearpage

\clearpage
 
\clearpage
 
\clearpage
 
\clearpage
 
\clearpage
 
\clearpage
 
\clearpage
 
\clearpage
 
\clearpage
 
\clearpage
 
\clearpage
 
\clearpage
 
\begin{figure}
\begin{center}
 \renewcommand{\baselinestretch}{1}
 \includegraphics{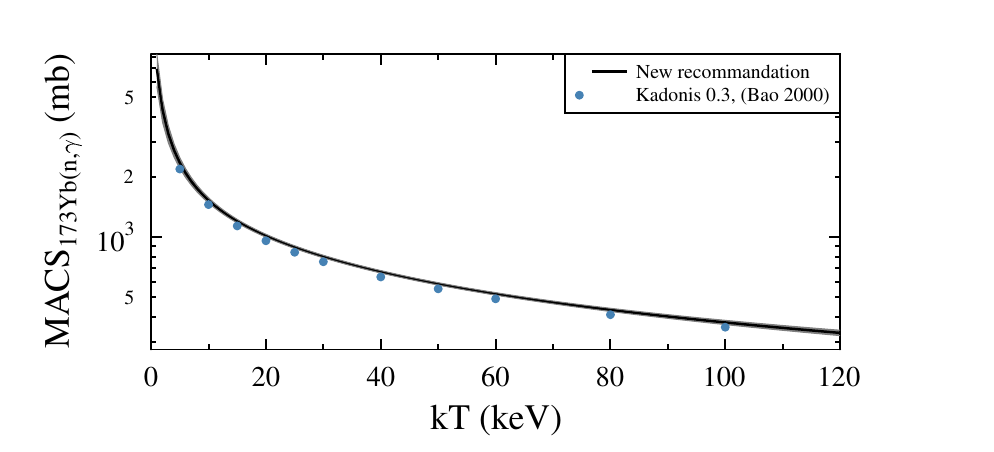}
 \includegraphics{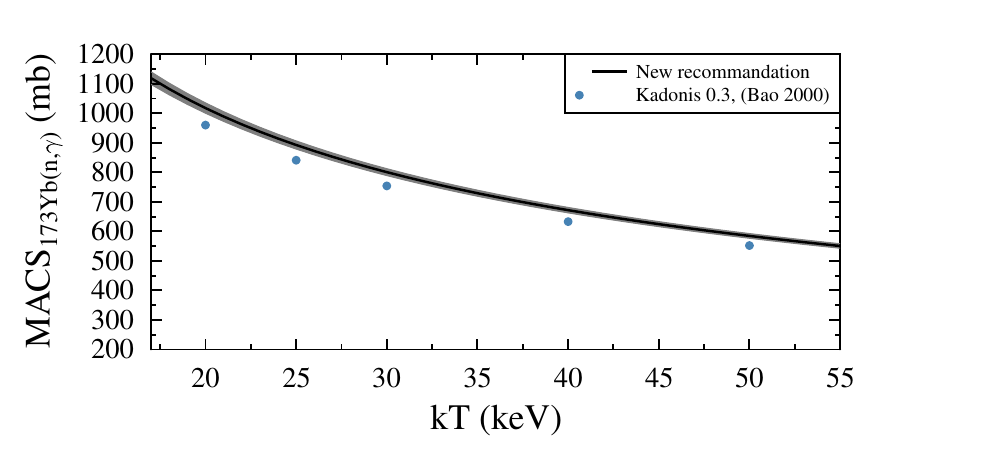}
 \caption{\label{fig:173yb_comparison} New and old \cite{WVA00a,BBK00,DHK06}  recommendation for the $MACS$ of $^{173}$Yb.}
\end{center}
\end{figure}
 
\clearpage
 
\begin{figure}
\begin{center}
 \renewcommand{\baselinestretch}{1}
 \includegraphics{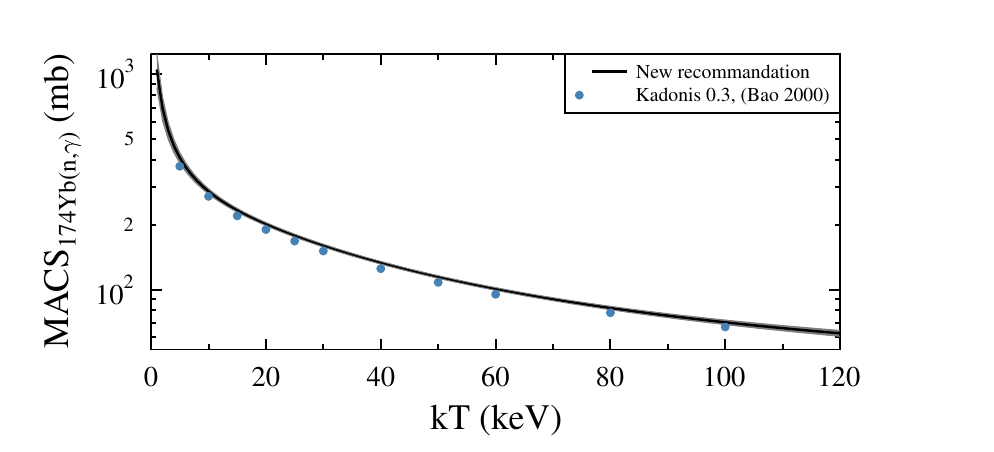}
 \includegraphics{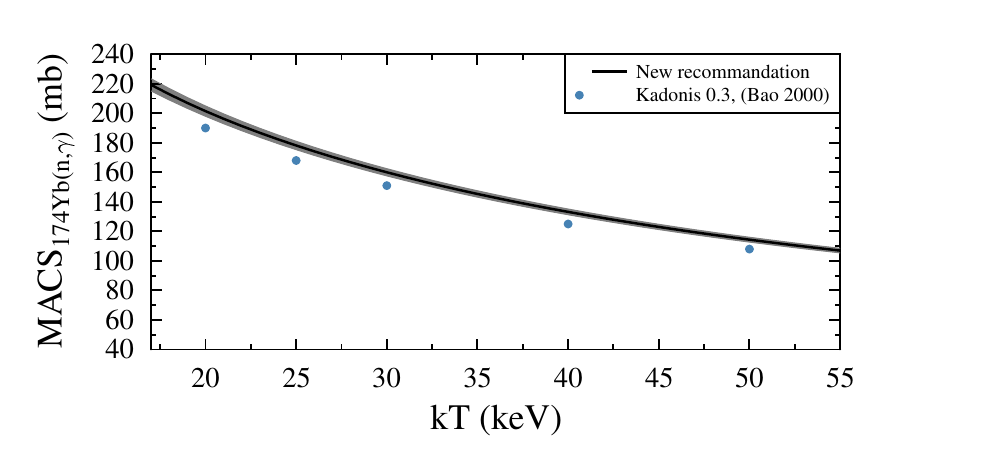}
 \caption{\label{fig:174yb_comparison} New and old \cite{WVA00a,BBK00,DHK06}  recommendation for the $MACS$ of $^{174}$Yb.}
\end{center}
\end{figure}
 
\begin{figure}
\begin{center}
 \renewcommand{\baselinestretch}{1}
 \includegraphics{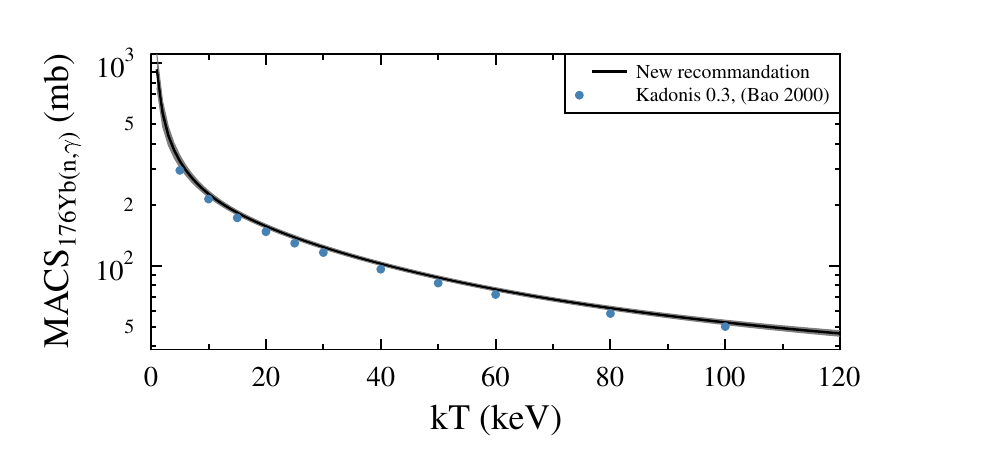}
 \includegraphics{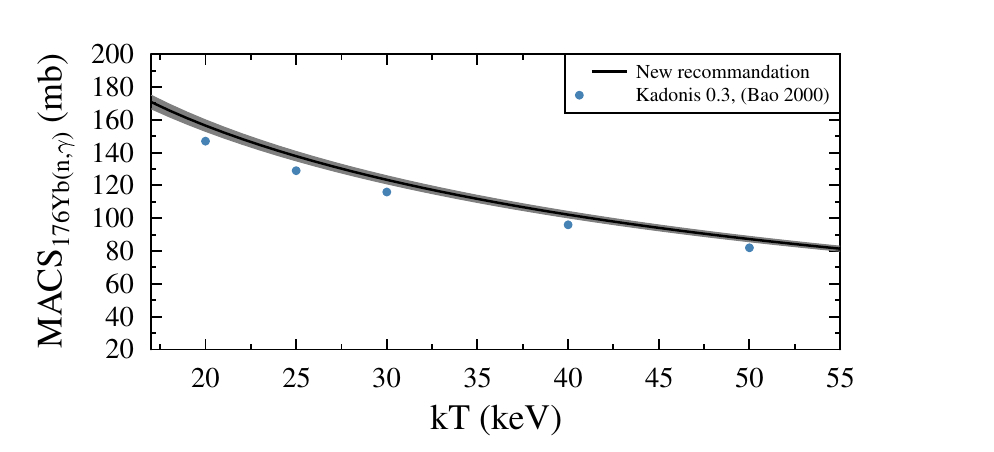}
 \caption{\label{fig:176yb_comparison} New and old \cite{WVA00a,BBK00,DHK06}  recommendation for the $MACS$ of $^{176}$Yb.}
\end{center}
\end{figure}
 
\begin{figure}
\begin{center}
 \renewcommand{\baselinestretch}{1}
 \includegraphics{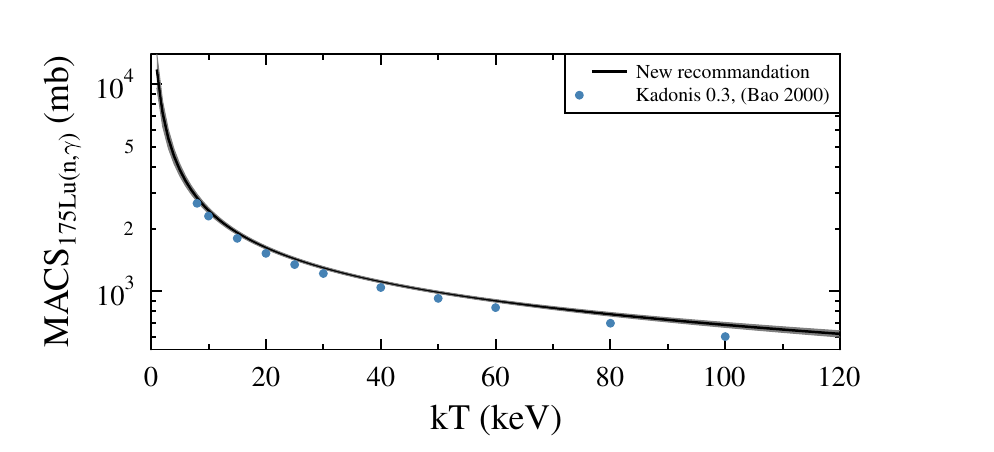}
 \includegraphics{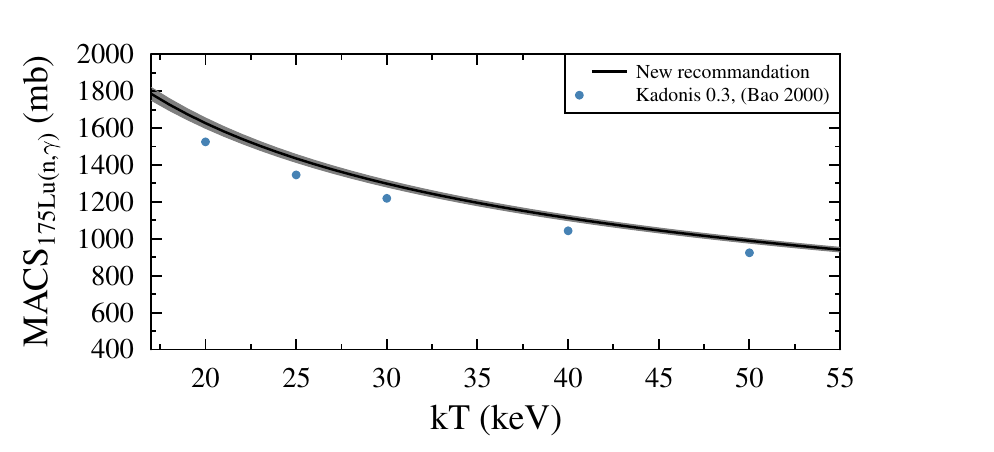}
 \caption{\label{fig:175lu_comparison} New and old \cite{WVK06b,BBK00,DHK06}  recommendation for the $MACS$ of $^{175}$Lu.}
\end{center}
\end{figure}

\begin{figure}
\begin{center}
 \renewcommand{\baselinestretch}{1}
 \includegraphics{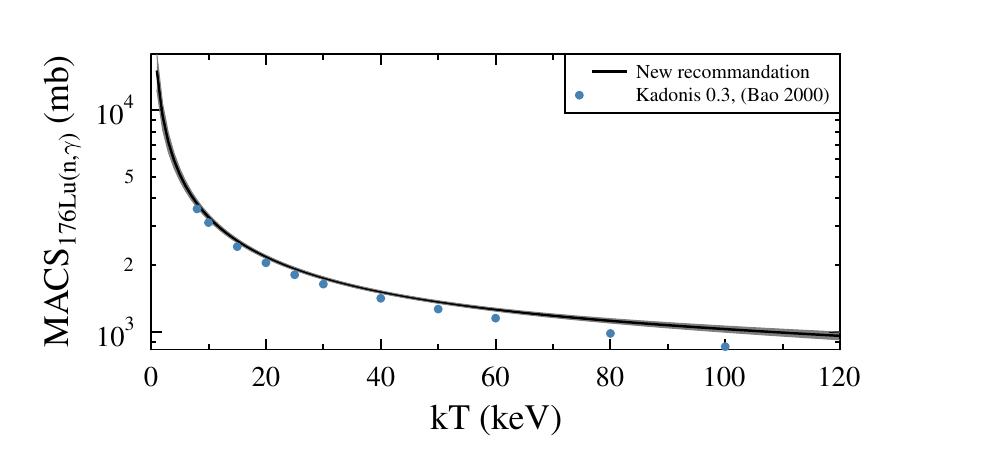}
 \includegraphics{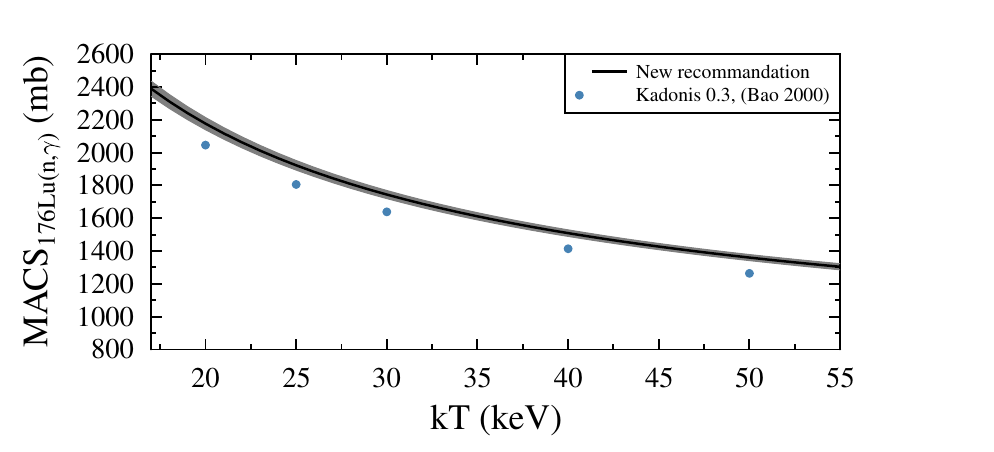}
 \caption{\label{fig:176lu_comparison} New and old \cite{WVK06b,BBK00,DHK06}  recommendation for the $MACS$ of $^{176}$Lu.}
\end{center}
\end{figure}
 
\clearpage
 
\begin{figure}
\begin{center}
 \renewcommand{\baselinestretch}{1}
 \includegraphics{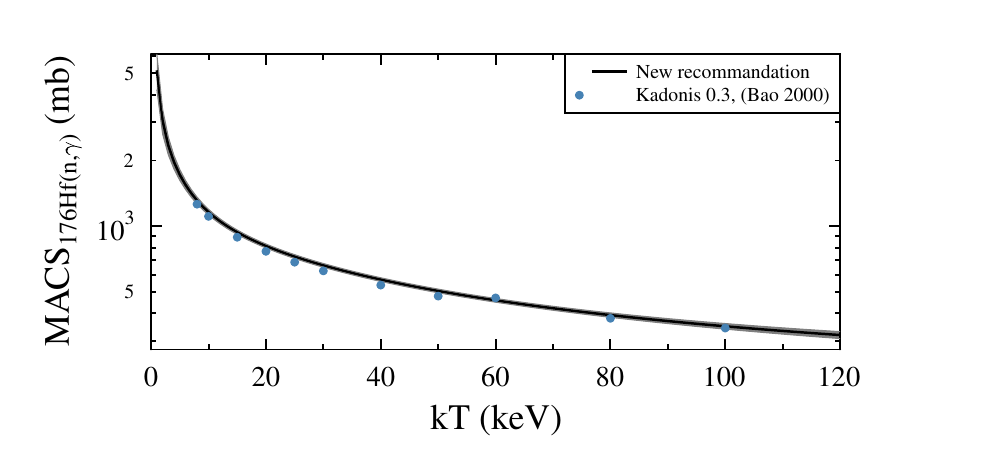}
 \includegraphics{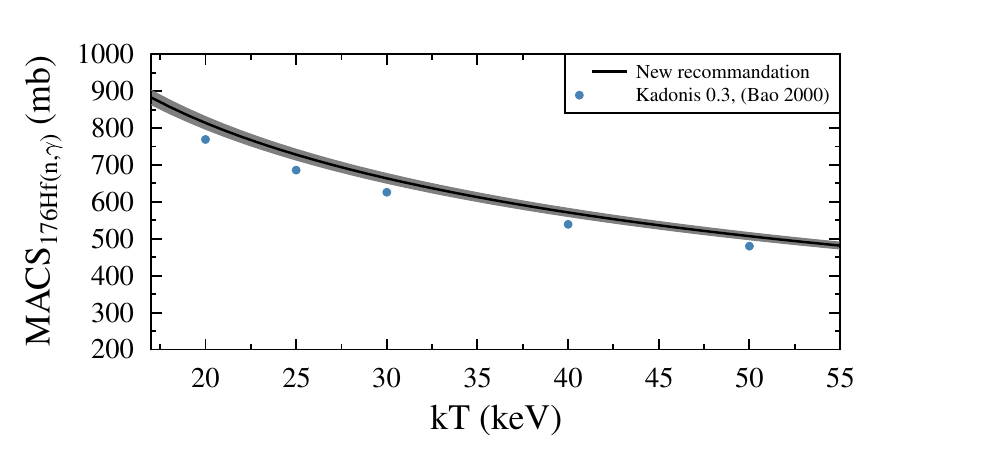}
 \caption{\label{fig:176hf_comparison} New and old \cite{WVK06c,BBK00,DHK06}  recommendation for the $MACS$ of $^{176}$Hf.}
\end{center}
\end{figure}
 
\begin{figure}
\begin{center}
 \renewcommand{\baselinestretch}{1}
 \includegraphics{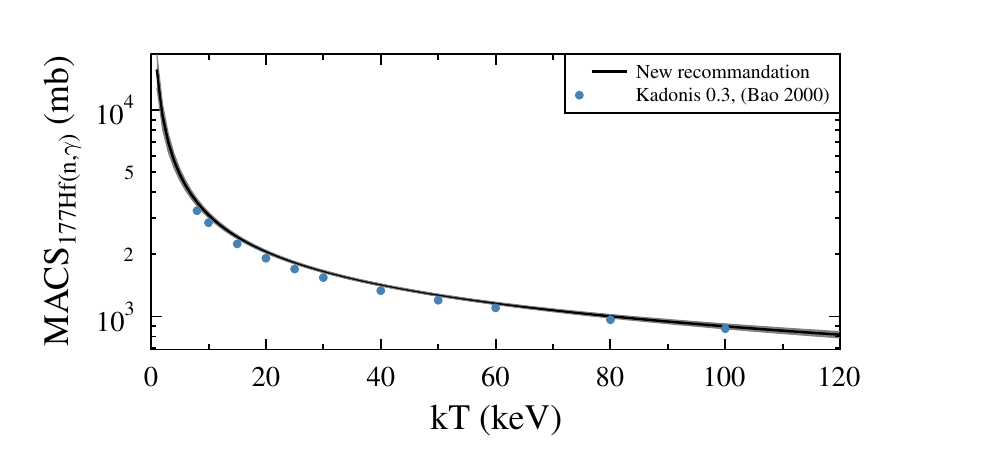}
 \includegraphics{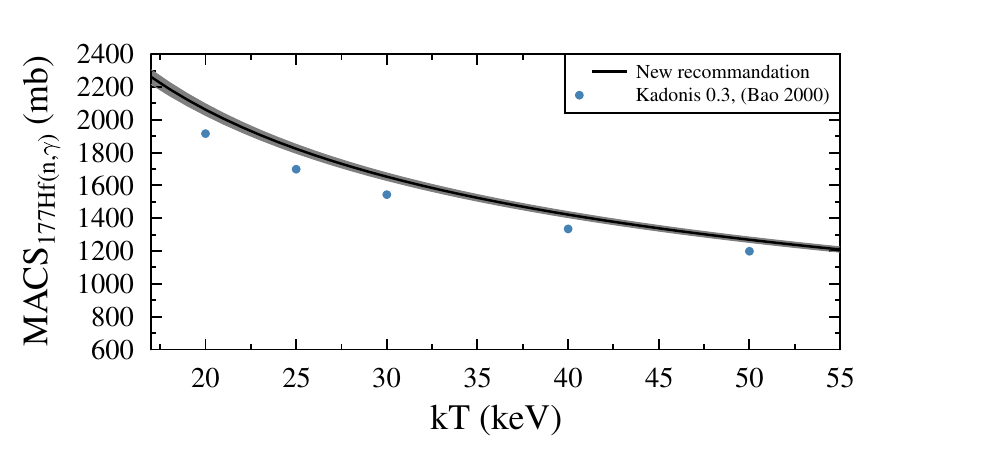}
 \caption{\label{fig:177hf_comparison} New and old \cite{WVK06c,BBK00,DHK06}  recommendation for the $MACS$ of $^{177}$Hf.}
\end{center}
\end{figure}
 
\begin{figure}
\begin{center}
 \renewcommand{\baselinestretch}{1}
 \includegraphics{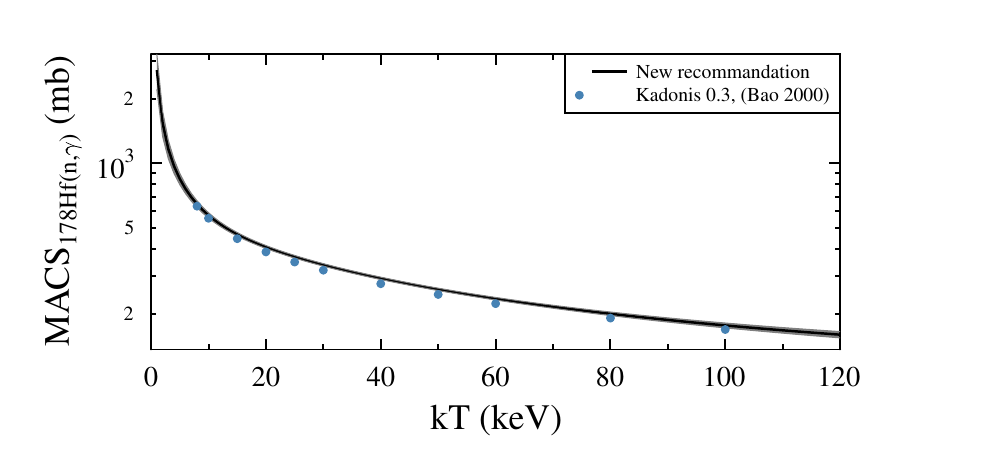}
 \includegraphics{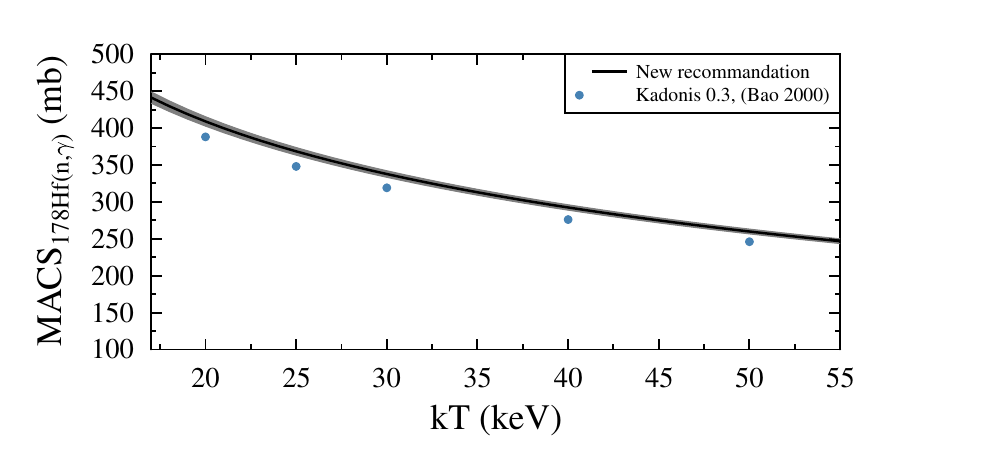}
 \caption{\label{fig:178hf_comparison} New and old \cite{WVK06c,BBK00,DHK06}  recommendation for the $MACS$ of $^{178}$Hf.}
\end{center}
\end{figure}

\begin{figure}
\begin{center}
 \renewcommand{\baselinestretch}{1}
 \includegraphics{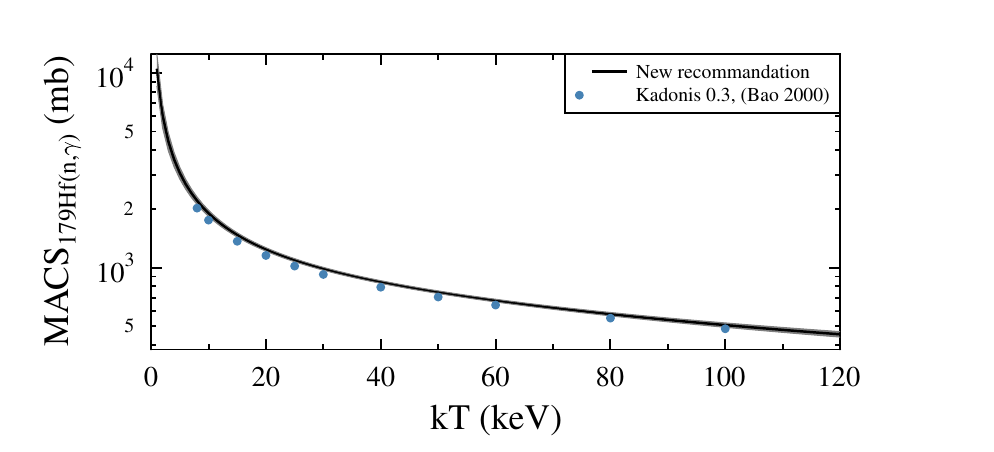}
 \includegraphics{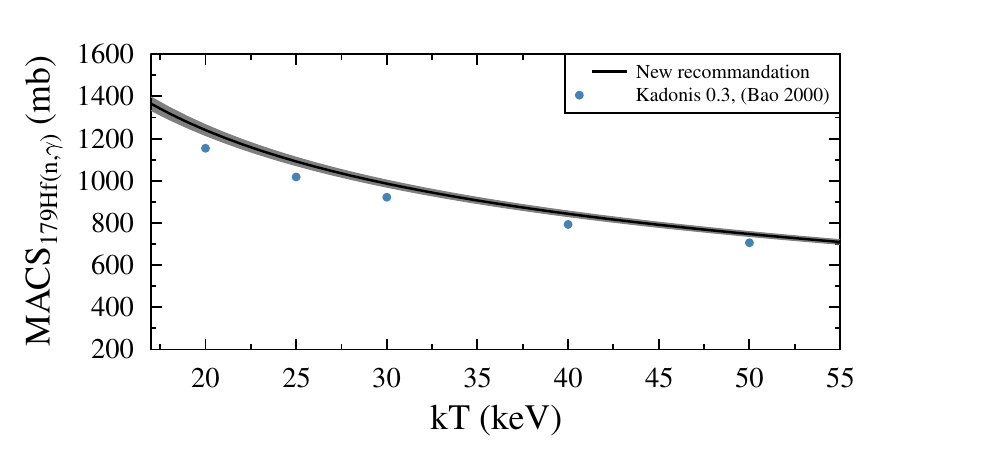}
 \caption{\label{fig:179hf_comparison} New and old \cite{WVK06c,BBK00,DHK06}  recommendation for the $MACS$ of $^{179}$Hf.}
\end{center}
\end{figure}
 
\clearpage
 
\begin{figure}
\begin{center}
 \renewcommand{\baselinestretch}{1}
 \includegraphics{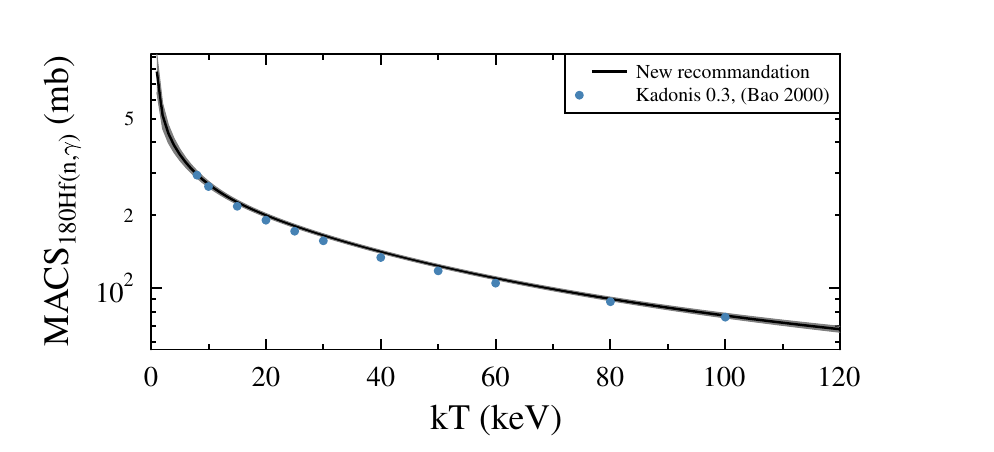}
 \includegraphics{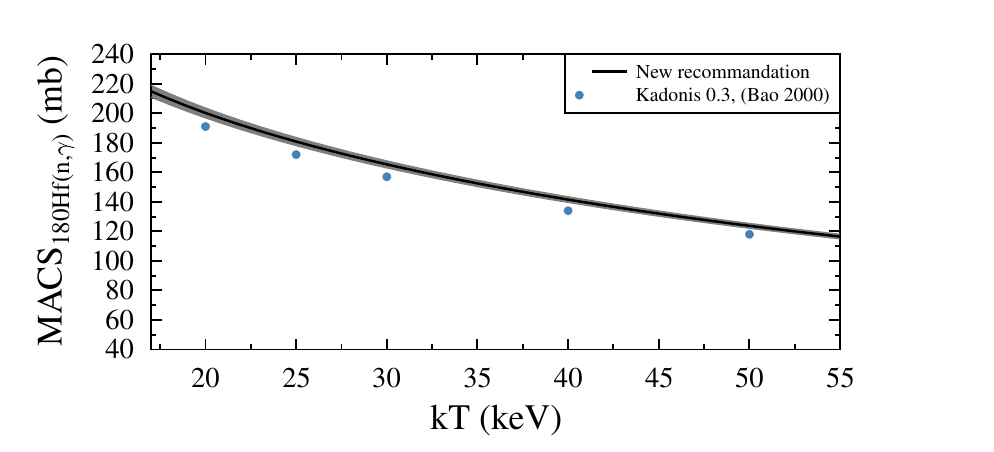}
 \caption{\label{fig:180hf_comparison} New and old \cite{WVK06c,BBK00,DHK06}  recommendation for the $MACS$ of $^{180}$Hf.}
\end{center}
\end{figure}
 
\begin{figure}
\begin{center}
 \renewcommand{\baselinestretch}{1}
 \includegraphics{180ta_MACS_log_auto}
 \includegraphics{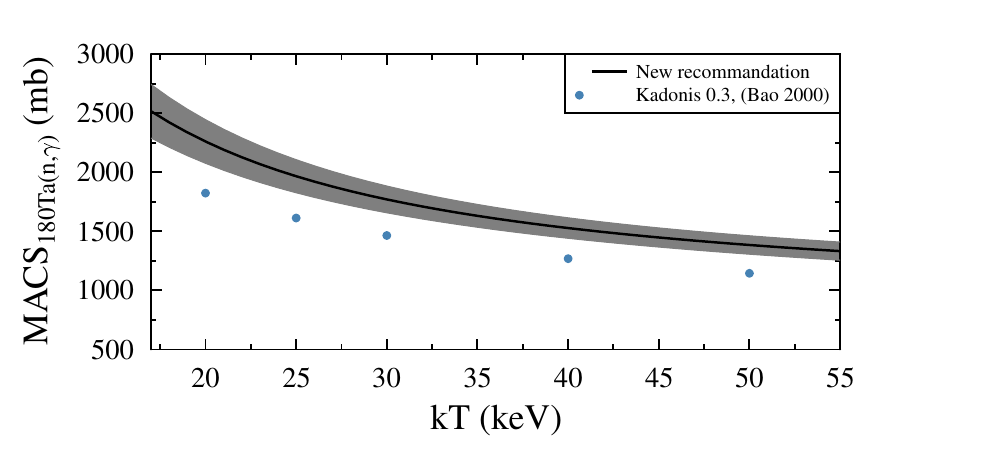}
 \caption{\label{fig:180ta_comparison} New and old \cite{WVA04,BBK00,DHK06}  recommendation for the $MACS$ of $^{180}$Ta.}
\end{center}
\end{figure}
 
\begin{figure}
\begin{center}
 \renewcommand{\baselinestretch}{1}
 \includegraphics{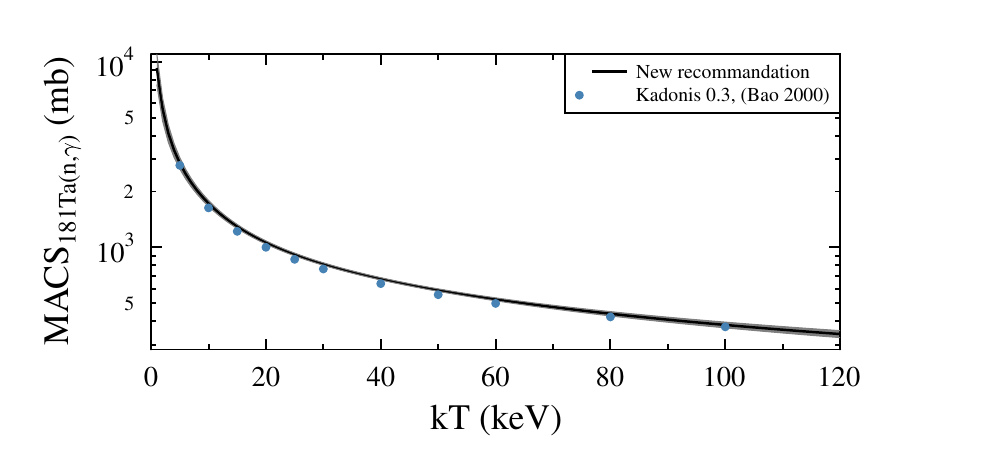}
 \includegraphics{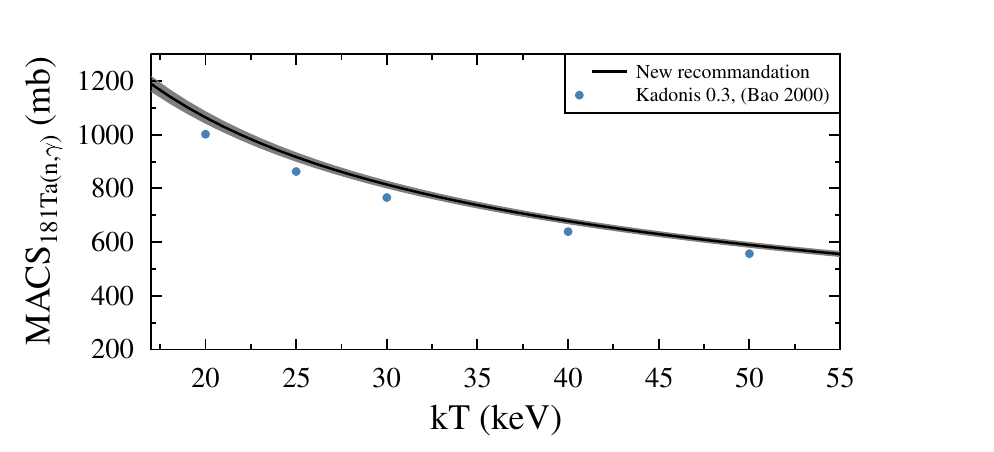}
 \caption{\label{fig:181ta_comparison} New and old \cite{WVK90,BBK00,DHK06}  recommendation for the $MACS$ of $^{181}$Ta.}
\end{center}
\end{figure}
 
\end{document}